\documentclass[a4paper,12pt]{article}
\pdfoutput=1

\usepackage{amsmath,amssymb}
\usepackage{paralist}
\usepackage{slashed}
\usepackage{graphicx}
\usepackage{subfig}
\usepackage{color}
\usepackage{fullpage}
\usepackage{multirow}
\usepackage{hyperref}
\usepackage{mcite}

\begin{document}
\begin{titlepage}

\begin{flushright}
SHEP-12-18\\
\today
\end{flushright}

\begin{center}
{ \sffamily \Large \bf 
Exploring  Drell-Yan signals from the\\
  4D Composite Higgs Model at the LHC
}
\\[8mm]
D.~Barducci$^{a}$
\footnote{E-mail: \texttt{d.barducci@soton.ac.uk}},
A.~Belyaev$^{a}$
\footnote{E-mail: \texttt{a.belyaev@soton.ac.uk}},
S.~De~Curtis$^{b}$
\footnote{E-mail: \texttt{decurtis@fi.infn.it}},
S.~Moretti$^{a}$
\footnote{E-mail: \texttt{s.moretti@soton.ac.uk}}
and
G.M.~Pruna$^{c}$
\footnote{E-mail: \texttt{giovanni$_{-}$marco.pruna@tu-dresden.de}}
\\[3mm]
{\small\it
$^a$ School of Physics and Astronomy, University of Southampton,\\
Southampton SO17 1BJ, U.K.\\[2mm]
$^b$ INFN, Sezione di Firenze,\\
Via G. Sansone 1, 50019 Sesto Fiorentino, Italy\\[2mm]
$^c$ TU Dresden, Institut ur Kern- und Teilchenphysik, \\
Zellescher Weg 19, D-01069 Dresden, Germany
}
\\[1mm]
\end{center}
\vspace*{0.5cm}

\begin{abstract}
\noindent
We study the phenomenology of Drell-Yan processes at the Large Hadron Collider  for the case of both the neutral and  charged current  channels within a recently proposed 4-Dimensional formulation of the Minimal Composite Higgs Model. We estimate the integrated and differential event rates at the CERN machine, assuming 14 TeV and data samples of ${\cal O}(100~{\rm{fb}}^{-1})$, as at lower energy and/or luminosity event rates are prohibitively small. We pay particular attention to the presence of multiple resonances in either channel, by showing that in certain region of parameter space some of these can be distinguishable and experimentally accessible in the invariant and/or transverse mass distribution, sampled in either the cross section, the forward-backward asymmetry or both. At the same time, we assess the indirect impact onto the line-shape of the emerging gauge boson resonances, both neutral and charged, of additional heavy fermionic states present in the spectrum of the model. Finally, we show how to exploit in the kinematic selection the fact that the extra neutral and charged gauge boson resonances in composite Higgs models are correlated in mass.
Such results rely on a parton level study including a statistical error analysis.
\end{abstract}

\end{titlepage}
\newpage

\section{Introduction}
\label{sec:intro}

The Drell-Yan (DY) mechanism is one of the most important probes in the search for new vector boson resonances associated to possible physics Beyond the Standard Model (BSM). It consists of di-lepton production from hadron-hadron scattering via Neutral Current (NC) or Charged Current (CC) processes. From the theoretical point of view, such a mechanism is
well under control as higher order effects from both Electro-Weak (EW) interactions and 
Quantum Chromo-Dynamics (QCD) are known up to one- and two-loop contributions, respectively (see, e.g., Ref.~\cite{Campbell:2006wx} for a review). From the experimental point of view, the directions and energies of the particles of such final state are well reconstructed in a generic detector at an hadron machine, especially when consisting of electrons or muons ($e$ or $\mu$) and/or even their related neutrinos. For all such reasons, it is clear that this class of processes is ideal also for identifying the mass of the intermediate bosons being produced and studying their properties.

Nowadays, the Large Hadron Collider (LHC) offers an unprecedented chance to test DY phenomenology in high energy 
proton-proton scattering. Therefore, we currently have the opportunity to test many choices of models with additional vector bosons (generically denoted as  $Z'$ and $W'$) in the particle spectrum, being their phenomenology of primary interest provided that the new gauge sector is considerably coupled to the SM fermions. Concerning this kind of models, the community is showing a growing interest to bottom-up approaches to extra $Z'$ and $W'$  models (see \cite{Cacciapaglia:2006pk,*Langacker:2008yv,*Salvioni:2009mt,*Accomando:2010fz,*Basso:2011na}) as well as to top-down schemes (e.g., exploiting Supersymmetry \cite{Athron:2011wu,*Athron:2012sq,*Athron:2012pw,Basso:2012gz,*O'Leary:2011yq,*Hirsch:2012kv}, little Higgs inspired \cite{Schmaltz:2005ky,*Perelstein:2005ka,*Cheng:2007bu}, based on
extra-dimensions \cite{Contino:2003ve,Agashe:2004rs,Contino:2010rs}, etc.).

In this paper, we adopt the latter approach and
 focus on the recently proposed 4-Dimensional Composite Higgs Model (4DCHM) of Ref.~\cite{DeCurtis:2011yx}. Amongst the many alternative (to the SM) EW Symmetry Breaking (EWSB) scenarios proposed over the years, the one with a  Higgs as a pseudo Nambu Goldstone Boson (PNGB) associated to the spontaneous breaking of a symmetry $G$ to $H$ can  perhaps give one of the most natural solutions to the hierarchy problem of the SM.  The idea goes back to the '80s \cite{Kaplan:1983fs,*Georgi:1984ef,*Georgi:1984af,*Dugan:1984hq}. However, one modern ingredient is the mechanism of `partial compositness', wherein
(some of) the SM gauge fields and fermions mix with new force and matter states emerging such an alternative EWSB. The simplest example, based on the symmetry pattern $SO(5)/SO(4)$, was considered in \cite{Agashe:2004rs} in the context of Randall-Sundrum scenarios. The 4D effective description of this scenario, proposed in \cite{DeCurtis:2011yx}, is a highly deconstructed version of the 5D theory and  presents several exceptional features. 
In all generality, an extreme deconstruction of  5D composite Higgs models with partial compositness leads to just two 4D sectors: the elementary one, that contains the SM structure $SU(2)_L \otimes U(1)_Y$, and the composite one, which include the extra (bosonic and fermionic) resonances. They are mixed in order to realise the partial compositness mechanism and EWSB is induced by means of a composite 4D Higgs state which is a  PNGB stemming from the latter. 
From the phenomenological point of view,  the only degrees of freedom which might be accessible at the LHC are represented by the lowest lying resonances described by the extreme deconstruction in the 4DCHM of Ref.~\cite{DeCurtis:2011yx}. 
Here  a minimal choice for the fermionic sector is assumed, as one includes two $SO(5)$ multiplets of resonances per family in the composite sector.  This is indeed the minimal matter content that allows for a finite Higgs potential calculable via the Coleman-Weinberg technique.  From the explicit expression of the Higgs potential  one can extract  the Higgs Vacuum Expectation Value (VEV) and mass in terms of the model parameters. The peculiarity of this model is that, for a natural choice of the free parameters, both in  the gauge and fermion sectors, the Higgs mass value that one obtains is compatibile with the 
most recent LHC results by the ATLAS \cite{:2012gk} and CMS \cite{:2012gu} experiments.

As intimated, in the composite sector of the 4DCHM, besides the Higgs state and extra heavy fermions (with both standard and exotic quantum numbers), 
one also has extra composite spin-1 resonances associated with the $SO(5)\otimes U(1)_X$ local symmetry: five $Z'$ states and three  $W'$ states. These objects are weakly yet sizably coupled to the first and second generations of fermionic matter and this makes the 4DCHM an excellent candidate for a phenomenological analysis of DY processes at the LHC.

For the sake of completeness, we also remark that analogous EWSB scenarios have been already studied with respect to DY processes (see, e.g., Refs.~\cite{Agashe:2007ki} and \cite{Agashe:2008jb}, for the NC and CC case, respectively).  However, the purpose of this paper is to surpass previous literature along four main directions:
\begin{enumerate}
\item we study the effects of the opening of fermionic decay channels onto the line shape of the DY resonances, then we prove that this allows us to extrapolate information on the mass and coupling spectrum of the fermionic sector of the model;
\item we consider the case of nearly degenerate gauge boson resonances and show that two of these can be resolved (albeit  limited to the NC channel) over a large portion of the 4DCHM parameter space where decays of the additional gauge bosons into pairs of heavy fermions are forbidden;
\item we emphasise the role of the di-lepton invariant mass in investigating the above phenomenology not only in the usual terms of the cross section distribution but also in terms of the Forward-Backward Asymmetry (AFB) one;
\item we exploit correlated searches in the CC and NC channels and we show that kinematic information can be used to improve the overall event selection, as the mass of the additional neutral and charged gauge bosons of the 4DCHM are correlated by the underlying gauge symmetry.
\end{enumerate}
 
The paper is structured as follows: in the next Section we describe the model, in Section~\ref{sec:implement} we illustrate its implementation in the numerical tools used for the analysis, in Section~\ref{sec:benchmarks} we study the parameter space of the model and we select some benchmark points, in Section~\ref{sec:results} we present our results and in Section~\ref{sec:summa} we state our conclusions.

\section{The 4DCHM}
\label{sec:4DHCM}
Let us  recall the main characteristics of the 4DCHM introduced in \cite{DeCurtis:2011yx}, to which we refer for further details througout this section.
The 4DCHM is an effective low-energy Lagrangian approximation of the deconstructed Minimal Composite Higgs Model (MCHM) of \cite{Agashe:2004rs}  based on the coset $SO(5)/SO(4)$. It represents the framework where to study, in a computable way, the effects of the lowest lying resonances, both bosonic and fermionic, beyond the leading order chiral Lagrangian approach (this was also done in \cite{Panico:2011pw}
with a different construction). As intimated already, the 4DHCM  
can be schematised in two sectors, the elementary and the composite one, arising from the extreme deconstruction of the  5D theory\footnote{This follows the lines of \cite{Contino:2006nn} where, however, the full gauge/Goldstone boson
 structure of the theory is not incorporated.}.
The reason for this two-site truncation is that it  describes only the new states which might be accessible to the LHC while capturing all the relevant features of the composite Higgs models with the Higgs state as a PNGB. The symmetry structure is $SO(5)/SO(4)$ with four Goldstone Bosons (GBs) in the vector representation of $SO(4)$, containing the Higgs boson. The breaking is parametrised by the Vacuum Expectation Value (VEV) of a vector of $SO(5)$.

\subsection{Gauge sector}

\begin{figure}[!h]
\begin{center}
\includegraphics[width=10cm]{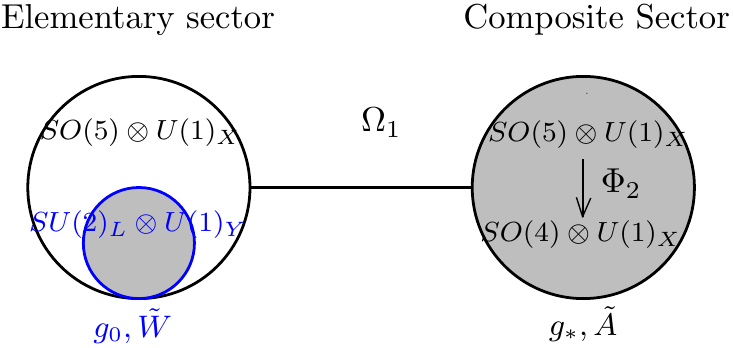}
\end{center}
\caption{Gauge sector of the 4DCHM. The first site represents the elementary sector, the second  the composite sector with $SO(5)\otimes U(1)_X$ gauge fields. The grey circles correspond to gauge symmetries.}
\label{fig:gauge}
\end{figure}
The elementary sector of the 4DCHM is associated with the SM gauge symmetry $SU(2)_L \otimes U(1)_Y$ while the composite sector  with the gauge symmetry $SO(5)\otimes U(1)_X$ spontaneously broken to $SO(4)\otimes U(1)_X$, as shown in Fig. \ref{fig:gauge}. 
Indicated in a generalised way with $(g_0,\tilde{W})$ and $(g_*,\tilde{A})$, respectively, are the couplings and fields of the elementary and composite site\footnote{With $g_0$ and $\tilde{W}$ we indicate the couplings and fields of $SU(2)_L$ and $U(1)_Y$, so that $g_0=  \{ g_0,g_{0Y} \}$ and $\tilde{W}= \{ \tilde{W_i},\tilde{Y} \}, i=1,2,3$.
In the same way with $\tilde{A}$ we indicate all the gauge fields of $SO(5)\otimes U(1)_X$.}. The gauge Lagrangian for the 4DCHM turns out to be:
\begin{equation}
\label{eq:laggauge}
\mathcal{L}_{gauge}=\frac{f_1^2}{4}Tr|D_{\mu}\Omega_1|^2+\frac{f_2^2}{2}(D_{\mu}\Phi_2)(D_{\mu}\Phi_2)^T-\frac{1}{4}\rho_{\mu\nu}^A\rho^{A\mu\nu}-\frac{1}{4}F_{\mu\nu}^A F^{A\mu\nu}
\end{equation}

where 
\begin{itemize}
\item the covariant derivatives are defined by:
\begin{eqnarray}
D^{\mu}\Omega_1&=&\partial^{\mu}\Omega_1-i g_{0}\tilde{W}\Omega_1+i g_*\Omega_1\tilde{A},\\
D_{\mu}\Phi_2&=&\partial_{\mu}\Phi_2-i g_*\tilde{A}\Phi_2;
\end{eqnarray}
\item the fields $\Omega_{n}$, $n=1,2$,  are  link fields responsible for the symmetry breaking
and in the unitary gauge they are given by:
\begin{equation}
\Omega_n=\textbf{1}+i\frac{s_n}{h}\Pi+\frac{c_n-1}{h^2}\Pi^2,\quad s_n=\sin(f h/f_n^2),\quad c_n=\cos(f h/f_n^2),\quad   h=\sqrt{h^{\hat{a}}h^{\hat{a}}}
\end{equation}
where $\Pi=\sqrt{2}h^{\hat{a}}T^{\hat{a}}$  is the GB matrix with $T^{\hat{a}}$ the broken generators ($\hat a=1,2,3,4$),
\begin{equation}
\Pi=\sqrt{2}h^{\hat{a}}T^{\hat{a}}=-i\left( \begin{array}{cc}
0_4 & \textbf{h}  \\
-\textbf{h}^T & 0  \\ 
\end{array}\right),\quad \textbf{h}^T=\left( h_1, h_2, h_3, h_4 \right),
\end{equation} 
wherein $\textbf{h}$ is related to the usual Higgs doublet, $H$, via the relation
\begin{equation}
H=\frac{1}{\sqrt{2}}\left( \begin{array}{c}
-i h_1 - h_2 \\
-i h_3 + h_4  \\ 
\end{array}\right),
\end{equation}
where the
$f_{i}$'s are the link coupling constants and  the strong sector scale $f$ is given by
\begin{equation}
\sum_{n=1}^2\frac{1}{f_n^2}=\frac{1}{f^2};
\end{equation}
\item the field $\Phi_2$ is a vector of $SO(5)$  that describes the spontaneous symmetry breaking of $SO(5)\otimes U(1)_X \rightarrow SO(4)\otimes U(1)_X$ and is defined as
\begin{equation}
\Phi_2=\phi_0\Omega_2^T \quad \text{where} \quad \phi_0^i=\delta^{i5};
\end{equation}
\item the last two terms are the kinetic terms of  the composite and elementary gauge fields, respectively.
\end{itemize}

\subsection{Fermionic sector}

\begin{figure}[!h]
\begin{center}
\includegraphics[width=5cm]{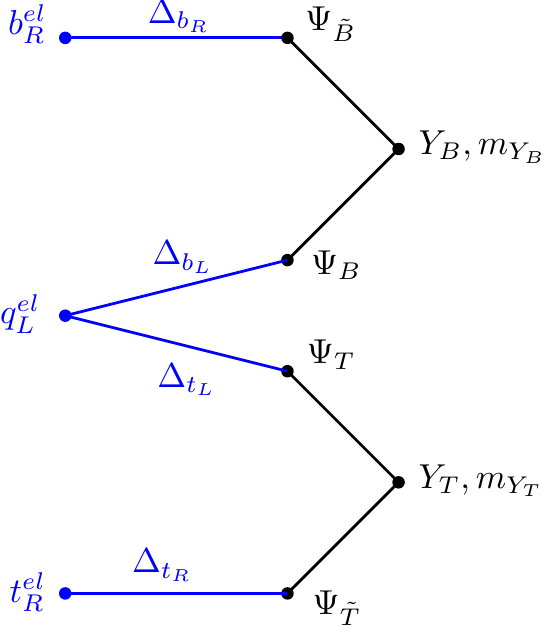}
\end{center}
\caption{Fermionic sector of the 4DCHM.  The elementary sector is on the left,   the composite sector with the fermions embedded in fundamental representations of $SO(5)\otimes U(1)_X$ is on the right. Symbolically showed are the mixing and Yukawa terms.}
\label{fig:ferm}
\end{figure}

In the 4DCHM the SM fermions  couple to fermionic operators in the {\bf 5} of $SO(5)$. This choice of representation is a realistic scenario compatible with precision EW measurements and represents a discretisation to two sites of the model in \cite{Contino:2006qr} (see Fig. \ref{fig:ferm}).
The new heavy fermions are embedded in the fundamental representation of $SO(5)\otimes U(1)_X$: the spectrum contains four {\bf 5} representations
 indicated with $\Psi_{T,\tilde{T}/B,\tilde{B}}$ in the composite top/bottom sector, respectively.
The SM third generation quarks, both for the left-handed doublet, $q_L^{el}$, and the two right-handed singlets, $b_R^{el}$ and $t_R^{el}$, are embedded in an incomplete representation of $SO(5)\otimes U(1)_X$ in such a way that their correct quantum numbers under $SU(2)_L\otimes U(1)_X$ are reproduced via the relation $Y=T^{3R}+X$. Here,
for illustration purposes, we are considering only the third generation quarks, which are relevant for the computation of the effective potential and for the upcoming phenomenological analysis.
The fermionic Lagrangian of the 4DCHM considered in \cite{DeCurtis:2011yx} is (for simplicity we take $m_T=m_{\tilde{T}}=m_B=m_{\tilde{B}}=m_*$):
\begin{equation}
\label{eq:lagferm}
\begin{split}
\mathcal{L}_{fermions}&=L_{fermions}^{el}+ (\Delta_{t_L}\bar{q}^{el}_L\Omega_1\Psi_T+\Delta_{t_R}\bar{t}^{el}_R\Omega_1\Psi_{\tilde{T}}+h.c.)\\
&+\bar{\Psi}_T(i\hat{D}^{\tilde{A}}-m_*)\Psi_T+\bar{\Psi}_{\tilde{T}}(i\hat{D}^{\tilde{A}}-m_*)\Psi_{\tilde{T}}\\
&-(Y_T\bar{\Psi}_{T,L}\Phi_2^T\Phi_2\Psi_{\tilde{T},R}+m_{Y_T}\bar{\Psi}_{T,L}\Psi_{\tilde{T},R}+h.c.)\\
&+(T\rightarrow B),
\end{split}
\end{equation}
where with $D^{\tilde{A}}$ we indicate the covariant derivative related to the composite gauge fields $\tilde{A}$,
$\Delta_{t_L,t_R,b_L,b_R}$ are the mixing parameters relating the elementary and the composite sector whilst $Y_{T,B}$ and $m_{Y_{T,B}}$ are the Yukawas of the composite sector.

{{For the analytical expressions of the masses of the gauge bosons, their couplings to the quarks and leptons and also the masses of top, bottom and heavy fermionic resonances, we refer to Appendix A.}}

\subsection{Higgs sector}
In the 4DCHM the Higgs boson arises as a PNGB through the breaking of $SO(5)\otimes U(1)_X$ to $SO(4)\otimes U(1)_X$.
The gauging of the EW interaction breaks, weakly but explicitly, the strong sector symmetry, as a consequence the Higgs is only approximately a GB and a potential is generated starting at one-loop.
With   our choice of the fermionic sector, namely, with complete multiplets of the new fermions, the potential turns out to be finite \cite{DeCurtis:2011yx}. From the location of the minimum of the potential one extracts the expression for the mass of the Higgs boson, $m_H$, and its VEV, $\langle h \rangle$,  in terms of the parameters of the model.
Also, by defining:
\begin{equation}
\Omega_n=\textbf{1}+\delta \Omega_n,
\end{equation}
and using eqs. (\ref{eq:laggauge}) and (\ref{eq:lagferm}),  we can extract the Higgs interactions with the gauge and fermion fields. They are described by the following Lagrangians:
\begin{equation}
\begin{split}
\mathcal{L}_{gauge,H}&=-\frac{f_1^2}{2} g_0 g_* Tr\left[\tilde{W}\delta\Omega_1 \tilde{A}+\tilde{W}\tilde{A}\delta\Omega_1^T+\tilde{W}\delta\Omega_1 \tilde{A} \delta\Omega_1^T   \right]\\
&+\frac{f_2^2}{2} g_*^2\phi_0^T\delta\Omega_2^T\tilde{A}\tilde{A}\phi_0+\frac{f_2^2}{2}g_*^2\phi_0^T\tilde{A}\tilde{A}\delta\Omega_2\phi_0\\
&+\frac{f_2^2}{2}g_*^2\phi_0^T\delta\Omega_2^T\tilde{A}\tilde{A}\delta\Omega_2\phi_0,
\end{split}
\end{equation}

\begin{equation}
\begin{split}
\mathcal{L}_{ferm,H}&=\Delta_{t_L}\bar{q}^{el}_L\delta\Omega_1\Psi_T+\Delta_{t_R}\bar{t}^{el}_R\delta\Omega_1\Psi_{\tilde{T}}\\
&-Y_T\bar{\Psi}_{T,L}(\phi_0^T\phi_0\delta\Omega_2^T+\delta\Omega_2\phi_0\phi_0^T+\delta\Omega_2\phi_0^T\phi_0\delta\Omega_2^T)\Psi_{\tilde{T},R}\\
&+(T\rightarrow B)+h.c.
\end{split}
\end{equation}

The theoretical setup of the 4DCHM allows one to compute a finite Higgs potential with the minimal number of degrees of freedom (in this sense it is the most economical one in the class of composite Higgs model), the latter being dictated by the symmetries of the theory.
{{From the expression of the potential it is possible to extract both the VEV and the mass of the Higgs boson as shown in \cite{DeCurtis:2011yx} and \cite{Redi:2012ha}}}. For this reason, we will adopt the effective description of the 4DCHM for our phenomenological analysis of DY processes at the LHC.

\section{Model implementation}
\label{sec:implement}

The 4DCHM presents a large number of new particles of both kinds,  bosons and fermions. In the following\footnote{As clear from our previous Section, hereafter we will distinguish between SM first and second quark families and the third one, the latter being formed by quasi-composite particles.} we summarise the particle content.
\begin{itemize}
\item{Standard Model  particles}
\begin{itemize}
\item SM leptons: $l_{1,2,3}$ and $\nu_{1,2,3}$ (the labels 1, 2, 3 correspond to the flavours $e,\mu$, $\tau$, respectively).
\item SM light quarks: $u_{1,2}$ and $d_{1,2}$ (the labels 1,2 correspond to the flavours $u$, $c$ and $d$, $s$,
respectively).
\item SM heavy quarks: $t$ and $b$.
\item SM gauge bosons: $\gamma$, $Z^0$ and $W$.
\end{itemize}
\item{New particles\footnote{An increasing number labelling the 4DCHM bosonic and fermionic states corresponds to their increasing masses.}}
\begin{itemize}
\item Neutral gauge bosons: $Z_i$ (with $i=1, ... 5$).
\item Charged gauge bosons: $W_i$ (with $i=1, ... 3$).
\item Charged $+2/3$ fermions: $T_i$  (with $i=1, ...8$).
\item Charged $-1/3$ fermions: $B_i$ (with $i=1,...8$).
\item Charged $+5/3$ fermions: $\tilde{T}_{i}$ (with $i=1,2$).
\item Charged $-4/3$ fermions: $\tilde{B}_{i}$ (with $i=1,2$).
\item The Higgs boson: $H$.
\end{itemize}
\end{itemize}

We have used the LanHEP package~\cite{Semenov:2010qt}
to automatically derive Feynman rules for  the
4DCHM in CalcHEP format~\cite{Pukhov:1999gg,Belyaev:2012qa}.
Because of the elaborate off-diagonal structure of the mass-matrices of the gauge fields, we have made use of the SLHAplus
library \cite{Belanger:2010st}.
The correspondence between 4DCHM particle names in this paper and their implementation in the CalcHEP model is presented in Tab.~\ref{table:calchpar}.
\begin{table}[htb]
\begin{center}
\begin{tabular}{|l|l|l|l|}
\hline
Model &CalcHEP\\
\hline
$l_1$,$l_2$,$l_3$ & e1,e2,e3\\
$\nu_1$,$\nu_2$,$\nu_3$ & n1,n2,n3\\
$u_1$,$u_2$,$d_1$,$d_2$ & u1,u2,d1,d2\\
$t$,$b$ & t1,b1\\
$\gamma$,$Z^0$,$W^+$ &Z1,Z2,W1+ \\
$Z_1$,...,$Z_5$& Z3,...,Z7\\
$W_1^{+}$,$W_2^{+}$,$W_3^{+}$ & W2+,W3+,W4+\\
$T_1$,...,$T_8$&t2,...,t9\\
$B_1$,...,$B_8$&b2,...,b9\\
$\tilde{T}_1$,$\tilde{T}_2$ & nn1,nn2\\
$\tilde{B}_1$,$\tilde{B}_2$ & mm1,mm2\\
$H$ & H\\
\hline
\end{tabular}
\end{center}
\caption{The correspondence between 4DCHM particle names and their implementation in the CalcHEP implementation.
In the CalcHEP notation capital letters for fermions corresponds to antiparticles.}
\label{table:calchpar}
\end{table}
The parameter  space of the 4DCHM, as defined in Sect. 2, is defined in terms of 13 variables. Namely:
\begin{equation}
\label{eq:parpaper}
f,\; g_*,\; g_0,\; g_{0Y},\; m_*,\; \Delta_{t_L},\; \Delta_{t_R},\; Y_T,\; M_{Y_T},\; \Delta_{b_L},\; \Delta_{b_R},\; Y_B,\; M_{Y_B}.
\end{equation}
The notation of these parameters in the CalcHEP implementation is given in Tab.~\ref{table:calchpar1}.
\begin{table}[h!]
\begin{center}
\begin{tabular}{l l||l| l|}
\hline
\multicolumn{1}{|c|}{Model} & CalcHEP& Model &CalcHEP \\ 
\hline
\multicolumn{1}{|c|}{$f$} 	 			 &f	     & $m_*$ & mm\\
\multicolumn{1}{|c|}{$g_*$}	 			 &g		 &$\Delta_{t_L}$ &DTL \\
\multicolumn{1}{|c|}{$g_0$}  			 &g0 	 &$\Delta_{t_R}$ &DTR  \\
\multicolumn{1}{|c|}{$g_{0Y}$}  			 &g0y 	 &$Y_T$		  &YT \\
\multicolumn{1}{|c|}{$\langle h \rangle$} & h	 &$M_{Y_T}$ 	  &MYT 	\\
\cline{1-2}
& &$\Delta_{b_L}$			  &DBL \\
 & &$\Delta_{b_R}$&DBR\\
 & &$Y_B$ &YB\\
 & &$M_{Y_B}$ &MYB\\
 & & $m_H$ &MH\\
\cline{3-4}
\end{tabular}
\end{center}
\caption{The notation of the 4DCHM  parameters in the CalcHEP implementation.}
\label{table:calchpar1}
\end{table}
The  4DCHM  will be publicly available on the High Energy Physics Model Data-Base (HEPMDB)~\cite{Brooijmans:2012yi}
at https://hepmdb.soton.ac.uk/ by November 2012 under the ``4DCHM" name.

We will use as input the following physical quantities: $e$, $M_Z$, $G_F$, $m_{t}$, $m_{b}$, $m_{H}$.
In order to analyse the parameter space of the 4DCHM in presence of such constraints, we have written a standalone Mathematica program 
 \cite{mathematica} 
that is able to sort allowed points  by varying the model parameters in a fixed range.
Specifically, we have considered $f$ (the strong coupling scale) and $g_*$ (the gauge coupling constant of the extra gauge group) as free parameter while we have scanned over $m_*, \Delta_{t_L}, \Delta_{t_R}, Y_T, M_{Y_T}, \Delta_{b_L}, \Delta_{b_R}, Y_B$ and  $M_{Y_B}$.

We have accounted for the above constraints in the following way:
from the gauge sector, by diagonalising the charged and neutral mass matrix, we have constrained the value of $g_0,\; g_{0Y}$ and $\langle h \rangle$ with respect to the electric charge, the mass of the $Z$ boson and the value of the Fermi constant leaving, as intimated, $f$ and $g_*$  as free  parameters.
In the fermionic sector, due to the large number of parameters, we have performed a random scan, with the value of $f$ and $g_*$ chosen from the gauge sector, over $m_*, \Delta_{t_L}, \Delta_{t_R}, Y_T, M_{Y_T}, \Delta_{b_L}, \Delta_{b_R}, Y_B$ and  $M_{Y_B}$, imposing to reconstruct the mass of the top and bottom quarks and of the Higgs
boson.  Namely, by taking into account the $t$-quark data from LEP, SLC, Tevatron and LHC  we have required 165 GeV $\leq m_t\leq$ 175 GeV on the top quark running mass, 2 GeV $\leq m_b\leq$ 6 GeV on the bottom quark running mass and we have assumed a  conservative interval 124 GeV $\leq m_H\leq 126$ GeV for the Higgs mass,
 which is compliant with the observation of a SM-like Higgs boson as recently claimed by the ATLAS \cite{:2012gk} and CMS \cite{:2012gu} experiments\footnote{Tevatron \cite{Aaltonen:2012qt} has also reported results that are consistent with the LHC observations in this respect.}.

The gauge resonance masses, couplings and widths for the ensuing DY analysis have been computed using sets of parameters generated in the same way. The calculation of the masses, couplings and widths has of course been double-checked between CalcHEP and the Mathematica program.

As for  event generation, the codes exploited for our study of the LHC signatures is based on helicity amplitudes, defined through the HELAS subroutines~\cite{Murayama:1992gi} and assembled by means of MadGraph~\cite{Stelzer:1994ta}, which has been validated against CalcHEP.  The Matrix Elements (MEs) generated account for all off-shellness effects of the particles involved. Two different phase space implemetations were used, an `ad-hoc one' (based on Metropolis \cite{Kharraziha:1999iw}) and a `blind one' based on RAMBO \cite{Kleiss:1985gy}, checked one against the other. The latter was adopted eventually, as it proved the most unbiased one in sampling the multiple resonances existing in each (neutral and charged current) DY channel. Further, VEGAS~\cite{Lepage:1977sw,*Lepage:1980dq} was finally used for the multi-dimensional numerical integrations. All these additional subroutines were also validated against CalcHEP outputs. 

The MEs have been computed at Leading Order (LO). Clearly, in the LHC environment, QCD corrections are not
negligible and associated scale uncertainties may impact on the dynamics of $Z'$ and $W'$ production and decay
(see Ref. \cite{Adam:2008pc,*Adam:2008ge} for the case of the SM $Z$ and $W$ channels). In fact, EW corrections
may also be relevant \cite{Balossini:2007zzb,Balossini:2009sa}. However, the treatment we are adopting here
of the two DY channels is such that real radiation of gluons/photons would be treated inclusively (i.e., no selection is
enforced here that relies on the gluon/photon dynamics), so that we do not expect such QCD effects to impact on the distributions
that we will be considering, neither those of the cross sections nor those of the asymmetries, apart from an
overall rescaling. The latter, in particular, when implemented at large invariant/transverse mass, is affected by a 
residual uncertainty of 5\% at the most \cite{Balossini:2007zzb,Balossini:2009sa}.

The Parton Distribution Functions (PDFs) used were CTEQ5L~\cite{Lai:1999wy}, with factorisation/renormalisation scale set to $Q=\mu=\sqrt{\hat{s}}$.
(We have verified that later PDF sets do not generate any significant difference in the results we are going to 
present\footnote{Furthermore, we have estimated the theoretical uncertainty (at NLO) due to the PDFs by adopting 
NNPDF sets \cite{Ball:2011uy}, which yielded a 10\% effect at the most, rather independent of the $Z'$ and $W'$ masses involved and with negligible impact onto the shape of the differential distributions presented.}.)
 Initial state quarks have been taken as massless, just like the final state leptons and neutrinos. 

Furthermore we have checked that our 4DCHM parameter choices, for any sorted point, are compatible with  LHC direct searches for heavy gauge bosons \cite{Aad:2011fe,Chatrchyan:2012qk,Hayden:2012gc,Chatrchyan:2012it} and fermions \cite{925327,925338,Aad:2012bb,ATLAS:2012aw}.

Let  us now comment about the bounds from the EW Precision Tests (EWPTs) on the 4DCHM \cite{Marzocca:2012zn}.
As it is well known, extra gauge bosons give a positive contribution to the Peskin-Takeuchi $S$ parameter and the requirement of consistency with the EWPTs generally gives a bound on the mass of these resonances around few TeV. 
In contrast, the fermionic sector is quite irrelevant for $S$ since the extra fermions are weakly coupled to the SM gauge bosons. Either way, as noticed in \cite{Contino:2006qr}, when dealing with effective theories, one can only parametrise $S$ rather than calculating it. In other words, since we are dealing with a truncated theory describing only the lowest-lying resonances that may exist, we need to invoke an Ultra-Violet (UV) completion for the physics effects we are not including in our description. These effects could well compensate for $S$, albeit with some tuning. One example is given in \cite{Contino:2006qr} by considering the contribution of higher-order operators in the chiral expantion. Another scenario leading to a reduced $S$ parameter is illustrated in \cite{DeCurtis:2011yx}, by including non-minimal interactions in the 4DCHM.

Nevertheless, in the following phenomenological analysis, we will chose values for the gauge resonance masses around 2 TeV or larger in order to avoid too big contributions to the $S$ parameter. Recalling that, in the 4DCHM, the mass scale of the lightest gauge boson resonance is given by
\begin{equation}
M_{\rm lightest}=f g_*,
\label{eq:mass}
\end{equation}
we feel justified in choosing for $f$, the compositeness scale of the model, values around 1 TeV and for $g_*$, the coupling constant of the composite gauge sector, values around 2.

Concerning the fermionic sector, our setup is constructed in such a way as to avoid leading-order corrections to the $Z b \bar b$ coupling. As a consequence, ones does not experience substantial bounds on the fermion parameters, so that we will vary them in order to easily reconstruct the top and bottom quark masses (as previously detailed).

\section{Parameter space and benchmarks }
\label{sec:benchmarks}
\subsection{$Z'$ and $W'$ decays}
\label{sec:decays}

Other than the masses of the new gauge bosons, since the latter appear in DY processes at the LHC as resonances,
we are interested here also in their decay widths.
Due to the high number of parameters in the fermionic sector, we can have different regimes in analysing the widths of the new gauge resonances.
We can have a configuration, (i), in which the mass of the fermionic resonances is too heavy to have a decay of a gauge resonance in pairs of heavy fermions (so that, 
other than the SM-like decay channels also the mixed ones, 
involving an elementary
and a composite fermion, are open)
or the opposite configuration, (ii), in which the mass of the fermionic resonances (typically that of the lightest one) is 
small enough so that this is possible. In the former case the typical decay widths can be well below 100 GeV
whereas in the latter case they can grow significantly, to become even comparable to the masses themselves. 
So, typically, we can analyse the  Branching Ratios (BRs) of the gauge resonances in three well-defined and
distinct situations.

\begin{enumerate}
\item The threshold for the gauge boson decays in pairs of heavy fermions has not been reached, therefore
case (i) above is realised: hereafter, `small width' regime.
\item The threshold for the gauge boson decays in pairs of heavy fermions has just been reached, therefore
case (ii) above is realised, where however phase space effects are not yet predominant: 
hereafter, `medium width' regime.
\item The threshold for the gauge boson decays in pairs of heavy fermions has been abundantly surpassed, therefore
case (ii) above is realised, where indeed phase space effects are becoming important: 
hereafter, `{large} width' regime.
\end{enumerate}

Despite such configurations emerge for whichever combination of the aforementioned $f$
and $g_*$ parameters of the new gauge sector that we studied and we will use two setups to illustrate this
($f=0.8$ TeV and $g_*=2.5$  versus 
 $f=1.2$ TeV and $g_*=1.8$), 
we will study the emerging decay dynamics in detail only for one benchmark combination
of these two parameters
($f=1.2$ TeV and $g_*=1.8$). However, we will eventually present results not only for both these
benchmarks but also for various other pairings of 
$f$ and $g_*$ when we will study possible LHC observables. 

It is in order at this point to attempt quantifying the naturalness of the  various $f$ and $g_*$ combinations studied,
within the 4DCHM. In order to do so, we have run 4DCHM parameter scans with $f$ and $g_*$ fixed to the following
combinations:
  (a) $f=0.75$ TeV and $g^*=2$;
  (b) $f=0.8$ TeV and $g^*=2.5$;
  (c) $f=1$ TeV and $g^*=2$;
  (d) $f=1$ TeV and $g^*=2.5$;
  (e) $f=1.1$ TeV and $g^*=1.8$;
  (f) $f=1.2$ TeV and $g^*=1.8$. All other parameters (with reference to their definitions in section \ref{sec:4DHCM})
were varied over the following intervals:
$m_*$, $\Delta_{tL}$, $\Delta_{tR}$, $Y_T$, $M_{Y_T}$, $Y_B$ and $M_{Y_B}$ between 0.5 and 5 TeV while $\Delta_{bL}$ and $\Delta_{bR}$ between 0.05 and 
0.5 TeV (in the spirit of `partial compositness' \cite{DeCurtis:2011yx}).
All aforementioned theoretical and experimental constraints were implemented in each case (though we notice that those on the $b$- and $t$-quark masses and
couplings have minimal impact).
The total number of random points generated for each $f$ and $g_*$ combination was of order 
15 millions. As the scans were not optimised, i.e., all points were equi-probable, a simple yet 
good measure of naturalness is just the percentage of points surviving in each scans. 
The typical survival rate is of ${\cal O}(2\div3\times10^{-7})$, with variations amongst the various  $f$ and $g_*$ combinations of no more than $30\%$.
We therefore notice, on the one hand, that the 4DCHM is highly constrained already by current theoretical and experimental
bounds (hence
amenable to stringent phenomenological investigation) 
and, on the other hand, that none of our choices of $f$ and $g_*$ is overwhelmningly more unnatural than others (in fact, we also reassuringly  
reconfirm that with increasing $f$ we have a 
decreasing number of points found \cite{DeCurtis:2011yx}).



\subsubsection{Spectrum with $f=1.2\text{ TeV}$ and $g_*=1.8$}
With these values of the gauge parameters, the lightest gauge resonances are slightly heavier than $2$ TeV while the heaviest ones are very close to 3 TeV, for both the neutral and the charged sector, so as to not run afoul of recent Tevatron and LHC constraints in searching for generic $W'$ \cite{Aad:2011fe,Chatrchyan:2012qk} and $Z'$ \cite{Hayden:2012gc,Chatrchyan:2012it} states.

\begin{figure}[htb]
\centering
\includegraphics[width=0.65\linewidth,angle=0]{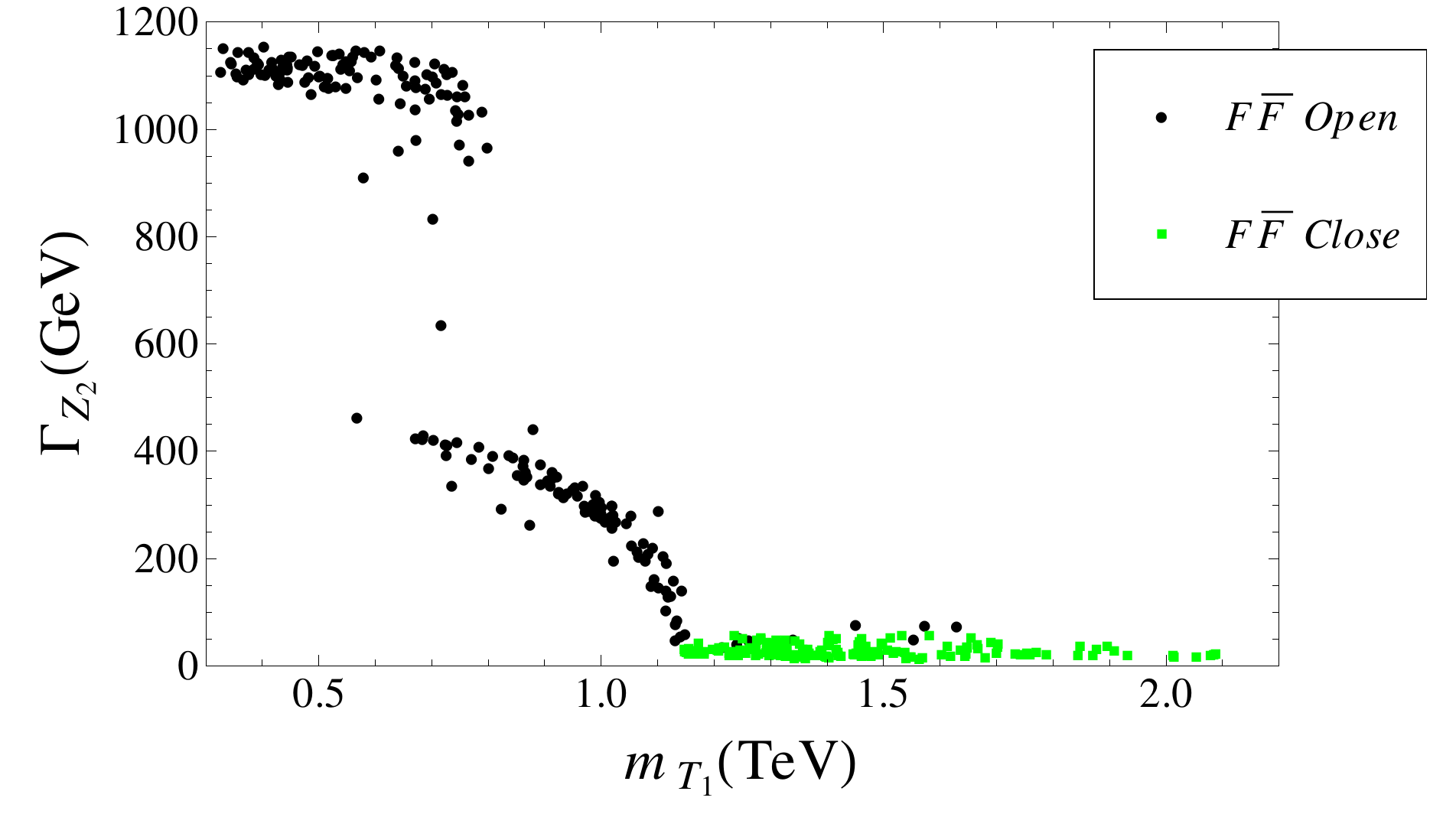}{(a)}
\includegraphics[width=0.65\linewidth,angle=0]{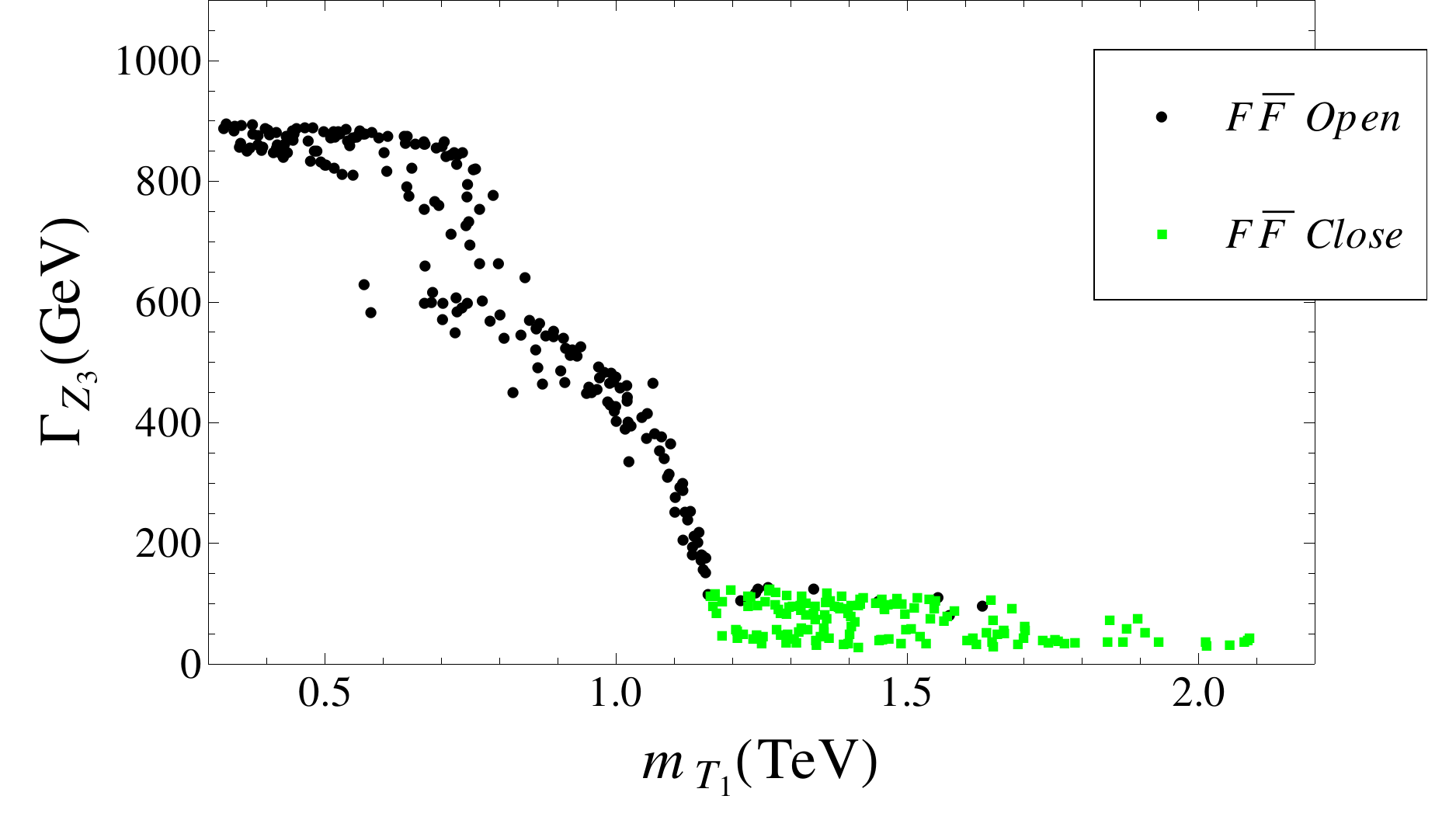}{(b)}
\includegraphics[width=0.65\linewidth,angle=0]{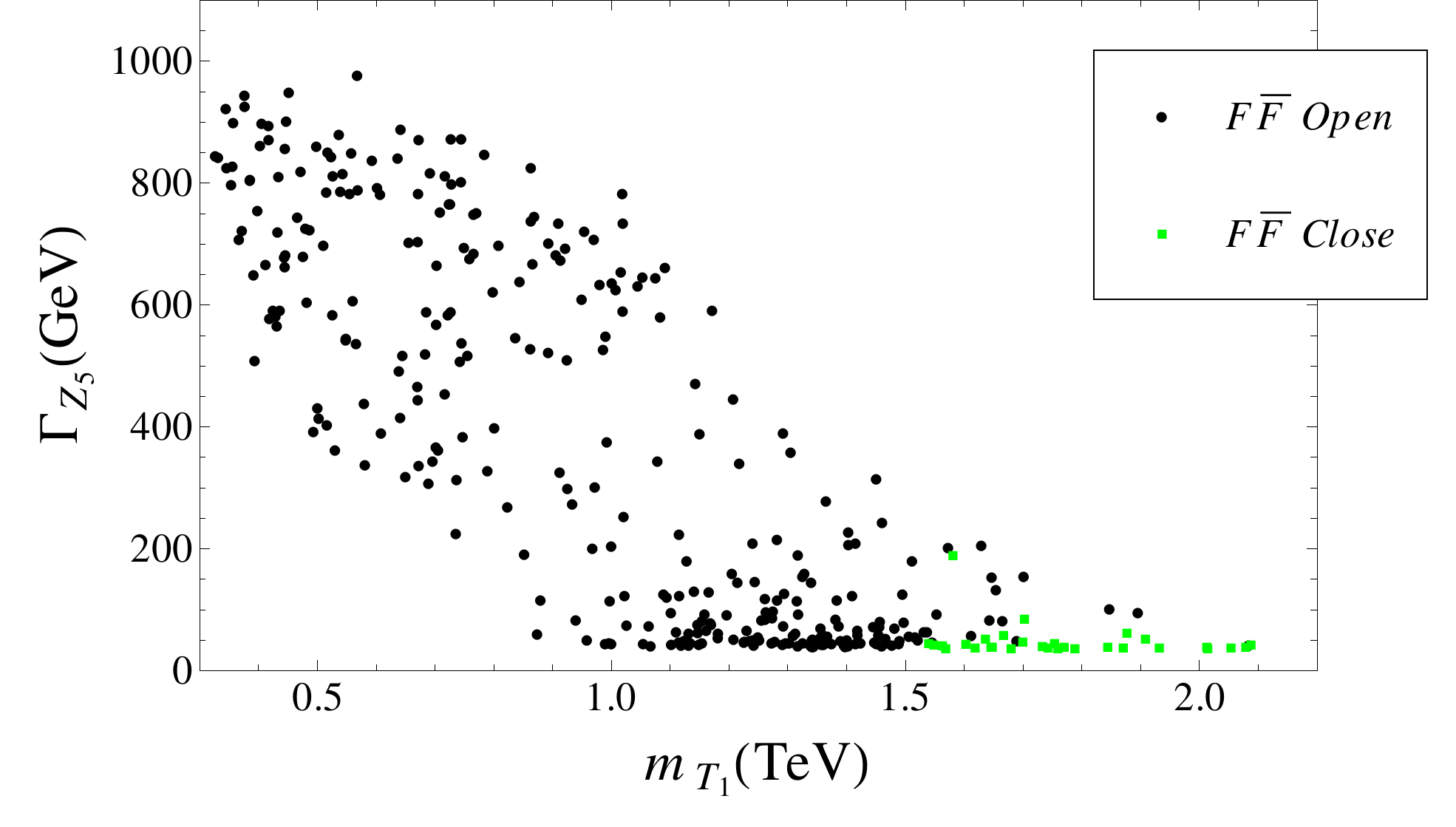}{(c)}
\caption{Width of the additional neutral gauge bosons of the 4DCHM, for the choice $f=1.2$ TeV and $g_*=1.8$, 
(a) for $Z_2$ (b) for $Z_3$ and (c) for $Z_5$, as a function of the mass of the lightest fermionic resonance of charge $+2/3$. The circle points in black are the ones where the decay in a pair of heavy fermions is permitted while 
the square points in 
green are the ones where this process is forbidden.
The fermionic parameters are varied between 0.5 and 5 TeV, except $\Delta_{bL}$ and $\Delta_{bR}$ that are varied between 0.05 and 0.5 TeV.}
\label{plot:neu}
\end{figure}

\begin{figure}[htb]
\centering
\includegraphics[width=0.65\linewidth,angle=0]{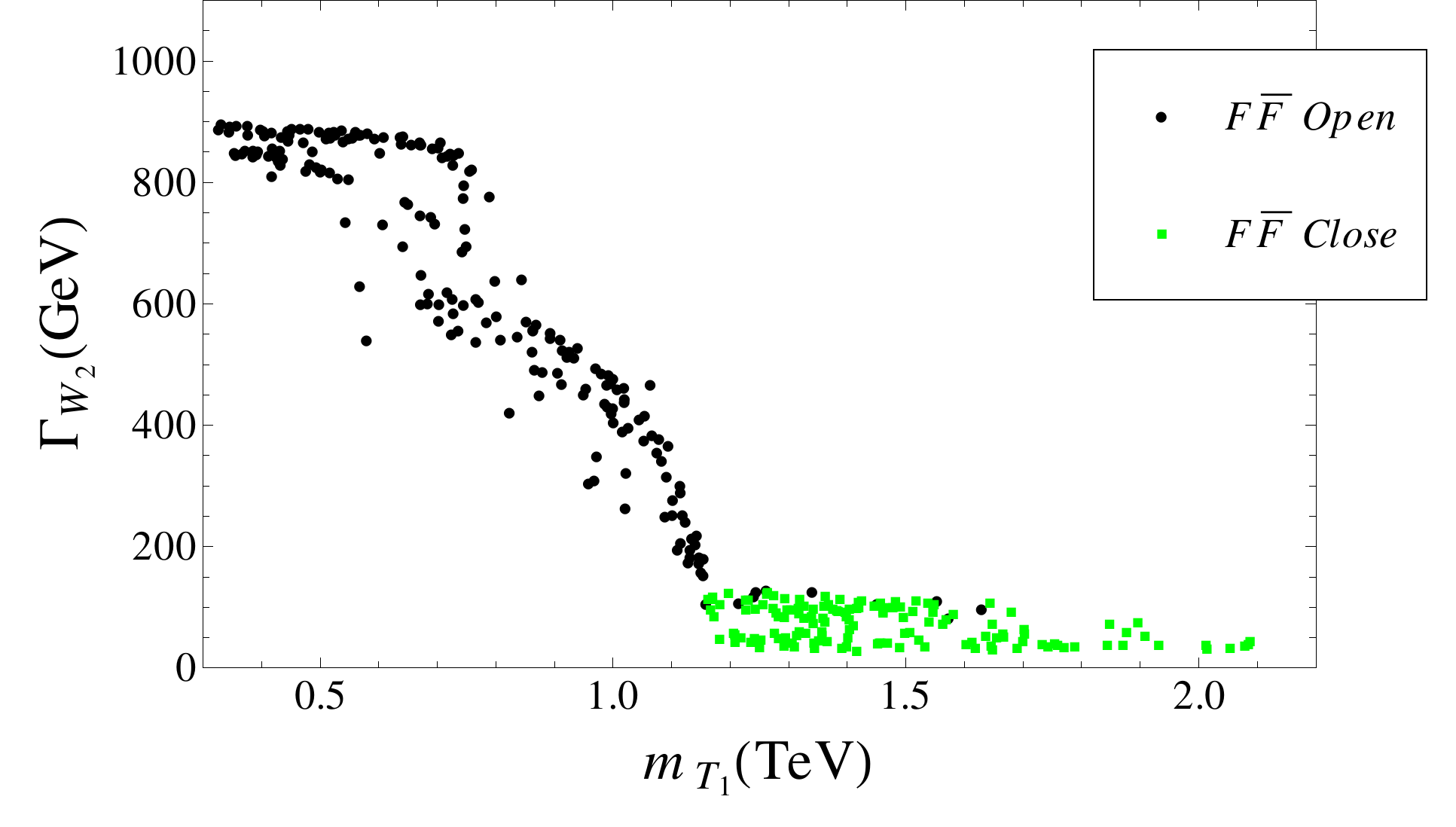}{(a)}
\includegraphics[width=0.65\linewidth,angle=0]{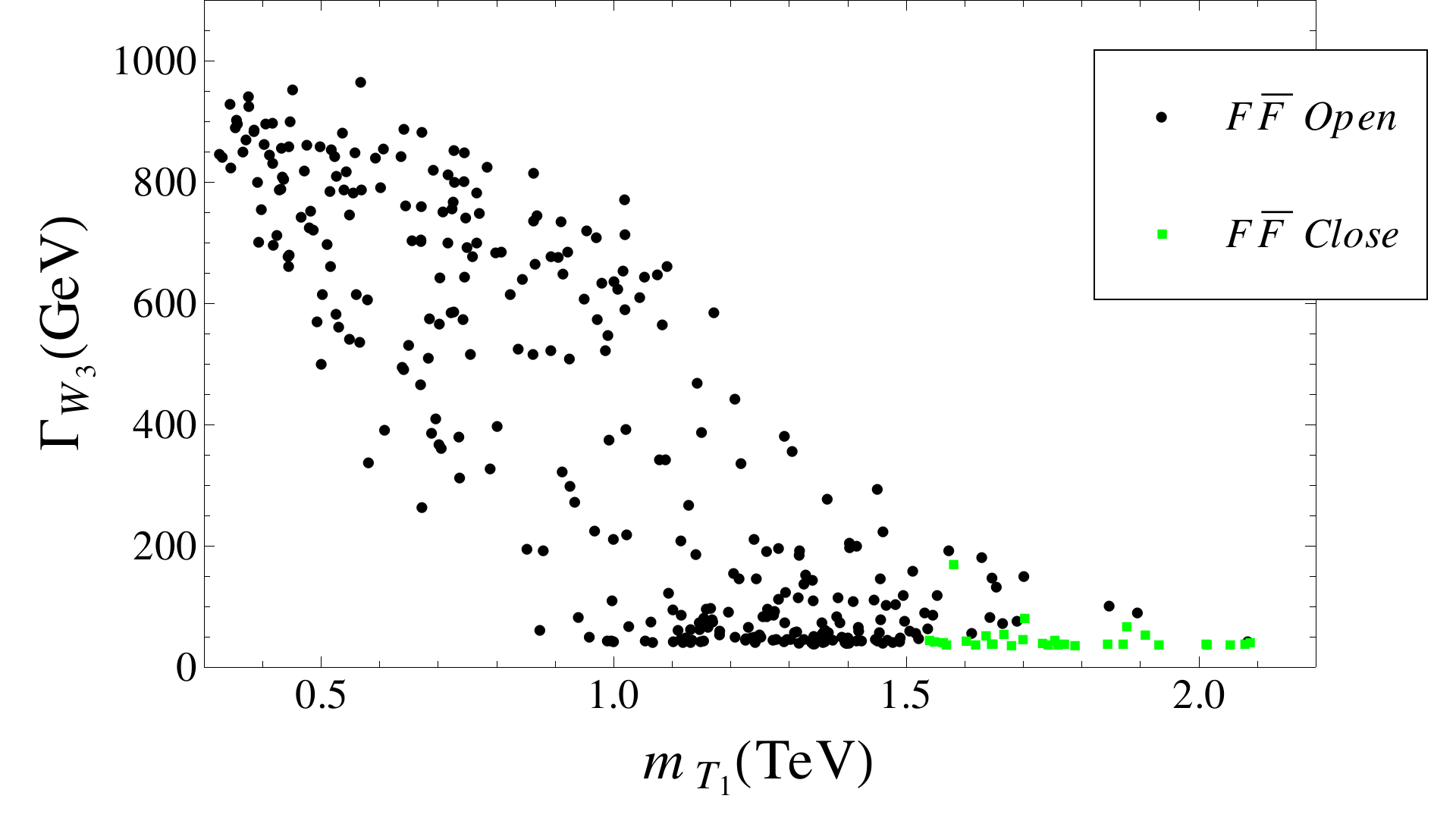}{(b)}
\caption{Width of the additional charged gauge bosons of the 4DCHM, for the choice $f=1.2$ TeV and $g_*=1.8$, 
(a) for $W_2$ and (b) for $W_3$, as a function of the mass of the lightest fermionic resonance of charge $+2/3$. The circle points in black are the ones where the decay in a pair of heavy fermions is permitted while the 
square points in 
green are the ones where this process is forbidden.
The fermionic parameters are varied between 0.5 and 5 TeV, except $\Delta_{bL}$ and $\Delta_{bR}$ that are varied between 0.05 and 0.5 TeV.}
\label{plot:car}
\end{figure}

As already noticed, the masses of the neutral and charged gauge resonances depend only on $f$ and $g_*$ so these are the same in the various width regimes. Recalling that two of the $Z'$ states and one of the $W'$ states
are coupled neither to light quarks nor to leptons,
i.e., $Z_1$ and $Z_4$ for the neutral sector plus $W_1$ for the charged sector\footnote{So that they are essentially
`inert' for the purpose of studying DY processes.}, we are only interested here in the cases $Z_i$ with $i=2,3,5$ 
and $W_i$ with $i=2,3$, whose masses  
are reported in Tab.~\ref{tab:f12g18}. In Figs.~\ref{plot:neu} and \ref{plot:car} we show instead the widths of the gauge bosons 
$Z_2$, $Z_3$, $Z_5$ and $W_2, W_3$, respectively, as a function of the lightest fermionic mass resonance of charge 2/3, that is $T_1$.
By definition this is the lightest non-exotic heavy fermion of the 4DCHM of charge $+2/3$, essentially degenerate in mass
with its charge $-1/3$ counterpart (i.e., $B_1$) and they are generally lighter than their exotic `cousins': that is, 
$\tilde T_1$ (charge $+5/3$) and $\tilde B_1$ (charge $-4/3$). The two different coloured regions in the plots correspond to the aforementioned
situations (i) and (ii): in green squares we have the points where the decay of the heavy gauge bosons in two heavy fermions $F\bar F$ is forbidden while in black circles we have the points where these decays are permitted. What is remarkable to notice here
(and this comment can be extended to the other $f$ and $g_*$ combinations studied) is that these two cases 
correspond to two almost distinct populations in the plots, 
separated roughly at the point $M_{\rm lightest}\approx 2m_{T_1}\approx 2 m_{B_1}$, with a minimum of `leakage' of one into the
other. Therefore, from a phenomenological point of view, to be able to measure a $Z'$ and/or $W'$ width below,
say, 100 GeV, would signify the existence of some heavy fermions with a mass scale not below ${\cal O}(fg_*/2)$
and up to  ${\cal O}(fg_*)$ (or indeed above) (case (i), regime 1.). Conversely, to measure widths in excess of this value would point towards the existence of such additional fermionic states at rather low mass. In fact, one may further argue,
albeit limited to the case of $Z_2$ or $Z_3$ (see top two plots in Fig.~\ref{plot:neu}) and $W_2$ (see left plot
in Fig.~\ref{plot:car}), that two sub-populations exist for case (ii), one corresponding to widths below 600 GeV or so,
and one above it (regimes 2. and 3. above, respectively).

\begin{table}[htb]
\begin{center}
\begin{tabular}{|l| l l|| l| l |}
\hline
	&	Mass (GeV)			& &		& Mass (GeV)			\\
\hline
$Z_2$	&	2249			& & $W_2$		&	2312	\\
$Z_3$	&	2312			& & $W_3$		&	3056	\\
$Z_5$	&	3056			& &			&			\\
\hline
\end{tabular}
\end{center}
\caption{Masses of the additional gauge boson resonances in the 4DCHM for the parameter point $f=1.2$ TeV and $g_*=1.8$. }
\label{tab:f12g18}
\end{table}

While such very distinctive spectrum configurations of the 4DCHM open up interesting phenomenological possibilities,
it remains to be seen whether these are accessible at the LHC via DY processes. We will address this point later on. 
In fact, in order to do so, we ought first to define some benchmarks in parameter space, representative of the three
width regimes already described.  We shall do so next. Further, we will at the same time list the main decay channels
of the gauge boson resonances in these various regimes of their widths.


\section*{\small{4.1.1.a ~Small width regime}}
In the small width regime the masses of the heavy fermions, the width for the (accessible) $Z'$ and $W'$ states and their BRs are reported in Tabs.~\ref{table:fermlow},  \ref{table:widthlow}, \ref{table:BRNlow} and 
\ref{table:BRClow} and are generated with the benchmark in Tab.~\ref{table:benchlow}. The characteristics of the benchmark point illustrated therein are representative of a typical pattern emerging from this populations of points.

\begin{table}[htb]
\begin{center}
\begin{tabular}{l l|l| l|}
\hline
\multicolumn{1}{|c|}{f (GeV)} & 1200 & $m_*$ (GeV) & 2219\\ 
\multicolumn{1}{|c|}{$g_*$} & 1.8& $\Delta_{t_L}$ (GeV)&2366\\
\multicolumn{1}{|c|}{$g_0$} &0.69 &  $\Delta_{t_R}$(GeV)&2245\\
\multicolumn{1}{|c|}{$g_{0y}$} &0.37 & $Y_T$ (GeV)&2824\\
\multicolumn{1}{|c|}{$\langle h \rangle$ (GeV)} & 248& $M_{Y_T}$(GeV) &$-1043$\\
\cline{1-2}
& & $\Delta_{b_L}$(GeV)&202\\
 & & $\Delta_{b_R}$(GeV)&284\\
 & & $Y_B$(GeV)&2543\\
 & & $M_{Y_B}$(GeV)&$-1378$\\
 & & $m_H$ (GeV)&  125 \\
\cline{3-4}
\end{tabular}
\end{center}
\caption{Small width benchmark}
\label{table:benchlow}
\end{table}

\begin{table}[h!]
\begin{center}
\begin{tabular}{|l l l|| l l || l l |l|}
\hline
	&	Mass (GeV)				& &			& Mass (GeV) & 					&	 Mass (GeV)			\\
\hline
$T_1$	&	1635				& & $B_1$		&	1635	  & $\tilde{T}_{1,2}$		& 	1758, 2801			\\
$T_2$	&	1751				& & $B_2$		&	1636	  & $\tilde{B}_{1,2}$	      &	1634, 3012			\\
$T_3$	&	2092				& & $B_3$		&	1726	  &					&						\\
$T_4$	&	2380				& & $B_4$		&	2245    &					&						\\
$T_5$	&	2809				& & $B_5$		&     2882    &					&						\\
$T_6$	&	3011				& & $B_6$		&     3011	  &					&						\\
$T_7$	&	3374				& & $B_7$		&	3013	  &					&						\\
$T_8$	&	3627				& & $B_8$		&     3397    &   				& 					 \\
\hline
\end{tabular}
\end{center}
\caption{Masses of the new fermionic resonances of the 4DCHM in the small width regime.}
\label{table:fermlow}
\end{table}
\begin{table}[h!]
\begin{center}
\begin{tabular}{|l| l l|| l| l |}
\hline
	&	Width (GeV)&&&Width (GeV)\\
\hline  
$Z_2$	&	32  &&  $W_2$ & 55	\\
$Z_3$	&	55  &&  $W_3$ & 54	\\
$Z_5$	&	54  &&    &	\\
\hline
\end{tabular}
\end{center}
\caption{Widths of the new gauge resonances of the 4DCHM in the small width regime.}
\label{table:widthlow}
\end{table}
\begin{table}[h!]
\begin{center}
\begin{tabular}{|l||l||l|}
\hline
$Z_2$			  				&	$Z_3$									 	& $Z_5$	\\
\hline
68\% in $t\bar{t}$				&33\% in $b\bar{b}$			 			&11\%in $W^+W_{1,2}^-$ and c.c.		\\
9\%  in $W^+W^-$	and 10\%  in $ Z^0 H$	&30\% in $t\bar{t}$						&11\%  in $Z_3 H$		\\
5\% in $b\bar{b}$					&11\%  in $W^+W^-$ and 10\% in $Z^0H$	 			&6.5\% in $t{{\bar T}_{4}}$ and c.c.		\\
1\% in $u_{1,2}{{\bar u}_{1,2}}$			&2.5\% in $d_{1,2}{{\bar d}_{1,2}}$ and $u_{1,2}{{\bar u}_{1,2}}$		&6\% in $Z_2 H$ and $Z_1 H$		\\
1\% in $l_{1,...3}{{\bar l}_{1,...3}}$		  		&1\% in  $l_{1,...3}{{\bar l}_{1,...3}}$									
&5\% in $t{{\bar T}_{3}}$ and c.c.	\\
0.5\% in $d_{1,2}{{\bar d}_{1,2}}$		&0.3\% in $t{{\bar T}_{2,3}}$ and c.c.					&3\% in $t{{\bar T}_{5}}$ and c.c.		\\
0.2\% in $b{{\bar B}_3}$ and c.c.				&								&1.5\% in $t\bar{t}$		\\
0.1\% in $t{{\bar T}_2}$ and c.c.				&											&		\\
\hline
\end{tabular}
\end{center}
\caption{BRs of the new neutral gauge bosons of the 4DCHM in the small width regime.}
\label{table:BRNlow}
\end{table}
\begin{table}[h!]
\begin{center}
\begin{tabular}{|l||l|}
\hline
		$W_2^+$						& $W_3^+$	\\
\hline
63\% in $t\bar{b}$					& 13\% in $t{{\bar B}_4}$	\\
10\% in $W^+Z^0$ and 10\% in $W^+H$			& 12\% in $W_1^+ H$	\\
5\% in $u_{1,2}{{\bar d}_{1,2}}$ 			& 11\% in $W^+ Z_3$	\\
1.5\% in $\nu_{1,...3}{{\bar l}_{1,...3}}$				& 11\% in $W_{1,2}^+ Z^0$	\\
0.6\% in $T_{3}\bar{b}$					& 11\% in $W_2^+ H$\\
0.4\% in $\tilde{T}_1 \bar{t}$			& 6.5\% in $T_3\bar{b}$	\\
								& 6\% in $W^+ Z_{1,2}$\\
								& 5.5\% in $\tilde{T}_2 \bar{t}$\\					
								& 4\% in $T_4\bar{b}$\\
\hline
\end{tabular}
\end{center}
\caption{BRs of the new charged gauge bosons of the 4DCHM in the small width regime.}
\label{table:BRClow}
\end{table}

In particular, in this regime we see that the main decay channels of the lightest $Z'$s
(i.e., $Z_2$ and $Z_3$) and $W'$ (i.e., $W_2$) are in SM fermions and gauge bosons, yet we can also have decay channels in one SM fermion and one heavy fermionic resonance, albeit 
at the percent level at the most. In the case of the
heavy $Z'$ (i.e., $Z_5$) and $W'$ (i.e., $W_3$) the corresponding rates grow significantly, up to the ten percent level
or more. Interestingly, both $Z_5$ and $W_3$ can also decay in lighter gauge (and Higgs) boson states and, whenever
they do, the corresponding rates are quite dominant. Further, $Z_2$ and $Z_3$ decays into $Z^0H$ plus $W_2$ decays into $W H$
are also active, at the 10\% level, possibly leading to new discovery modes of a SM-like Higgs boson which are peculiar to the
4DCHM. (Recall that we have fixed the Higgs mass around 125 GeV.)


\section*{\small{4.1.1.b ~Medium width regime}}
In the medium width regime the masses of the heavy fermions, the widths for the (accessible) $Z'$ and $W'$ states and BRs are reported in Tabs.~\ref{table:fermmedd},  \ref{table:widthmed},  \ref{table:BRNmed} and 
\ref{table:BRCmed} and are generated with the benchmark in Tab.~\ref{table:benchmedium}. Again, the point
illustrated here is representative of this sub-population.

\begin{table}[h!]
\begin{center}
\begin{tabular}{l l|l| l|}
\hline
\multicolumn{1}{|c|}{f (GeV)} & 1200 & $m_*$ (GeV) & 2216\\ 
\multicolumn{1}{|c|}{$g_*$} & 1.8& $\Delta_{t_L}$ (GeV)&2434\\
\multicolumn{1}{|c|}{$g_0$} &0.70 &  $\Delta_{t_R}$(GeV)&2362\\
\multicolumn{1}{|c|}{$g_{0y}$} &0.37 & $Y_T$ (GeV)&2771\\
\multicolumn{1}{|c|}{$\langle h \rangle$ (GeV)} & 248& $M_{Y_T}$(GeV) &$-1031$\\
\cline{1-2}
& & $\Delta_{b_L}$(GeV)&327\\
 & & $\Delta_{b_R}$(GeV)&299\\
 & & $Y_B$(GeV)&2815\\
 & & $M_{Y_B}$(GeV)&$-4093$\\
 & & $m_H$ (GeV)& 124   \\
\cline{3-4}
\end{tabular}
\end{center}
\caption{Medium width benchmark}
\label{table:benchmedium}
\end{table}

\begin{table}[h!]
\begin{center}
\begin{tabular}{|l l l|| l l| | l l |}
\hline
	&	Mass (GeV)			& &			& Mass (GeV) & 					&	 Mass (GeV)			\\
\hline
$T_1$	&	      986			& & $B_1$		&	971   & $\tilde{T}_{1,2}$			& 	1759, 2790			\\
$T_2$	&		1753			& & $B_2$		&	985   & $\tilde{B}_{1,2}$	      &	970, 5062			\\
$T_3$	&		2113			& & $B_3$		&	1683  &					&					\\
$T_4$	&		2408			& & $B_4$		&	2255  &					&					\\
$T_5$	&		2800			& & $B_5$		&     2950  &					&					\\
$T_6$	&		3408			& & $B_6$		&     3432  &					&					\\
$T_7$	&		3653			& & $B_7$		&	5062	&					&					\\
$T_8$	&		5065			& & $B_8$		&     5065  &   					& 					\\
\hline
\end{tabular}
\end{center}
\caption{Masses of the new fermionic resonances of the 4DCHM in the medium width regime.}
\label{table:fermmedd}
\end{table}
\begin{table}[h!]
\begin{center}
\begin{tabular}{|l| l l ||l| l |}
\hline
	&	Width (GeV)&&&Width (GeV)\\
\hline
$Z_2$	& 301	  &&  $W_2$ & 434	\\
$Z_3$	& 434	  &&  $W_3$ & 522	\\
$Z_5$	& 526	  &&    &	\\
\hline
\end{tabular}
\end{center}
\caption{Widths of the new gauge resonances of the 4DCHM in the medium width regime.}
\label{table:widthmed}
\end{table}
\begin{table}[h!]
\begin{center}
\begin{tabular}{|l||l||l|}
\hline
$Z_2$			  				&	$Z_3$									 	& $Z_5$	\\
\hline
46\% in $B_1{{\bar B}_1}$				&24\% in $\tilde{B_1}{\bar{\tilde{B}}}_1$					& 24\% in $B_1{\bar B_3}$ and c.c.	\\
40\%  in $\tilde{B_1}{\bar{\tilde{B}}}_1$	&	22\% in $T_1{\bar T_1}$							&20\% in 	$B_2{\bar B_3}$ and c.c						\\
8\% in $t\bar{t}$					&	20\% in $B_{2,3}{{\bar B}_{2,3}}$						&1\% in $W^+W_{1,2}^-$ and c.c.						\\
2\% in $B_2{\bar B_2}$				&	4\% in $t\bar{t}$ and $b\bar{b}$					& 1\% in $t{\bar T_4}$ and c.c.							\\
1\% in $T_1{\bar T_1}$		  		&	1\% in	$W^+W^-$ and $ Z^0H$					& 1\% in $Z_3H$							\\
1\% in $W^+W^-$ and $Z^0 H$			&	0.5\% in $u_{1,2}{{\bar u}_{1,2}}$ and $d_{1,2}{{\bar d}_{1,2}}$	&0.5\% in $ Z_{1,2}H$\\
0.6\% in $b\bar{b}$ and c.c.			&											&							\\
0.1\% in $u_{1,2}{{\bar u}_{1,2}}$ and $l_{1,...3}{\bar{l}}_{1,...3}$					&				&							
\\
\hline
\end{tabular}
\end{center}
\caption{BRs of the new neutral gauge bosons of the 4DCHM in the medium width regime.}
\label{table:BRNmed}
\end{table}
\begin{table}[h!]
\begin{center}
\begin{tabular}{|l||l|}
\hline
		$W_2^+$						& $W_3^+$	\\
\hline
44\% in $B_1{\bar{\tilde{B}}}_1$	& 45\% in $B_3\bar{{\tilde{B}}}_{1}$ 	\\
42\% in $T_1{\bar B_2}$			& 43\% in $T_1{\bar B_3}$	\\
9\% in $t\bar{b}$				& 1.6\% in  $t{\bar B_4}$	\\
1\% in $W^+ H$ and $W^+ Z^0$		& 1\% in $W_{1,2}^+ H$, $W_{1,2}Z^0$ and $W^+ Z_3$	\\
0.5\% in $u_{1,2}{{\bar d}_{1,2}}$	& 1\% in $t{\bar B_3}$ 	 \\
\hline
\end{tabular}
\end{center}
\caption{BRs of the new charged gauge bosons of the 4DCHM in the medium width regime.}
\label{table:BRCmed}
\end{table}

As compared to the previous case, we indeed notice here the appearance of decays in pairs of heavy fermions 
(including exotic states) and that these  are the main decay channels of the heavy gauge bosons, both neutral and charged. However, 
as clear from comparing the heavy
gauge boson and heavy fermion masses relatively, these decays occur just above threshold, so that the $Z'$ and $W'$ widths,
although increasing significantly, remain well below the corresponding mass values, so that these objects would still potentially 
appear as clear resonances over the SM background.


\section*{\small{4.1.1.c ~Large width regime}}
In the {large} width regime the masses of the heavy fermions, the width for the (accessible) $Z'$ and $W'$ states and their BRs are reported in Tabs.~\ref{table:fermhigh}, \ref{table:widthhigh},  \ref{table:BRNhigh} and  
\ref{table:BRChigh} and are generated with the benchmark in Tab.~\ref{table:benchlarge}. The dynamics of the
sub-population of which this point is representative is typical, as seen here. 

In this regime we see that once again the main decay channels of the $Z'$ and $W'$ are in heavy fermions 
(including exotic ones) and that we have the appearance of decays in pairs of heavy fermions different from the lightest
ones, unlike the previous case. Regarding this decay phenomenology in general, it is clear that both the new channels opening up 
and the phase space effects fully onsetting in the others contribute to render the $Z'$ and $W'$ widths here very substantial,
to the extent that it remains to be seen whether they would actually appear as resonant objects in the detectors.

\begin{table}[h!]
\begin{center}
\begin{tabular}{l l|l| l|}
\hline
\multicolumn{1}{|c|}{f (GeV)} & 1200 & $m_*$ (GeV) &1293 \\ 
\multicolumn{1}{|c|}{$g_*$} & 1.8& $\Delta_{t_L}$ (GeV)&4714\\
\multicolumn{1}{|c|}{$g_0$} &0.70 &  $\Delta_{t_R}$(GeV)&3402\\
\multicolumn{1}{|c|}{$g_{0y}$} &0.37 & $Y_T$ (GeV)&4165\\
\multicolumn{1}{|c|}{$\langle h \rangle$ (GeV)} & 248& $M_{Y_T}$(GeV) &$-1503$\\
\cline{1-2}
& & $\Delta_{b_L}$(GeV)&224\\
 & & $\Delta_{b_R}$(GeV)&480\\
 & & $Y_B$(GeV)&4260\\
 & & $M_{Y_B}$(GeV)&$-2835$\\
 & & $m_H$ (GeV)&  125   \\
\cline{3-4}
\end{tabular}
\end{center}
\caption{Large width benchmark}
\label{table:benchlarge}
\end{table}

\begin{table}[h!]
\begin{center}
\begin{tabular}{|l l l|| l l || l l |}
\hline
	&	Mass (GeV)				& &			& Mass (GeV) & 					&	 Mass (GeV)			\\
\hline
$T_1$	&	502				& & $B_1$		&	502	  & $\tilde{T}_{1,2}$		& 	744, 2247					\\
$T_2$	&	740				& & $B_2$		&	507	  & $\tilde{B}_{1,2}$	      &	501, 3336					\\
$T_3$	&	1910				& & $B_3$		&	865	  &					&						\\
$T_4$	&	2232				& & $B_4$		&	1936        &					&						\\
$T_5$	&	2663				& & $B_5$		&     2202        &					&						\\
$T_6$	&	3336				& & $B_6$		&     3336	  &					&						\\
$T_7$	&	3878				& & $B_7$		&	3336	  &					&						\\
$T_8$	&	4914				& & $B_8$		&     4913        &   				& 	\\
\hline				
\end{tabular}
\end{center}
\caption{Masses of the new fermionic resonances of the 4DCHM in the {large} width regime.}
\label{table:fermhigh}
\end{table}
\begin{table}[h!]
\begin{center}
\begin{tabular}{|l|| l l l|| l| }
\hline
	&	Width (GeV)&&&Width (GeV)\\
\hline
$Z_2$	& 1099 	  &&  $W_2$ & 820	\\
$Z_3$	& 827	  &&  $W_3$ & 614	\\
$Z_5$	&  413	  &&    &	\\
\hline
\end{tabular}
\end{center}
\caption{Widths of the new gauge resonances of the 4DCHM in the {large} width regime.}
\label{table:widthhigh}
\end{table}
\begin{table}[h!]
\begin{center}
\begin{tabular}{|l||l||l|}
\hline
$Z_2$			  				&	$Z_3$									 	& $Z_5$	\\
\hline
31\% in $T_2{\bar T_2}$				&	17\% in $\tilde{B_1}{\bar{\tilde{B}}}_1$	&24\% in $B_1{\bar B_3}$ and c.c.		\\
29\% in $\tilde{B_1}{\bar{\tilde{B}}}_1$	&	16\% in $T_1{\bar T_1}$				&6\% in $t\bar{T_2}$ and c.c.		\\
16\% in $\tilde{T_1}{\bar{\tilde{T}}}_1$	&	17\% in $\tilde{T_1}{\bar{\tilde{T}}}_1$    &6\% in $B_1{\bar B_5}$ and c.c.		\\
5\% in $t\bar{t}$					&	12\% in $T_2{\bar T_2}$				&5\% in $t{\bar T_4}$ and c.c.		\\
4\% in $B_1{\bar B_2}$ and c.c. 		&	11\% in $B_1{\bar B_2}$ and c.c.		&2\% in $b{\bar B_1}$ and c.c.		\\
4\% in $B_1{\bar B_1}$				&	7\% in $t\bar{t}$ and $b\bar{b}$		&1.5\% in $W^+W_{1,2}^-$ and c.c.		\\
2\% in $B_{2,3}{{\bar B}_{2,3}}$			&	0.7\% in 	$W^+W^-$ and $ Z^0H$	      &1.5\% in $Z_3H$		\\
\hline
\end{tabular}
\end{center}
\caption{BRs of the new neutral gauge bosons of the 4DCHM in the {large} width regime.}
\label{table:BRNhigh}
\end{table}
\begin{table}[h!]
\begin{center}
\begin{tabular}{|l||l|}
\hline
		$W_2^+$						& $W_3^+$	\\
\hline
29\% in $\tilde{T_1}{\bar T_1}$			& 22\% in $B_3{\bar{\tilde{B}}}_1$	\\
17\% in $T_1{\bar B_2}$					& 22\% in $T_1{\bar B_3}$	\\
15\% in $t\bar{b}$					& 15\% in $B_5{\bar{\tilde{B}}}_1$	\\
13\% in ${\bar{\tilde{B}}}_1B_{1,2}$		& 15\% in $T_1{\bar B_5}$	\\
10\% in $T_1{\bar B_1}$					& 9\% in $\tilde{T_1}\bar{t}$	 \\
0.7\% in $W^+Z^0$ and $W^+ H$				& 6\% in$\tilde{T_2}\bar{t}$	\\
								& 2\% in $t{\bar B_4}$	\\
								& 1\% in $b{\bar{\tilde{B}}}_1$ and $T_1\bar{b}$	\\
								& 1\% in $W_{1,2}^+Z^0$ and $W^+Z_3$\\
								& 1\% in $W_1^+ H$\\
\hline
\end{tabular}
\end{center}
\caption{BRs of the new charged gauge bosons of the 4DCHM in the {large} width regime.}
\label{table:BRChigh}
\end{table}

\section*{\small{4.1.1.d ~The role of the Higgs mass }}

Before investigating the phenomenology of $Z'$ and $W'$ production and decay in DY processes, we would like to show
explicitly the impact that the recent LHC (and Tevatron) data on a possible Higgs boson have on the 4DCHM, specifically on its fermionic sector. This is mandatory, if one recalls that  
Ref.~\cite{DeCurtis:2011yx} predates such data.

\begin{figure}
\centering
\includegraphics[width=0.75\linewidth,angle=0]{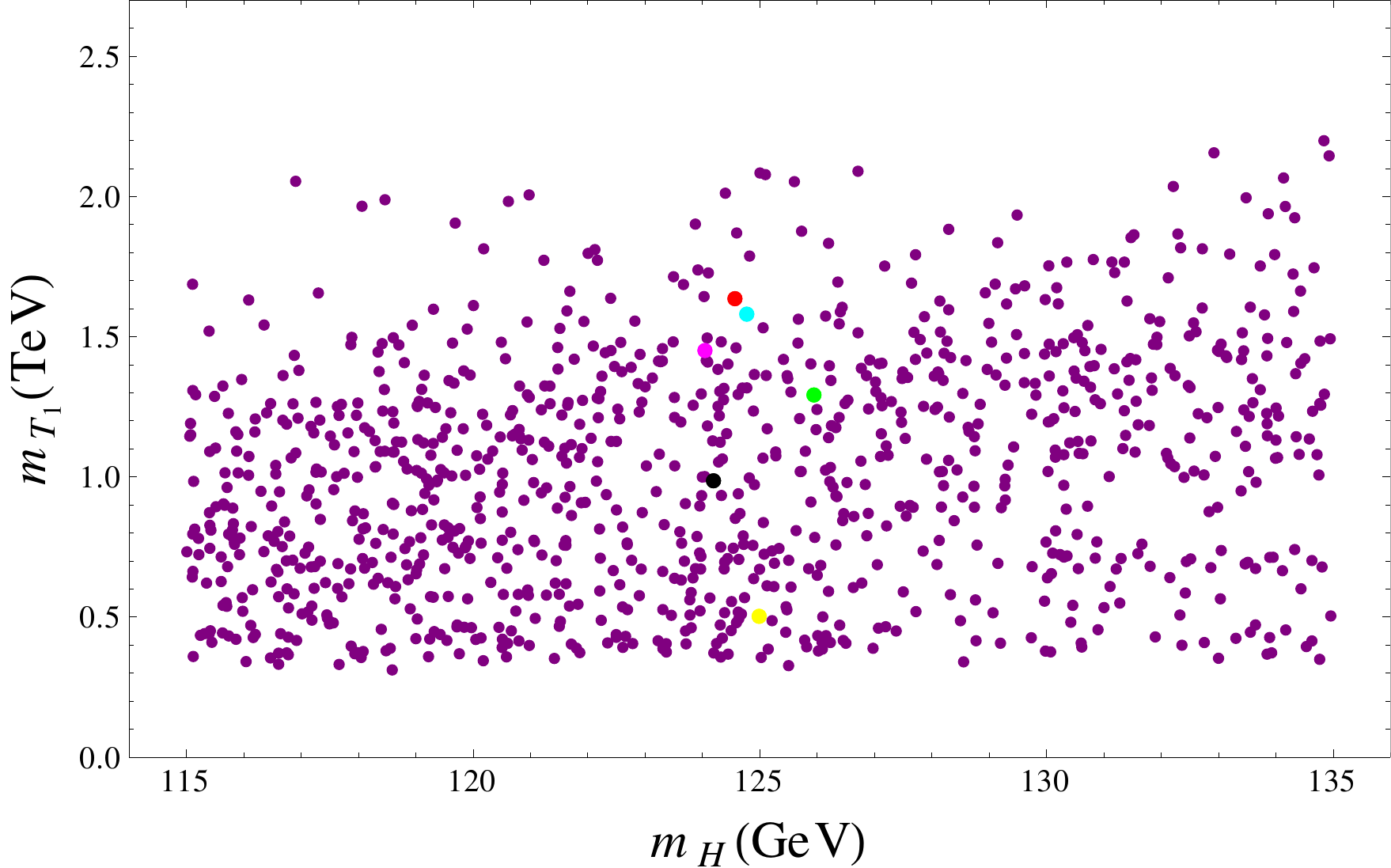}
\caption{Mass of ligthest additional fermion of charge +2/3 versus the mass of the Higgs boson
of the 4DCHM, for the choice $f=1.2$ TeV and $g_*=1.8$. 
The fermionic parameters are varied between 0.5 and 5 TeV, except $\Delta_{bL}$ and $\Delta_{bR}$ that are 
varied between 0.05 and 0.5 TeV. Amongst the majority of purple points,
we singled out six differently coloured ones, in (1) red, (2) green, (3) cyan, (4) magenta, (5) black and (6) yellow, which correspond to benchmarks that will be used in the forthcoming collider study.}
\label{plot:Hpotential}
\end{figure}

We therefore present in Fig.~\ref{plot:Hpotential} the allowed masses of the lightest additional fermion of charge
$+2/3$ in the window 115 GeV $\leq m_H\leq$ 135 GeV,
 for the choice $f=1.2$ TeV and $g_*=1.8$. (The pattern for the
counterpart with charge $-1/3$ is rather similar.) As already realised in Ref.~\cite{DeCurtis:2011yx}, such a
small Higgs mass is consistent with light fermionic partners, down to 400 GeV or so (hence well within the scope of
the LHC). In fact, the plot confirms the well established result within the 4DCHM that a light Higgs boson prefers light fermionic partners
\cite{Matsedonskyi:2012ym,Redi:2012ha}, thereby further corroborating our claim of a strong interplay that should be expected in DY processes
between the gauge and fermionic sectors of the 4DCHM.

We also use this plot to highlight some benchmark configurations. Amongst the generic points we have highlighted six in particular, differently coloured, that will define a convention that will be exploited in the upcoming phenomenological 
analysis.
Their use will be as follows.
\begin{itemize}
\item The points in red, green, cyan, magenta, black and yellow correspond to the benchmarks studied in 
Figs.~\ref{fig:NC-Widths}--\ref{fig:CC-Widths}, in which we will maintain the same colour scheme.
The exact numerical values of the parameters defining these are found in
Tab.~\ref{table:benchmark2}.
\item Amongst these, the points in red, black and yellow correspond to the small, medium and large width benchmarks
of the previous subsections.
\item In particular, the point in red corresponds to the benchmark investigated in detail in 
Figs.~\ref{fig:NC-Split}--\ref{fig:CC-Split}. 
\end{itemize}

\subsubsection{Spectrum with $f=0.8\text{ TeV}$ and $g_*=2.5$ (and the other benchmarks)}

\begin{figure}
\centering
\includegraphics[width=0.65\linewidth,angle=0]{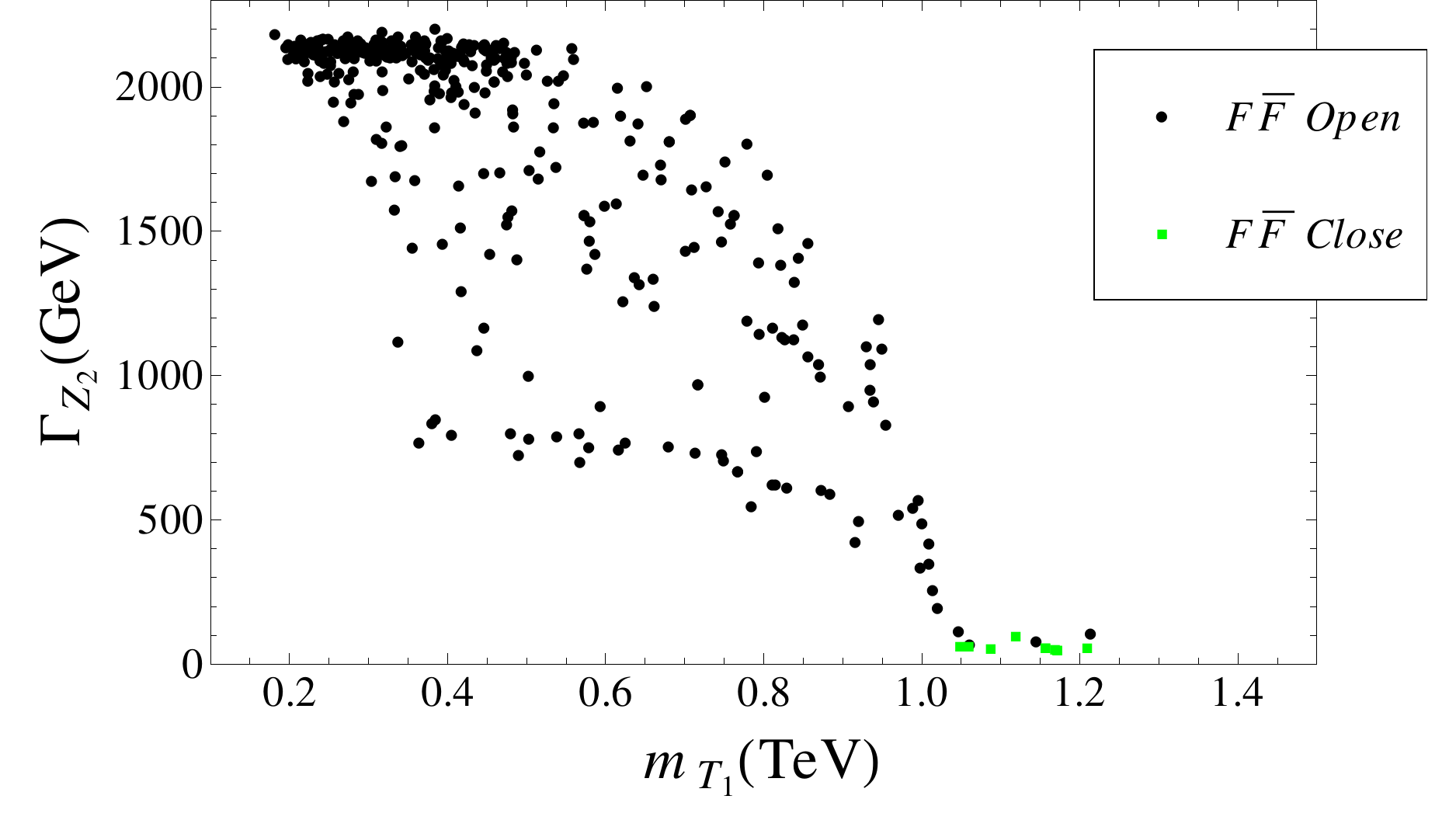}{(a)}
\includegraphics[width=0.65\linewidth,angle=0]{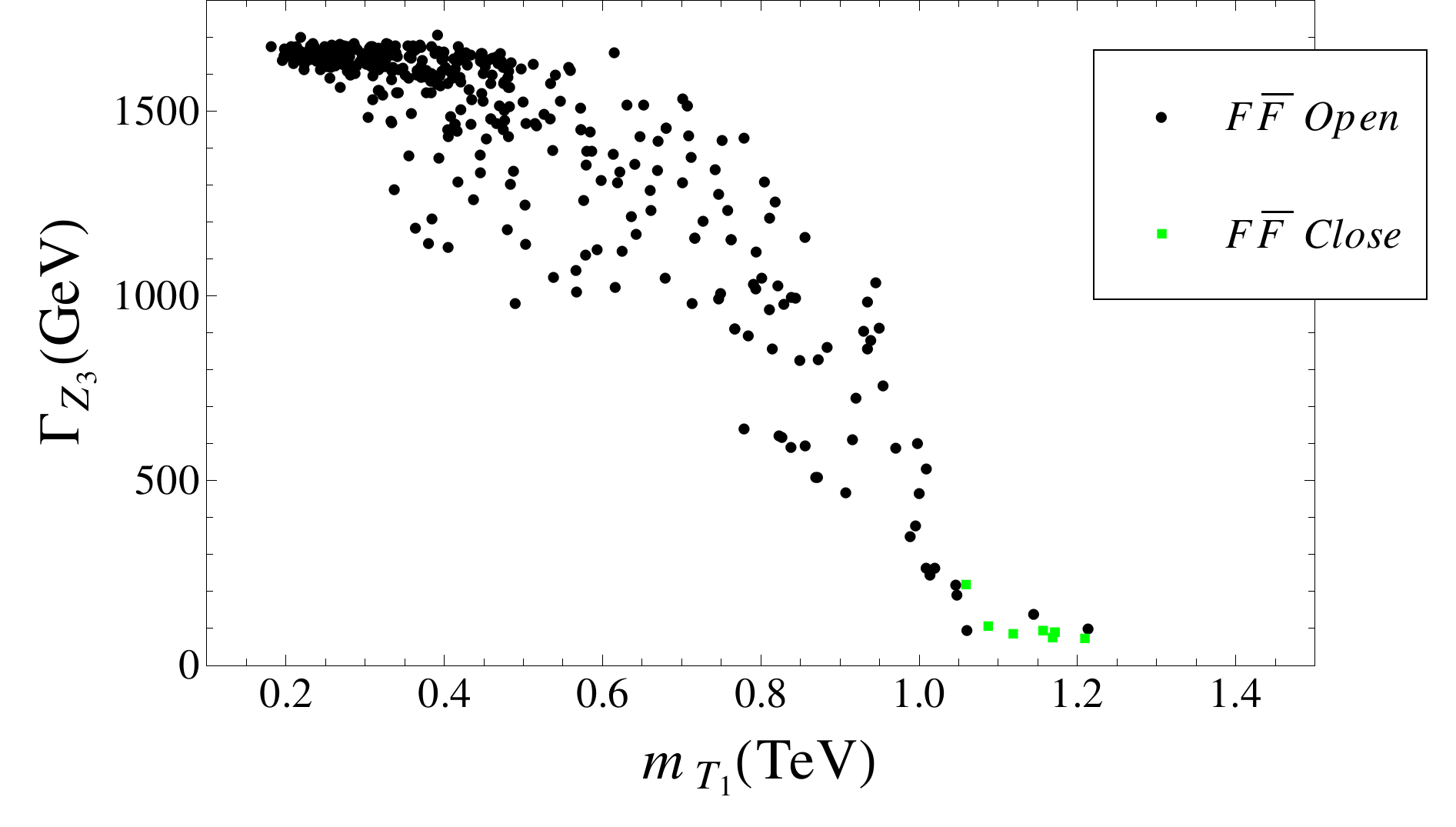}{(b)}
\includegraphics[width=0.65\linewidth,angle=0]{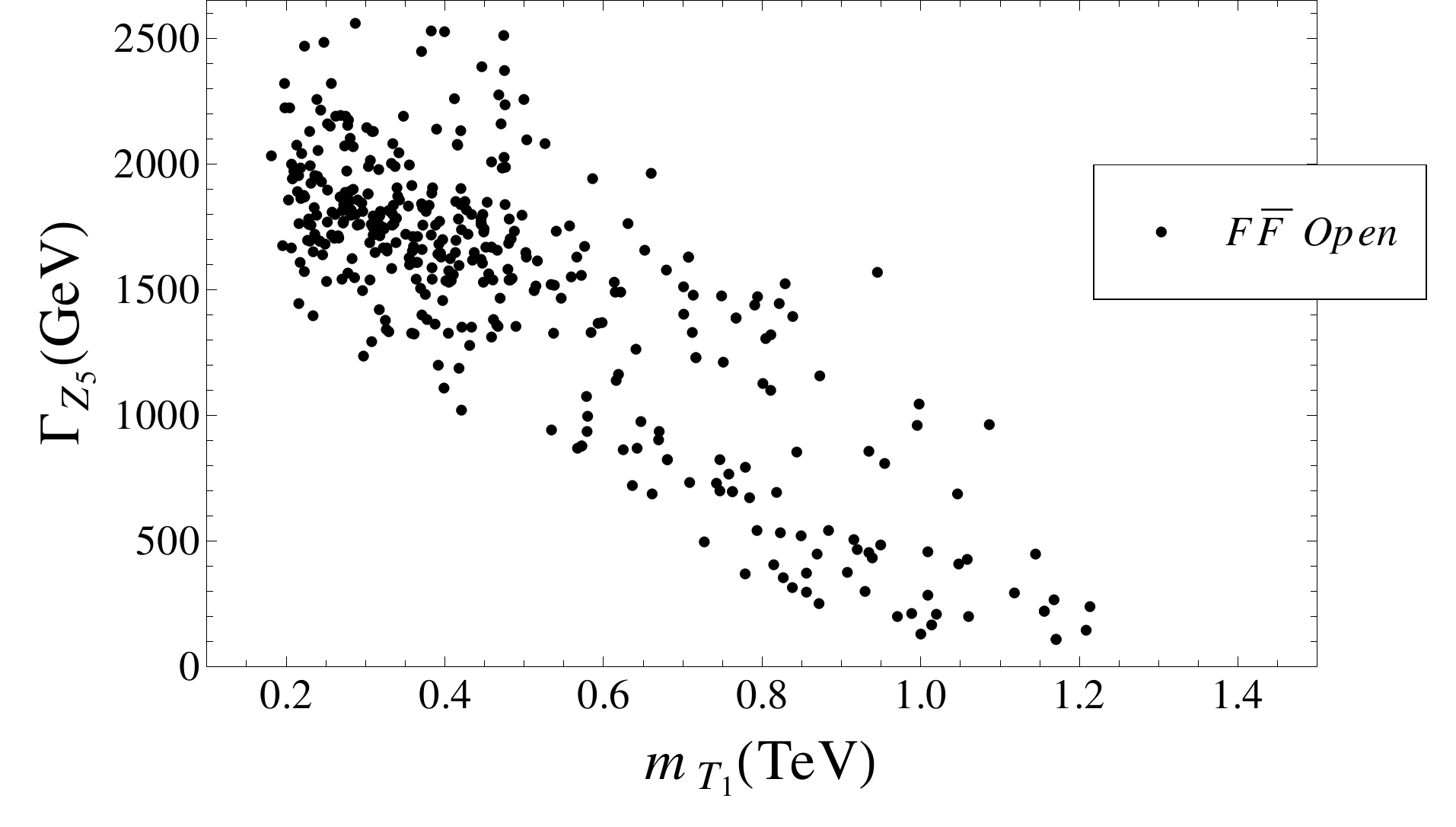}{(c)}
\caption{Width of the additional neutral gauge bosons of the 4DCHM, for the choice $f=0.8$ TeV and $g_*=2.5$, 
(a) for $Z_2$ (b) for $Z_3$ and (c) for $Z_5$, as a function of the mass of the lightest fermionic resonance of charge $+2/3$. The circle points in black are the ones where the decay in a pair of heavy fermions is permitted while 
the square points in 
green are the ones where this process is forbidden.
The fermionic parameters are varied between 0.5 and 5 TeV, except $\Delta_{bL}$ and $\Delta_{bR}$ that are varied between 0.05 and 0.5 TeV.}
\label{plot:neu-new}
\end{figure}

\begin{figure}
\centering
\includegraphics[width=0.65\linewidth,angle=0]{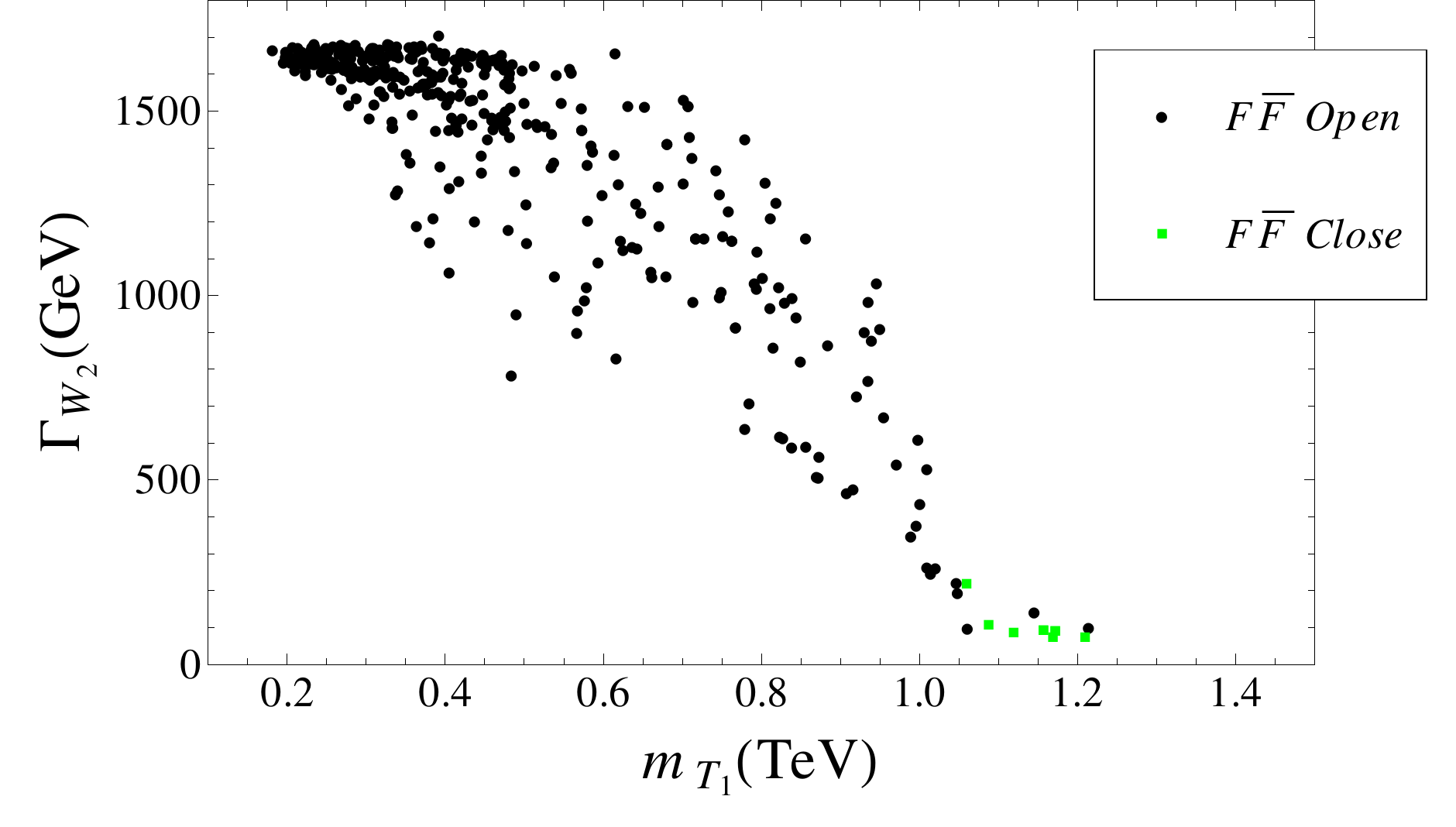}{(a)}
\includegraphics[width=0.65\linewidth,angle=0]{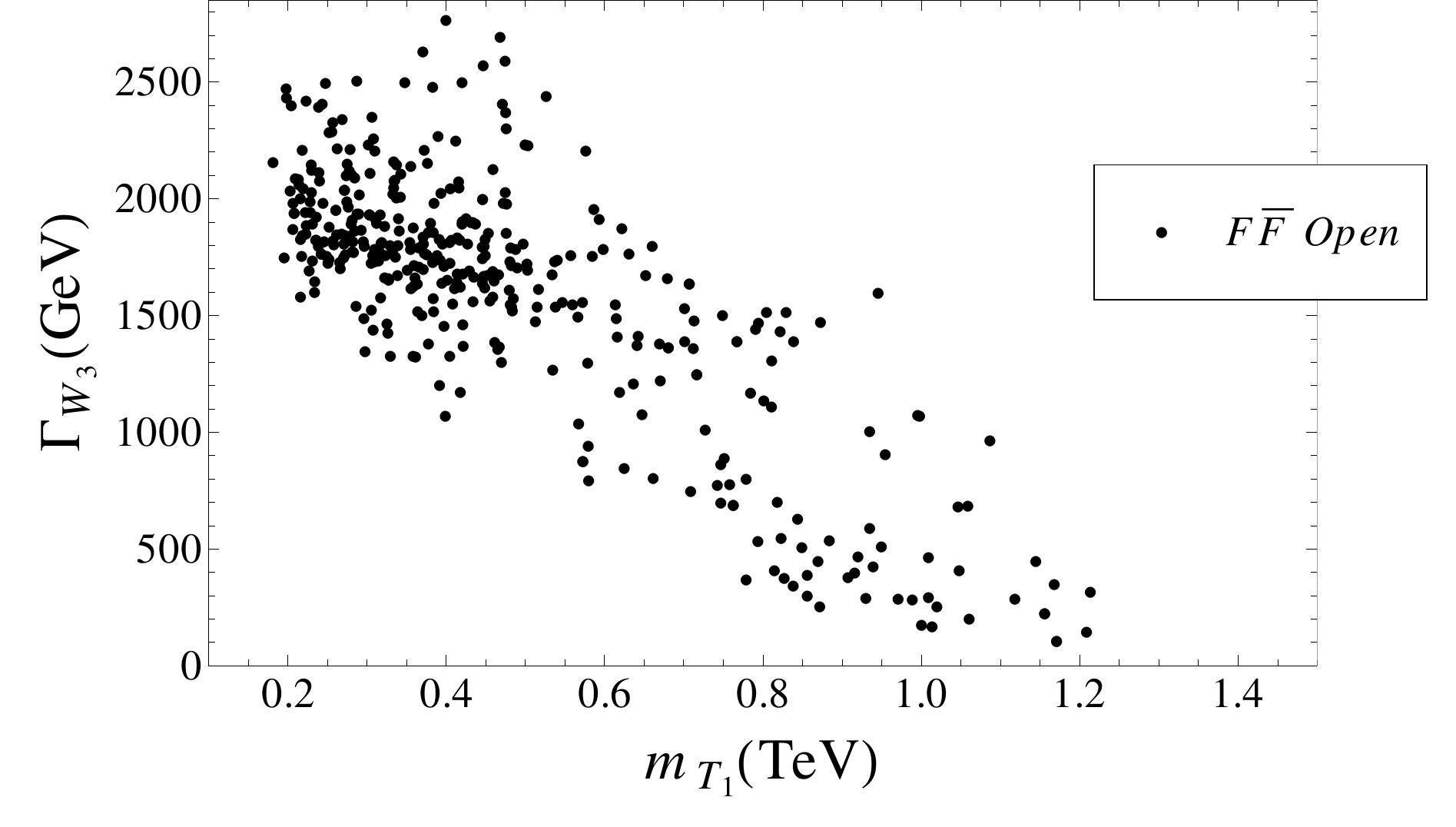}{(b)}
\caption{Width of the additional charged gauge bosons of the 4DCHM, for the choice $f=0.8$ TeV and $g_*=2.5$, 
(a) for $W_2$ and (b) for $W_3$, as a function of the mass of the lightest fermionic resonance of charge $+2/3$. The circle points in black are the ones where the decay in a pair of heavy fermions is permitted while the 
square points in 
green are the ones where this process is forbidden.
The fermionic parameters are varied between 0.5 and 5 TeV, except $\Delta_{bL}$ and $\Delta_{bR}$ that are varied between 0.05 and 0.5 TeV.}
\label{plot:car-new}
\end{figure}

This section is introduced merely to confirm that 
the presence of the width distributions  as seen in the case
$f=1.2\text{ TeV}$ and $g_*=1.8$  is repeated
also for the most extreme benchmark point, namely,
$f=0.8\text{ TeV}$ and $g_*=2.5$, in the sense that this choice (amongst all those adopted, see (a)--(f) above),
is the one for which the typical mass difference between gauge bosons (governed by the parameter $g_*$) and heavy fermions
(governed by $f$) is maximal. Even here, there exist the two populations (i)--(ii) specified above, though 
limited to the case of the lightest states $Z_2$, $Z_3$ and $W_2$ (the only accessible ones at the LHC in fact, see below), i.e.,
not for the case of the heaviest states $Z_5$ and $W_3$. Further, population (i) is very sparse while population (ii) does not 
obviously show the sudden rise in width size seen above, due to the opening of additional $F\bar F$ channels and/or the onsetting
of phase space effects onto pre-existing ones. All this is clearly exemplified in Figs.~\ref{plot:neu-new} and 
\ref{plot:car-new}. Finally, although not shown, we can also confirm that the typical
decay patterns of all the $Z'$ and $W'$ gauge bosons of the 4DCHM seen here is  
consistent with the results of the previous sub-sections, so we refrain here from presenting the corresponding results.

For all other benchmarks studied, i.e., (a), (c), (d) and (e), the $Z'$ and $W'$ decay dynamics is somewhere in between the cases (b) and (f), yet displaying
the same un-mistakable characteristics that we have just discussed.  The numerical values of the input and output
parameters of the 4DCHM used for the phenomenological analysis are found in Tab.~\ref{table:benchmark1}, 
which will refer to 
Figs.~\ref{fig:NC-Mass}--\ref{fig:CC-AFB}.

\begin{table}[h!]
\captionsetup[subfloat]{labelformat=empty,position=top}
\centering
\subfloat[(1) red]
  {\begin{tabular}{l l|l| l|}
\hline
\multicolumn{1}{|c|}{f (GeV)}				& 1200		 & $m_*$ (GeV)	     	&	 2219	 \\ 
\multicolumn{1}{|c|}{$g_*$} 				&	1.8  	 & $\Delta_{t_L}$ (GeV)	&     	 2366	\\
\multicolumn{1}{|c|}{$g_0$} 				&0.69		 &  $\Delta_{t_R}$(GeV)	& 2245		\\
\multicolumn{1}{|c|}{$g_{0Y}$} 			& 0.37 		 & $Y_T$ (GeV)	     	& 2824		\\
\multicolumn{1}{|c|}{$\langle h \rangle$ (GeV)} &   248  	 & $M_{Y_T}$(GeV) 	&	 $-1043$	\\
\cline{1-2}
								&		 & $\Delta_{b_L}$(GeV)  &	 202	\\
 								&		 & $\Delta_{b_R}$(GeV)  & 284		\\
 								& 		 & $Y_B$(GeV)		& 2543		\\
 								& 		 & $M_{Y_B}$(GeV)		&   $-1378$    	\\
 								& 		 & $m_H$ (GeV)		&   125   \\
\cline{3-4}
\end{tabular}}
\subfloat[(2) green]
  {\begin{tabular}{l l|l| l|}
\hline
\multicolumn{1}{|c|}{f (GeV)}				&1200		 & $m_*$ (GeV)	     	&2014 		 \\ 
\multicolumn{1}{|c|}{$g_*$} 				&1.8	  	 & $\Delta_{t_L}$ (GeV)	& 3303     		\\
\multicolumn{1}{|c|}{$g_0$} 				&0.69		 &  $\Delta_{t_R}$(GeV)	&3158 		\\
\multicolumn{1}{|c|}{$g_{0Y}$} 			& 0.37		 & $Y_T$ (GeV)	     	&1907		\\
\multicolumn{1}{|c|}{$\langle h \rangle$ (GeV)} &  248     	 & $M_{Y_T}$(GeV) 	& $-647$ 		\\
\cline{1-2}
								&		 & $\Delta_{b_L}$(GeV)  &493 		\\
 								&		 & $\Delta_{b_R}$(GeV)  &366 		\\
 								& 		 & $Y_B$(GeV)		&1126 		\\
 								& 		 & $M_{Y_B}$(GeV)		& $-1884$      	\\
 								& 		 & $m_H$ (GeV)		& 126    \\
\cline{3-4}
\end{tabular}}
\\
\subfloat[(3) cyan]
  {\begin{tabular}{l l|l| l|}
\hline
\multicolumn{1}{|c|}{f (GeV)}				&1200		 & $m_*$ (GeV)	     	& 1908		 \\ 
\multicolumn{1}{|c|}{$g_*$} 				&1.8	  	 & $\Delta_{t_L}$ (GeV)	&    3328  		\\
\multicolumn{1}{|c|}{$g_0$} 				&0.69		 &  $\Delta_{t_R}$(GeV)	& 4585		\\
\multicolumn{1}{|c|}{$g_{0Y}$} 			& 0.37		 & $Y_T$ (GeV)	     	& 1762		\\
\multicolumn{1}{|c|}{$\langle h \rangle$ (GeV)} &  248    	 & $M_{Y_T}$(GeV) 	& $-715$ 		\\
\cline{1-2}
								&		 & $\Delta_{b_L}$(GeV)  &	340 	\\
 								&		 & $\Delta_{b_R}$(GeV)  &414		\\
 								& 		 & $Y_B$(GeV)		& 999		\\
 								& 		 & $M_{Y_B}$(GeV)		&  $-725$      	\\
 								& 		 & $m_H$ (GeV)		&125      \\

\cline{3-4}
\end{tabular}}
\subfloat[(4) magenta]
  {\begin{tabular}{l l|l| l|}
\hline
\multicolumn{1}{|c|}{f (GeV)}				&1200		 & $m_*$ (GeV)	     	&2031		 \\ 
\multicolumn{1}{|c|}{$g_*$} 				&1.8	  	 & $\Delta_{t_L}$ (GeV)	&  4423    		\\
\multicolumn{1}{|c|}{$g_0$} 				&0.69		 &  $\Delta_{t_R}$(GeV)	& 4419		\\
\multicolumn{1}{|c|}{$g_{0Y}$} 			& 0.37		 & $Y_T$ (GeV)	     	& 1636 		\\
\multicolumn{1}{|c|}{$\langle h \rangle$ (GeV)} &   248    	 & $M_{Y_T}$(GeV) 	&$-558$		\\
\cline{1-2}
								&		 & $\Delta_{b_L}$(GeV)  & 127 		\\
 								&		 & $\Delta_{b_R}$(GeV)  &286 		\\
 								& 		 & $Y_B$(GeV)		&	4543	\\
 								& 		 & $M_{Y_B}$(GeV)		&  $-1394 $    	\\
 								& 		 & $m_H$ (GeV)		&   124   \\
\cline{3-4}
\end{tabular}}
\\
\subfloat[(5) black]
  {\begin{tabular}{l l|l| l|}
\hline
\multicolumn{1}{|c|}{f (GeV)} & 1200 & $m_*$ (GeV) & 2216\\ 
\multicolumn{1}{|c|}{$g_*$} & 1.8& $\Delta_{t_L}$ (GeV)&2434\\
\multicolumn{1}{|c|}{$g_0$} &0.70 &  $\Delta_{t_R}$(GeV)&2362\\
\multicolumn{1}{|c|}{$g_{0Y}$} &0.37 & $Y_T$ (GeV)&2771\\
\multicolumn{1}{|c|}{$\langle h \rangle$ (GeV)} & 248& $M_{Y_T}$(GeV) &$-1031$\\
\cline{1-2}
& & $\Delta_{b_L}$(GeV)&327\\
 & & $\Delta_{b_R}$(GeV)&299\\
 & & $Y_B$(GeV)&2815\\
 & & $M_{Y_B}$(GeV)&$-4093$\\
 & & $m_H$ (GeV)& 124   \\
\cline{3-4}
\end{tabular}}
\subfloat[(6) yellow]
  {\begin{tabular}{l l|l| l|}
\hline
\multicolumn{1}{|c|}{f (GeV)} & 1200 & $m_*$ (GeV) &1293 \\ 
\multicolumn{1}{|c|}{$g_*$} & 1.8& $\Delta_{t_L}$ (GeV)&4714\\
\multicolumn{1}{|c|}{$g_0$} &0.70 &  $\Delta_{t_R}$(GeV)&3402\\
\multicolumn{1}{|c|}{$g_{0Y}$} &0.37 & $Y_T$ (GeV)&4165\\
\multicolumn{1}{|c|}{$\langle h \rangle$ (GeV)} & 248& $M_{Y_T}$(GeV) &$-1503$\\
\cline{1-2}
& & $\Delta_{b_L}$(GeV)&224\\
 & & $\Delta_{b_R}$(GeV)&480\\
 & & $Y_B$(GeV)&4260\\
 & & $M_{Y_B}$(GeV)&$-2835$\\
 & & $m_H$ (GeV)&  125   \\
\cline{3-4}
\end{tabular}}
\caption{Benchmark points for the generation of Figs.~\ref{fig:NC-Widths} and  \ref{fig:CC-Widths}. 
All points assume $f=1.2$ TeV and $g_*=1.8$.
Figs.~\ref{fig:NC-pT} and \ref{fig:CC-pT} are generated only with the benchmark (4) magenta. The numbering and colour scheme is as in Fig.~\ref{plot:Hpotential}.  }
\label{table:benchmark2}
\end{table}

\begin{table}[h!]
\captionsetup[subfloat]{labelformat=empty,position=top}
\centering
\subfloat[(a) $f=0.75$ TeV and $g_*=2$]
  {\begin{tabular}{l l|l| l|}
\hline
\multicolumn{1}{|c|}{f (GeV)}				&	750	 & $m_*$ (GeV)	     	&1673		 \\ 
\multicolumn{1}{|c|}{$g_*$} 				&	  2  	 & $\Delta_{t_L}$ (GeV)	&    974 		\\
\multicolumn{1}{|c|}{$g_0$} 				&	0.68	 &  $\Delta_{t_R}$(GeV)	&1780		\\
\multicolumn{1}{|c|}{$g_{0Y}$} 			&	0.37	 & $Y_T$ (GeV)	     	&2442		\\
\multicolumn{1}{|c|}{$\langle h \rangle$ (GeV)} &      251	 & $M_{Y_T}$(GeV) 	&	$-1231$	\\
\cline{1-2}
		&	 	 & $\Delta_{b_L}$(GeV)  &77		\\
		&		 & $\Delta_{b_R}$(GeV)  &238		\\
		& 		 & $Y_B$(GeV)		&2884		\\
		& 		 & $M_{Y_B}$(GeV)		&  $-1878 $   	\\
		& 		 & $m_H$ (GeV)		&   126  \\
\cline{3-4}
\end{tabular}}
\subfloat[(b) $f=0.8$ TeV and $g_*=2.5$]
  {\begin{tabular}{l l|l| l|}
\hline
\multicolumn{1}{|c|}{f (GeV)}				&	800	 & $m_*$ (GeV)	     	&	1700	 \\ 
\multicolumn{1}{|c|}{$g_*$} 				&	2.5     	 & $\Delta_{t_L}$ (GeV)	&     1225		\\
\multicolumn{1}{|c|}{$g_0$} 				&	0.67	 &  $\Delta_{t_R}$(GeV)	&1391		\\
\multicolumn{1}{|c|}{$g_{0Y}$} 			&0.36		 & $Y_T$ (GeV)	     	&	2770	\\
\multicolumn{1}{|c|}{$\langle h \rangle$ (GeV)} &	250	 & $M_{Y_T}$(GeV) 	&	$-1339$	\\
\cline{1-2}
								&		 & $\Delta_{b_L}$(GeV)  & 222		\\
 								&		 & $\Delta_{b_R}$(GeV)  &	99	\\
 								& 		 & $Y_B$(GeV)		&2485		\\
 								& 		 & $M_{Y_B}$(GeV)		&   $-1185$   	\\
 								& 		 & $m_H$ (GeV)		&   125  \\
\cline{3-4}
\end{tabular}}
\\
\subfloat[(c) $f=1$ TeV and $g_*=2$]
  {\begin{tabular}{l l|l| l|}
\hline
\multicolumn{1}{|c|}{f (GeV)}				&	1000	 & $m_*$ (GeV)	     	&	1915	 \\ 
\multicolumn{1}{|c|}{$g_*$} 				&	   2 	 & $\Delta_{t_L}$ (GeV)	&     1503		\\
\multicolumn{1}{|c|}{$g_0$} 				&	0.69 	 &  $\Delta_{t_R}$(GeV)	&1972		\\
\multicolumn{1}{|c|}{$g_{0Y}$} 			&0.37		 & $Y_T$ (GeV)	     	&2901		\\
\multicolumn{1}{|c|}{$\langle h \rangle$ (GeV)} &	249	 & $M_{Y_T}$(GeV) 	&	$-1303	$\\
\cline{1-2}
								&		 & $\Delta_{b_L}$(GeV)  &196		\\
 								&		 & $\Delta_{b_R}$(GeV)  &187		\\
 								& 		 & $Y_B$(GeV)		&2662		\\
 								& 		 & $M_{Y_B}$(GeV)		&$ -984$     	\\
 								& 		 & $m_H$ (GeV)		& 126    \\
\cline{3-4}
\end{tabular}
}
\subfloat[(d) $f=1$ TeV and $g_*=2.5$]
  {\begin{tabular}{l l|l| l|}
\hline
\multicolumn{1}{|c|}{f (GeV)}				&1000		 & $m_*$ (GeV)	     	&2027 		 \\ 
\multicolumn{1}{|c|}{$g_*$} 				&2.5	     	 & $\Delta_{t_L}$ (GeV)	&   1677  		\\
\multicolumn{1}{|c|}{$g_0$} 				&0.67 		 &  $\Delta_{t_R}$(GeV)	& 2209		\\
\multicolumn{1}{|c|}{$g_{0Y}$} 			&0.36 		 & $Y_T$ (GeV)	     	& 2578		\\
\multicolumn{1}{|c|}{$\langle h \rangle$ (GeV)} &	249 	 & $M_{Y_T}$(GeV) 	&	 $-1146$ 	\\
\cline{1-2}
								&		 & $\Delta_{b_L}$(GeV)  &223		\\
 								&		 & $\Delta_{b_R}$(GeV)  &299		\\
 								& 		 & $Y_B$(GeV)		&2319		\\
 								& 		 & $M_{Y_B}$(GeV)		&  $-1104 $   	\\
 								& 		 & $m_H$ (GeV)		&    125 \\
\cline{3-4}
\end{tabular}}
\\
\subfloat[(e) $f=1.1$ TeV and $g_*=1.8$]
  {\begin{tabular}{l l|l| l|}
\hline
\multicolumn{1}{|c|}{f (GeV)}				&1100		 & $m_*$ (GeV)	     	& 2318 		 \\ 
\multicolumn{1}{|c|}{$g_*$} 				&1.8	     	 & $\Delta_{t_L}$ (GeV)	& 2440 		\\
\multicolumn{1}{|c|}{$g_0$} 				&0.69  		 &  $\Delta_{t_R}$(GeV)	&  1877 		\\
\multicolumn{1}{|c|}{$g_{0Y}$} 			& 0.37		 & $Y_T$ (GeV)	     	& 2909 		\\
\multicolumn{1}{|c|}{$\langle h \rangle$ (GeV)} &  248 	 & $M_{Y_T}$(GeV) 	&$-636$	  	\\
\cline{1-2}
								&		 & $\Delta_{b_L}$(GeV)  &	 272	\\
 								&		 & $\Delta_{b_R}$(GeV)  & 208 		\\
 								& 		 & $Y_B$(GeV)		&2435		\\
 								& 		 & $M_{Y_B}$(GeV)		& $ -1429$    	\\
 								& 		 & $m_H$ (GeV)		& 124\\
\cline{3-4}
\end{tabular}}
\subfloat[(f) $f=1.2$ TeV and $g_*=1.8$]
  {\begin{tabular}{l l|l| l|}
\hline
\multicolumn{1}{|c|}{f (GeV)}				&	1200	 & $m_*$ (GeV)	     	&  	2219 	 \\ 
\multicolumn{1}{|c|}{$g_*$} 				&	1.8     	 & $\Delta_{t_L}$ (GeV)	&  2366 		\\
\multicolumn{1}{|c|}{$g_0$} 				&  0.69		 &  $\Delta_{t_R}$(GeV)	& 2245  		\\
\multicolumn{1}{|c|}{$g_{0Y}$} 			& 0.37 		 & $Y_T$ (GeV)	     	&  2824		\\
\multicolumn{1}{|c|}{$\langle h \rangle$ (GeV)} & 248 	 & $M_{Y_T}$(GeV) 	& $-1043$	  	\\
\cline{1-2}
								&		 & $\Delta_{b_L}$(GeV)  & 202 	\\
 								&		 & $\Delta_{b_R}$(GeV)  & 284		\\
 								& 		 & $Y_B$(GeV)		& 2543		\\
 								& 		 & $M_{Y_B}$(GeV)		& $-1378$    	\\
 								& 		 & $m_H$ (GeV)		& 125\\
\cline{3-4}
\end{tabular}}
\caption{Benchmark points for the generation of Figs.~\ref{fig:NC-Mass},   \ref{fig:CC-Mass},  \ref{fig:NC-AFB} and  \ref{fig:CC-AFB}. Figs.~\ref{fig:NC-Split} and \ref{fig:CC-Split} are generated only with benchmark (f).
The labelling of the benchmarks is as described in the text.}
\label{table:benchmark1}
\end{table}

\section{Phenomenology of $Z'$ and $W'$ production and decay}
\label{sec:results}

We consider the two tree-level processes
\begin{equation}
\label{NC}
pp\to l^+l^-\qquad\qquad{\rm{(NC)}}
\end{equation}
and
\begin{equation}
\label{CC}
pp\to l^+\nu_l~+~{\rm{c.c.}}\qquad\qquad{\rm{(CC)}}
\end{equation}
where $l=e$ or $\mu$. The initial state in both cases includes all possible quark-antiquark subchannels, the latter being dominated (especially at large invariant masses) by contributions involving valence quarks\footnote{In particular,
the $b\bar b$ contribution in the NC case amounts to the percent level.}.

Notice that in the forthcoming analysis we will present the rates for either of the lepton flavours, not the sum of the two.
This will enable us to discuss mass reconstruction efficiencies (resolutions) separately. The latter in fact change significantly from electron to muons. 
Considering the NC, $e^+e^-$ is much better than $\mu^+\mu^-$, as the mass resolution for the former is about 1\% while for the 
latter is about 10\%, assuming a resonance  at 
2 TeV or so (the typical lightest mass of our benchmarks). Here, the mass concerned, reconstructing the $Z'$ resonances, is the invariant one, 
$M_{l^+l^-}\equiv \sqrt{(p_{l^+}+p_{l^-})^2}$. In the case of the CC, the mass resolution is typically 20\% for both and is dominated by the uncertainty
in reconstructing the missing transverse energy/momentum (due to the neutrino escaping detection). Here, the mass concerned, reconstructing the $W'$ resonances,
is the transverse one, 
$M_T\equiv\sqrt{(E^T_l+E^T_{\rm miss})^2
             -(p^x_{l}+p^x_{\rm miss})^2
             -(p^y_{l}+p^y_{\rm miss})^2}$,
where $E^T$ represents missing energy/momentum (as we consider the electron and muon massless) in the transverse plane and $p_{x,y}$ are the two components therein 
(assuming that the proton beams are directed along the $z$ axis). Finally, notice that the upcoming plots we will use the benchmarks defined in the previous Section, i.e.,
the set (a)--(f) of Tab.~\ref{table:benchmark1} and 
the set (1)--(6) of Tab.~\ref{table:benchmark2} which correspond to the gauge boson masses and widths of
Tabs.~\ref{table:benchmark1bis} and \ref{table:benchmark2bis}, respectively. 

\begin{table}[htb]
\captionsetup[subfloat]{labelformat=empty,position=top}
\centering
\subfloat[(a) $f=0.75$ TeV and $g_*=2$]
 {\begin{tabular}{|l|l|l||}
\hline
 & $M$ (GeV) & $\Gamma$ (GeV) \\
\hline
$Z_2$ &1549 &28 \\
$Z_3$ &1581 &26 \\
$Z_5$ &2124 &34 \\
\hline
$W_2$ &1581 &26 \\
$W_3$ &2123 &33 \\
\hline
\hline
\multicolumn{1}{|c|}{} & $l^+ l^-$& $l^+\nu$ + c.c. \\
\hline
\multicolumn{1}{|c|}{$\sigma$ (fb)} & 7.44[5.46]& 13.22[6.96] \\
\hline
\multicolumn{1}{|c|}{$p_l^T$(GeV)} &$>20$ &$>20$ \\
\multicolumn{1}{|c|}{$|\eta_l|$} &$<2.5$ & $<2.5$ \\
\multicolumn{1}{|c|}{$M_{l^+l^-/T}$(TeV)} & $>1$ &$>1$  \\
\cline{1-3}
\end{tabular}}
\subfloat[(b) $f=0.8$ TeV and $g_*=2.5$]
  {\begin{tabular}{|l|l|l||}
\hline
 & $M$ (GeV) & $\Gamma$ (GeV) \\
\hline
$Z_2$ &2041 &61 \\
$Z_3$ &2068 &98 \\
$Z_5$ &2830 &223 \\
\hline
$W_2$ &2067 &98 \\
$W_3$ &2830 &221 \\
\hline
\hline
\multicolumn{1}{|c|}{} & $l^+ l^-$& $l^+\nu$ + c.c. \\
\hline
\multicolumn{1}{|c|}{$\sigma$ (fb)} & 0.90[0.91]& 1.19[1.06]\\
\hline
\multicolumn{1}{|c|}{$p_l^T$(GeV)} &$>20$ &$>20$ \\
\multicolumn{1}{|c|}{$|\eta_l|$} &$<2.5$ & $<2.5$ \\
\multicolumn{1}{|c|}{$M_{l^+l^-/T}$(TeV)} & $>1.5$ &$>1.5$  \\
\cline{1-3}
\end{tabular}}
\\
\subfloat[(c) $f=1$ TeV and $g_*=2$]
  {\begin{tabular}{|l|l|l||}
\hline
 & $M$ (GeV) & $\Gamma$ (GeV) \\
\hline
$Z_2$ &2066 &39 \\
$Z_3$ &2111 &52 \\
$Z_5$ &2830 &71 \\
\hline
$W_2$ &2111 &52 \\
$W_3$ &2830 &50 \\
\hline
\hline
\multicolumn{1}{|c|}{} & $l^+ l^-$& $l^+\nu$ + c.c. \\
\hline
\multicolumn{1}{|c|}{$\sigma$ (fb)} & 1.24[0.91]& 2.04[1.06] \\
\hline
\multicolumn{1}{|c|}{$p_l^T$(GeV)} &$>20$ &$>20$ \\
\multicolumn{1}{|c|}{$|\eta_l|$} &$<2.5$ & $<2.5$ \\
\multicolumn{1}{|c|}{$M_{l^+l^-/T}$(TeV)} & $>1.5$ &$>1.5$  \\
\cline{1-3}
\end{tabular}}
\subfloat[(d) $f=1$ TeV and $g_*=2.5$]
  {\begin{tabular}{|l|l|l||}
\hline
 & $M$ (GeV) & $\Gamma$ (GeV) \\
\hline
$Z_2$ &2552 &81 \\
$Z_3$ &2586 &115 \\
$Z_5$ &3537 &332 \\
\hline
$W_2$ &2586 &114 \\
$W_3$ &3537 &328 \\
\hline
\hline
\multicolumn{1}{|c|}{} & $l^+ l^-$& $l^+\nu$ + c.c. \\
\hline
\multicolumn{1}{|c|}{$\sigma$ (fb)} & 0.21[0.21]& 0.28[0.23] \\
\hline
\multicolumn{1}{|c|}{$p_l^T$(GeV)} &$>20$ &$>20$ \\
\multicolumn{1}{|c|}{$|\eta_l|$} &$<2.5$ & $<2.5$ \\
\multicolumn{1}{|c|}{$M_{l^+l^-/T}$(TeV)} & $>2$ &$>2$  \\
\cline{1-3}
\end{tabular}}
\\
\subfloat[(e) $f=1.1$ TeV and $g_*=1.8$]
  {\begin{tabular}{|l|l|l||}
\hline
 & $M$ (GeV) & $\Gamma$ (GeV) \\
\hline
$Z_2$ &2061 &24 \\
$Z_3$ &2119 &45 \\
$Z_5$ &2802 &42 \\
\hline
$W_2$ &2119 &45 \\
$W_3$ &2802 &42 \\
\hline
\hline
\multicolumn{1}{|c|}{} & $l^+ l^-$& $l^+\nu$ + c.c. \\
\hline
\multicolumn{1}{|c|}{$\sigma$ (fb)} & 1.37[0.21]& 1.33[0.23] \\
\hline
\multicolumn{1}{|c|}{$p_l^T$(GeV)} &$>20$ &$>20$ \\
\multicolumn{1}{|c|}{$|\eta_l|$} &$<2.5$ & $<2.5$ \\
\multicolumn{1}{|c|}{$M_{l^+l^-/T}$(TeV)} & $>2$ &$>2$  \\
\cline{1-3}
\end{tabular}}
\subfloat[(f) $f=1.2$ TeV and $g_*=1.8$]
  {\begin{tabular}{|l|l|l||}
\hline
 & $M$ (GeV) & $\Gamma$ (GeV) \\
\hline
$Z_2$ &2249 &32 \\
$Z_3$ &2312 &55 \\
$Z_5$ &3056 &54 \\
\hline
$W_2$ &2312 &55 \\
$W_3$ &3056 &54 \\
\hline
\hline
\multicolumn{1}{|c|}{} & $l^+ l^-$& $l^+\nu$ + c.c. \\
\hline
\multicolumn{1}{|c|}{$\sigma$ (fb)} & 0.78[0.21]& 1.11[0.23] \\
\hline
\multicolumn{1}{|c|}{$p_l^T$(GeV)} &$>20$ &$>20$ \\
\multicolumn{1}{|c|}{$|\eta_l|$} &$<2.5$ & $<2.5$ \\
\multicolumn{1}{|c|}{$M_{l^+l^-/T}$(TeV)} & $>2$ &$>2$  \\
\cline{1-3}
\end{tabular}}
\caption{Gauge boson masses and width in GeV arising from the benchmarks of Tab.~\ref{table:benchmark1}.
The labelling of the benchmarks is as described in the text. Here, $\sigma$ represents the 4DCHM cross section after the cuts described
herein, with the value in square brakets being the reference SM result.}
\label{table:benchmark1bis}
\end{table}

\begin{table}[htb]
\captionsetup[subfloat]{labelformat=empty,position=top}
\centering
\subfloat[(1) red]
{
\begin{tabular}{|l|l|l|}
\hline
 &$M$ (GeV) &$\Gamma$ (GeV) \\
\hline
$Z_2$&2249&32 \\
$Z_3$&2312&55 \\
$Z_5$&3056&54 \\
\hline
$W_2$&2312&55 \\
$W_3$&3056&54 \\
\hline
\hline
\multicolumn{1}{|c|}{} & $l^+ l^-$& $l^+\nu$ + c.c. \\
\hline
\multicolumn{1}{|c|}{$\sigma$ (fb)} & 0.78 & 1.11 \\
\hline
\end{tabular}
}
\subfloat[(2) green]
{\begin{tabular}{|l|l|l|}
\hline
 &$M$ (GeV) &$\Gamma$ (GeV) \\
\hline
$Z_2$&2249& 48\\
$Z_3$&2312& 86\\
 $Z_5$&3056& 389\\
\hline
 $W_2$&2312& 86\\
 $W_3$&3056& 381\\
\hline
\hline
\multicolumn{1}{|c|}{} & $l^+ l^-$& $l^+\nu$ + c.c. \\
\hline
\multicolumn{1}{|c|}{$\sigma$ (fb)} & 0.56 & 0.79 \\
\hline
\end{tabular}
}
\\
\subfloat[(3) cyan]
{\begin{tabular}{|l|l|l|}
\hline
 &$M$ (GeV) &$\Gamma$ (GeV) \\
\hline
 $Z_2$&2249&60 \\
 $Z_3$&2312&92 \\
 $Z_5$&3056& 192\\
\hline
 $W_2$&2312&91 \\
 $W_3$&3056& 172\\
\hline
\hline
\multicolumn{1}{|c|}{} & $l^+ l^-$& $l^+\nu$ + c.c. \\
\hline
\multicolumn{1}{|c|}{$\sigma$ (fb)} & 0.52 & 0.76 \\
\hline
\end{tabular}
}
\subfloat[(4) magenta]
{\begin{tabular}{|l|l|l|}
\hline
 &$M$ (GeV) &$\Gamma$ (GeV) \\
\hline
 $Z_2$&2249& 75\\
 $Z_3$&2312& 104\\
 $Z_5$&3056& 313\\
\hline
 $W_2$&2312& 104\\
 $W_3$&3056& 293\\
\hline
\hline
\multicolumn{1}{|c|}{} & $l^+ l^-$& $l^+\nu$ + c.c. \\
\hline
\multicolumn{1}{|c|}{$\sigma$ (fb)} & 0.47 & 0.70 \\
\hline
\end{tabular}
}
\\
\subfloat[(5) black]
{\begin{tabular}{|l|l|l|}
\hline
 &$M$ (GeV) &$\Gamma$ (GeV) \\
\hline
 $Z_2$&2249&301 \\
 $Z_3$&2312&434 \\
 $Z_5$&3056&526 \\
\hline
 $W_2$&2312& 434\\
 $W_3$&3056& 522\\
\hline
\hline
\multicolumn{1}{|c|}{} & $l^+ l^-$& $l^+\nu$ + c.c. \\
\hline
\multicolumn{1}{|c|}{$\sigma$ (fb)} & 0.26 & 0.36 \\
\hline
\end{tabular}
}
\subfloat[(6) yellow]
{\begin{tabular}{|l|l|l|}
\hline
 &$M$ (GeV) &$\Gamma$ (GeV) \\
\hline
$Z_2$&2249&1099 \\
$Z_3$&2312& 827\\
$Z_5$&3056& 413\\
\hline
$W_2$&2312& 820\\
$W_3$&3056& 614\\
\hline
\hline
\multicolumn{1}{|c|}{} & $l^+ l^-$& $l^+\nu$ + c.c. \\
\hline
\multicolumn{1}{|c|}{$\sigma$ (fb)} & 0.23 & 0.30 \\
\hline
\end{tabular}
}
\caption{Gauge boson masses and widths in GeV arising from the benchmarks of Tab.~\ref{table:benchmark2}. 
All points assume $f=1.2$ TeV and $g_*=1.8$. 
The numbering and colour scheme of the benchmarks is as in Fig.~\ref{plot:Hpotential}.
Here, $\sigma$ represents the 4DCHM cross section after the cuts described
for benchmark (f) in Tab.~{\protect{\ref{table:benchmark1bis}}}, where also the SM reference rates can be found.}
\label{table:benchmark2bis}
\end{table}

\begin{figure}[htb]
\centering
\includegraphics[width=0.35\linewidth,angle=90]{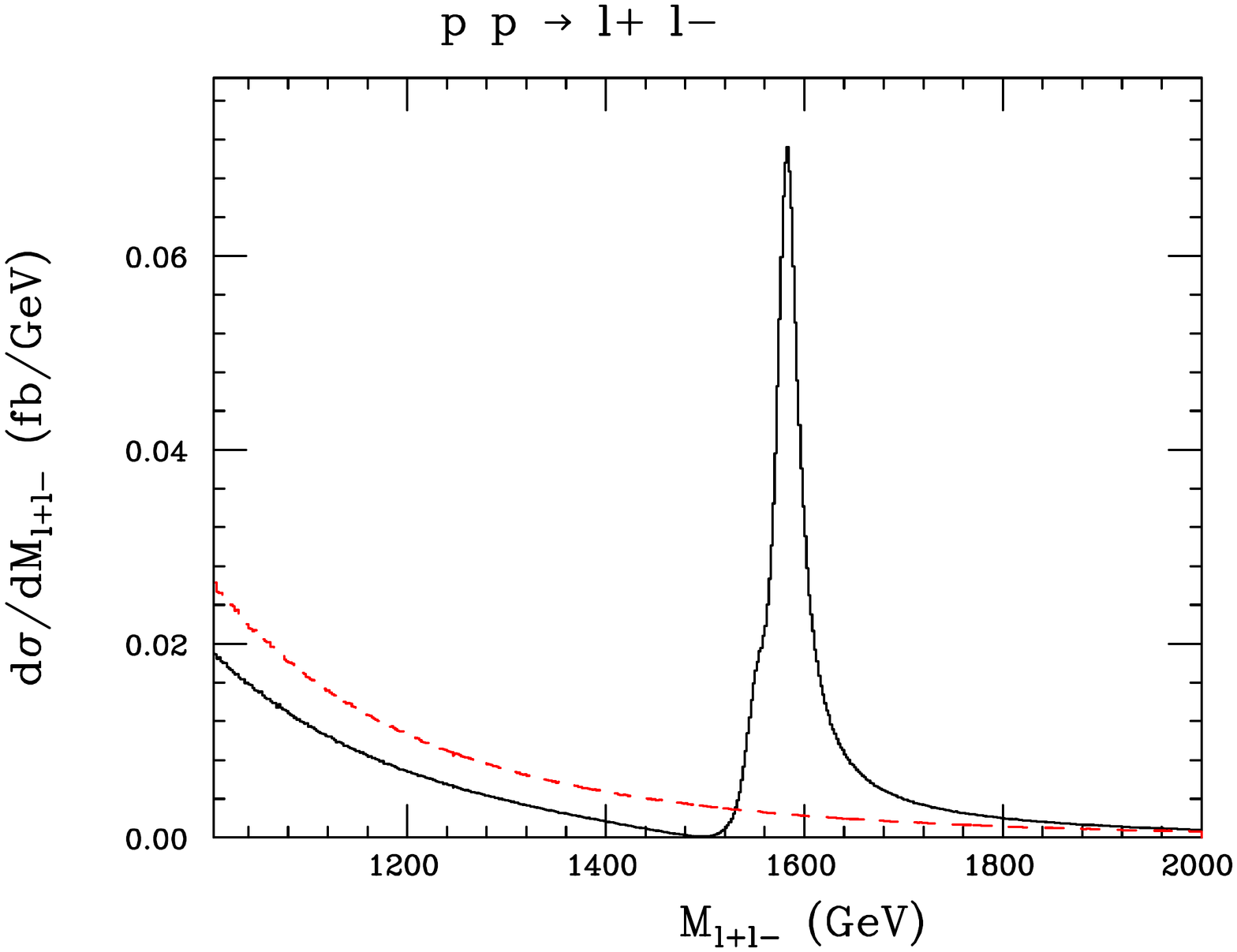}{(a)}
\includegraphics[width=0.35\linewidth,angle=90]{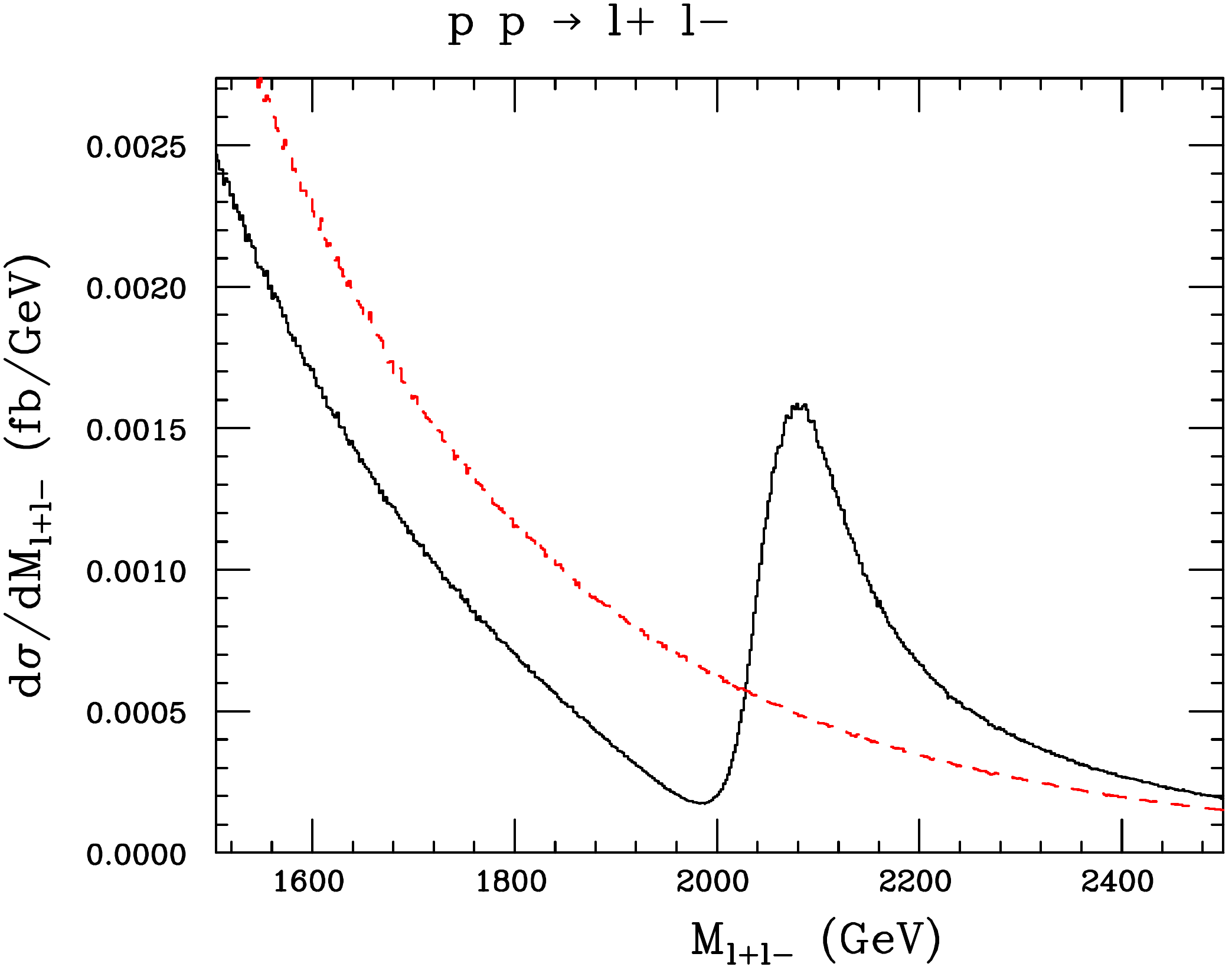}{(b)}
\includegraphics[width=0.35\linewidth,angle=90]{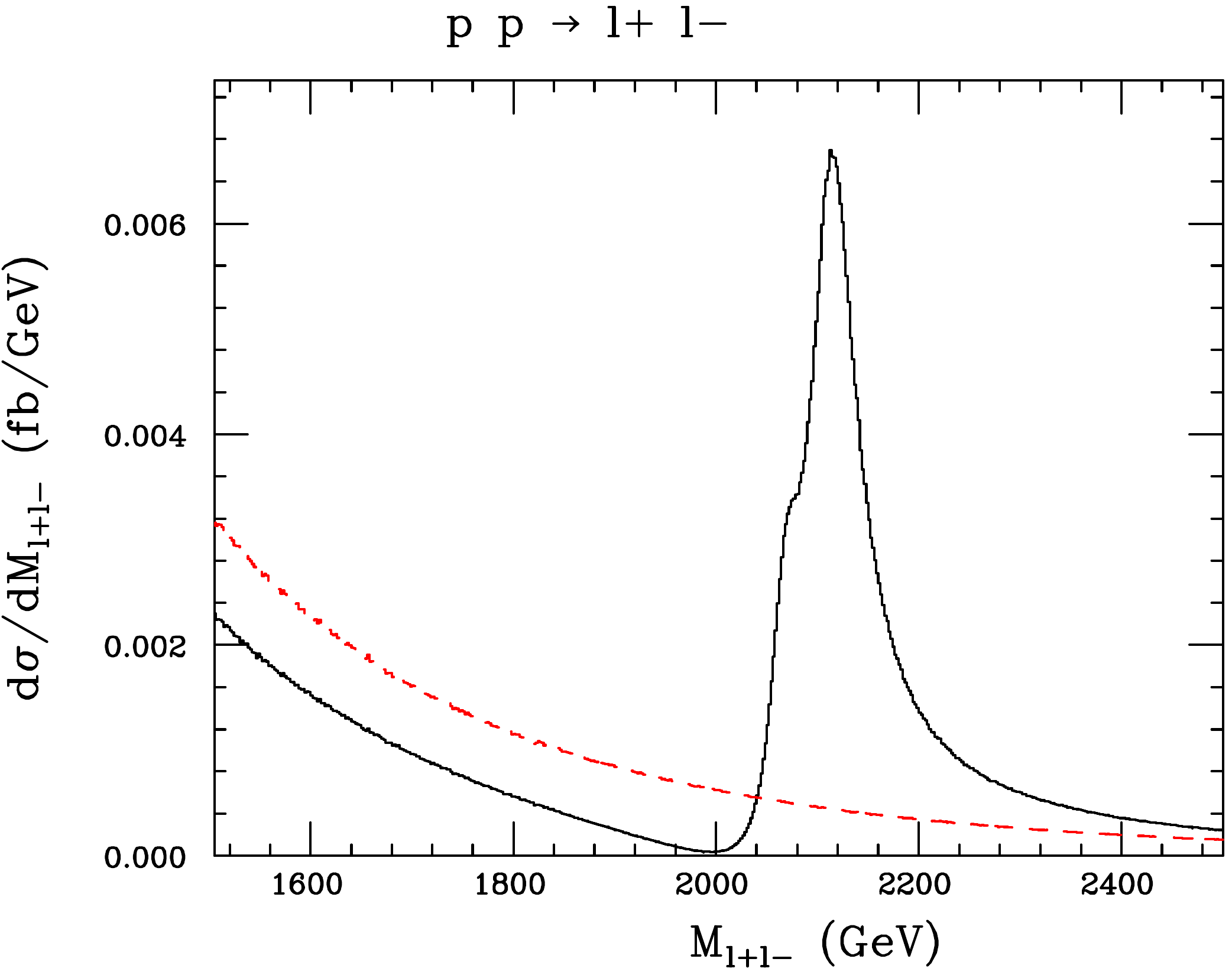}{(c)}
\includegraphics[width=0.35\linewidth,angle=90]{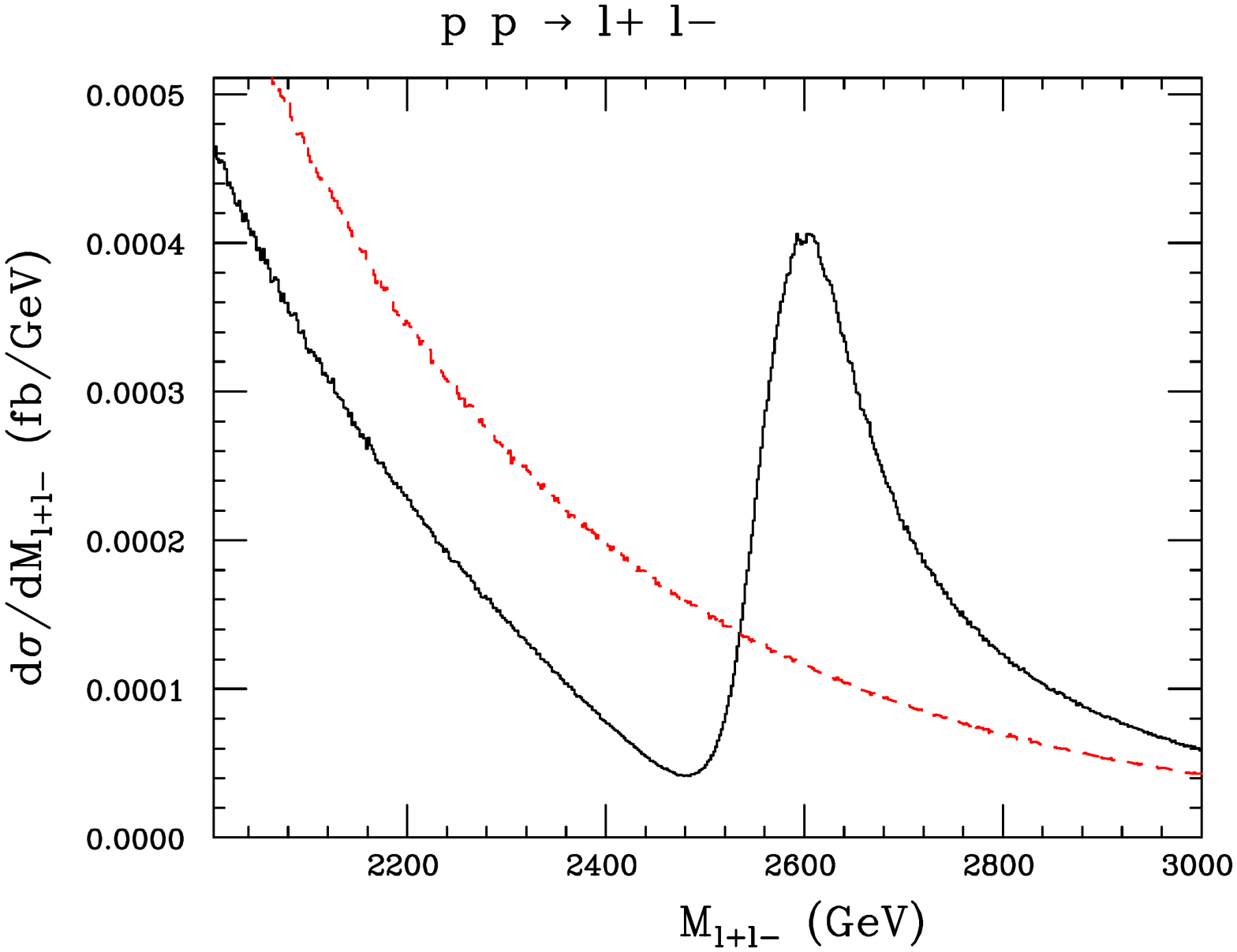}{(d)}
\includegraphics[width=0.35\linewidth,angle=90]{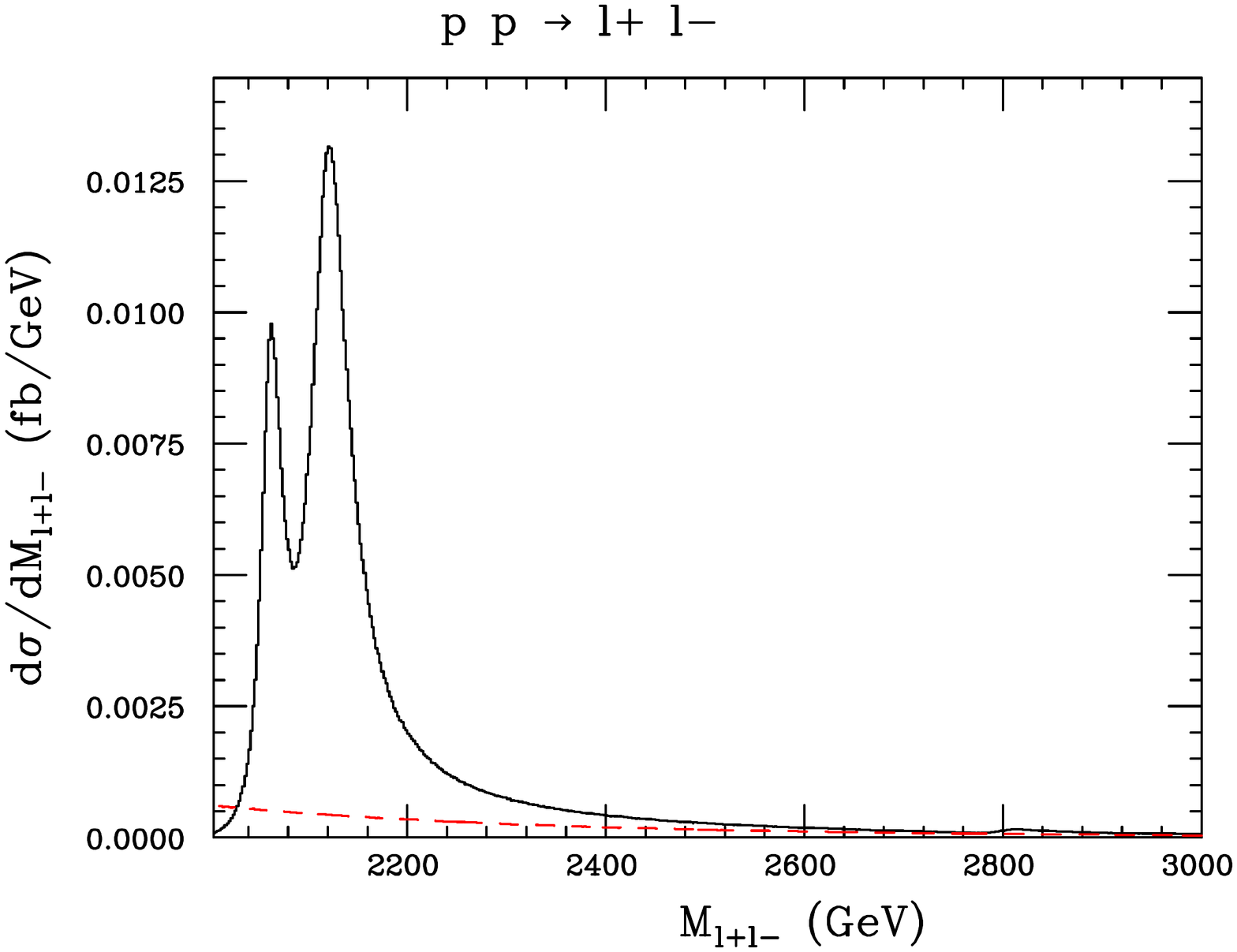}{(e)}
\includegraphics[width=0.35\linewidth,angle=90]{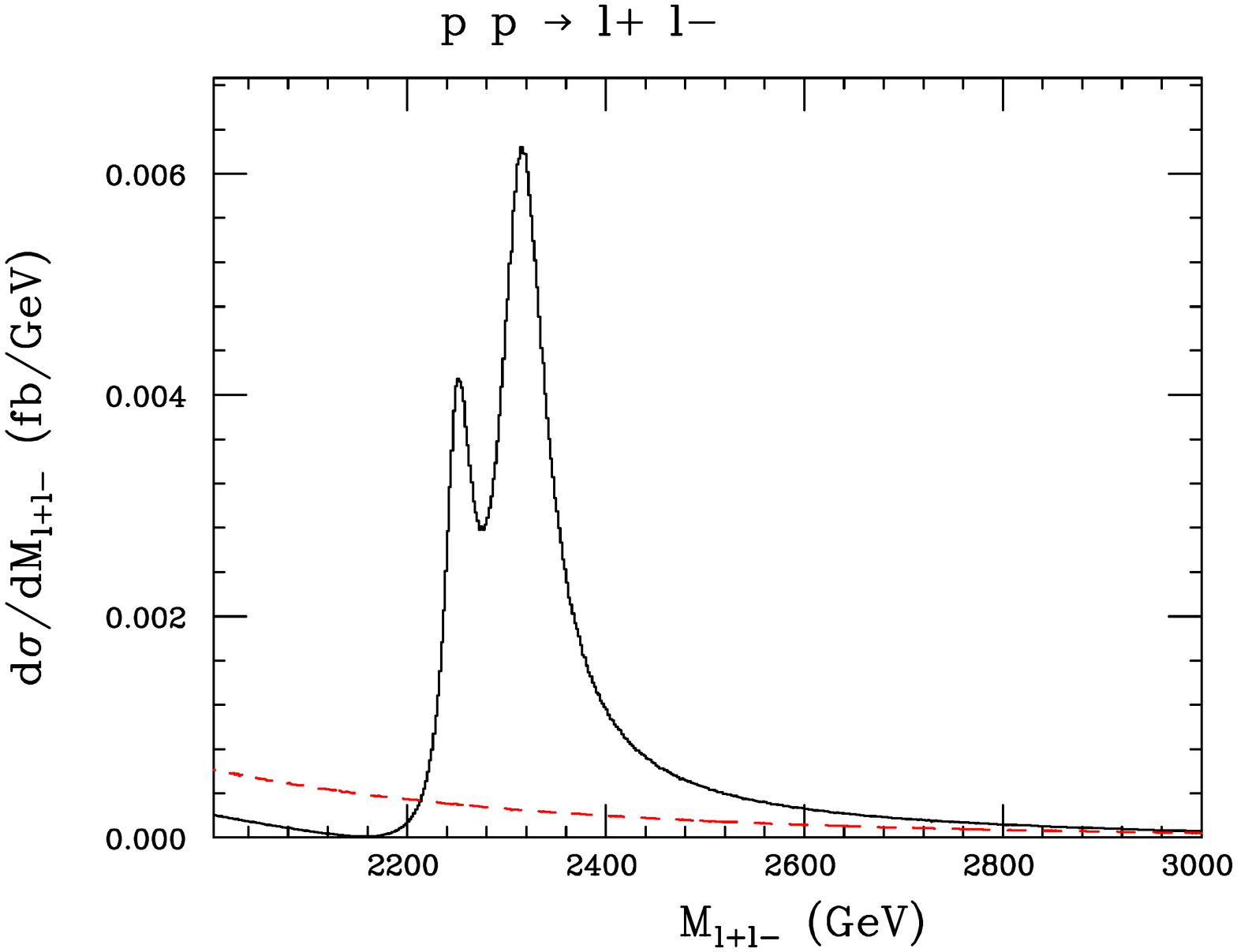}{(f)}
\caption{Differential distributions in invariant mass $M_{l^+l^-}$
for the cross section
at the 14 TeV LHC for NC DY in the 4DCHM (solid) for
(a) $f=0.75$ TeV and $g^*=2$ 
(b) $f=0.8$ TeV and $g^*=2.5$ 
(c) $f=1$ TeV and $g^*=2$ 
(d) $f=1$ TeV and $g^*=2.5$ 
(e) $f=1.1$ TeV and $g^*=1.8$ 
(f) $f=1.2$ TeV and $g^*=1.8$ 
and in the SM (dashed).
The integrated cross sections are
(a) 7.44[5.46] fb  
(b) 0.90[0.91] fb  
(c) 1.24[0.91] fb  
(d) 0.21[0.21] fb  
(e) 1.37[0.21] fb  
(f) 0.78[0.21] fb
for the 4DCHM[SM]
after the cuts 
$p^T_l>20$ GeV, $|\eta_l|<2.5$ and 
(a) $M_{l^+l^-}>1.0$ TeV
(b) $M_{l^+l^-}>1.5$ TeV
(c) $M_{l^+l^-}>1.5$ TeV
(d) $M_{l^+l^-}>2.0$ TeV
(e) $M_{l^+l^-}>2.0$ TeV
(f) $M_{l^+l^-}>2.0$ TeV.
Bin width is here 2 GeV.}
\protect{\label{fig:NC-Mass}}
\end{figure}

Fig.~\ref{fig:NC-Mass} shows the invariant mass distribution of the NC cross section for our six choices of the $f$ and $g_*$ parameters, as detailed in the caption.

While the two lightest $Z'$ resonances, $Z_2$ and $Z_3$,  are clearly accessible, this is obviously not the case for the third one, $Z_5$, which is much suppressed to remain most
probably invisible at the 14 TeV LHC, even with a ${\cal O}(300~{\rm{fb}}^{-1})$ luminosity\footnote{Its presence could however be ascertained at the so-called Super-LHC, an LHC
upgrade involving a tenfold increase in luminosity \cite{Gianotti:2002xx}. So is the case for the $W_3$ state (see further on).}. 
In each plot we also show the SM yield over the same mass range. Standard acceptance and selection cuts
(again, as detailed in the caption) have been adopted here. In all cases a clear excess is seen around the (very close) $Z_2$ and $Z_3$ masses. While in some cases the two resonances
are not resolvable, in others they can be separated, certainly in $e^+e^-$ final states, but not in $\mu^+\mu^-$ ones. Notice in fact that the bin width in this plots is 2 GeV,
so that even assembling 10 of these (in the case of electron-positron pairs) does not spoil the ability to establish the two resonances separately, while such a separation is not
possible for the case of muon-antimuon pairs (as this imply integrating over 100 bins).  Pure SM event rates are very small compared to those emerging
in the 4DCHM (the latter are the complete result, including not only the $Z'$ resonances individually, but also their interferences as well as
the SM contribution due to $\gamma$ and $Z^0$ exchange, together with the relative interferences with the resonant $Z'$ contributions),
to the extent that analyses should essentially be background free in several cases. Overall normalisations of the cross sections are typically of 
${\cal O}(1~{\rm{fb}})$  to ${\cal O}(10~{\rm{fb}})$, thereby rendering each of these parameter configurations of the 4DMCH accessible at the LHC
(albeit through the two lowest mass resonances only).

\begin{figure}[htb]
\centering
\includegraphics[width=0.35\linewidth,angle=90]{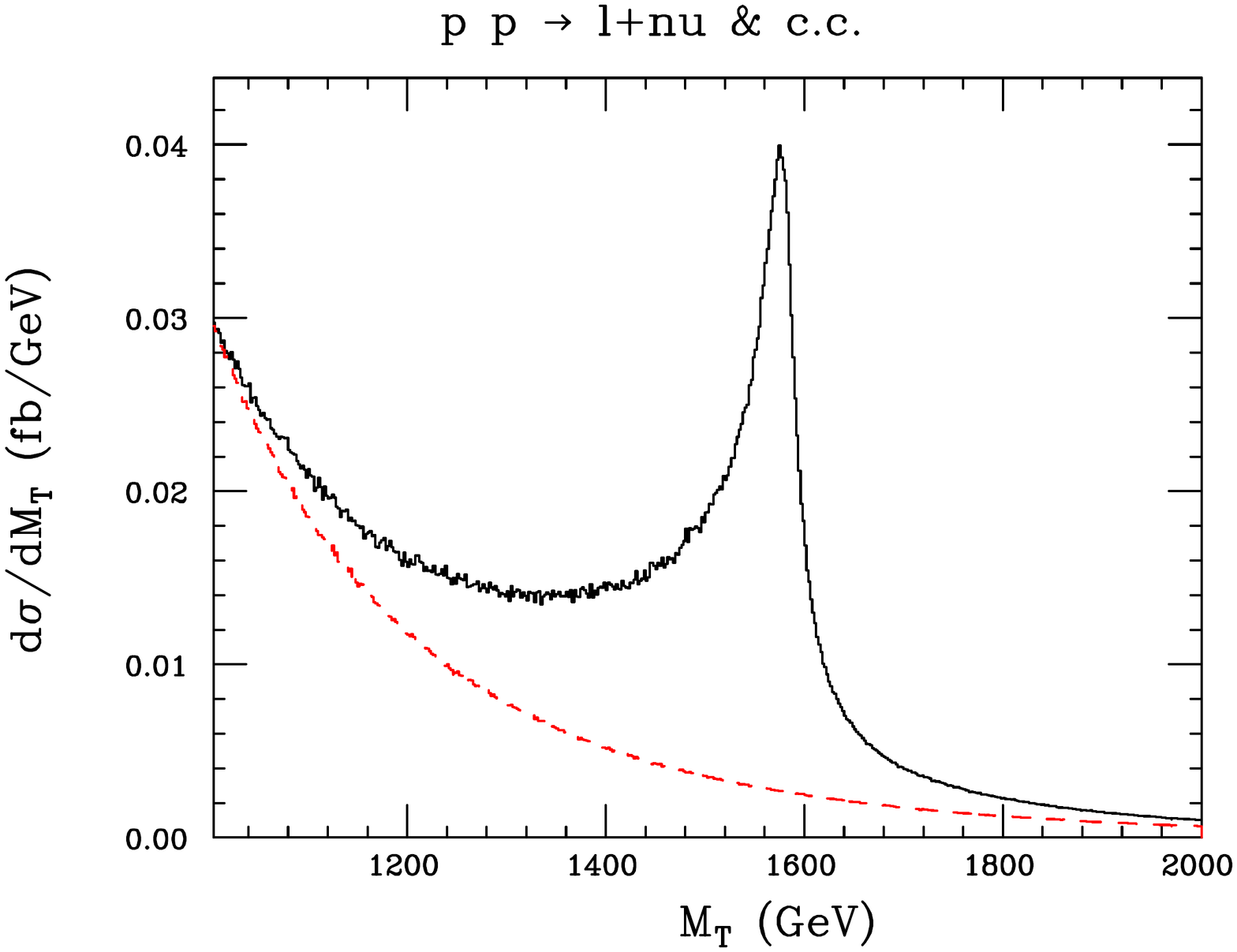}{(a)}
\includegraphics[width=0.35\linewidth,angle=90]{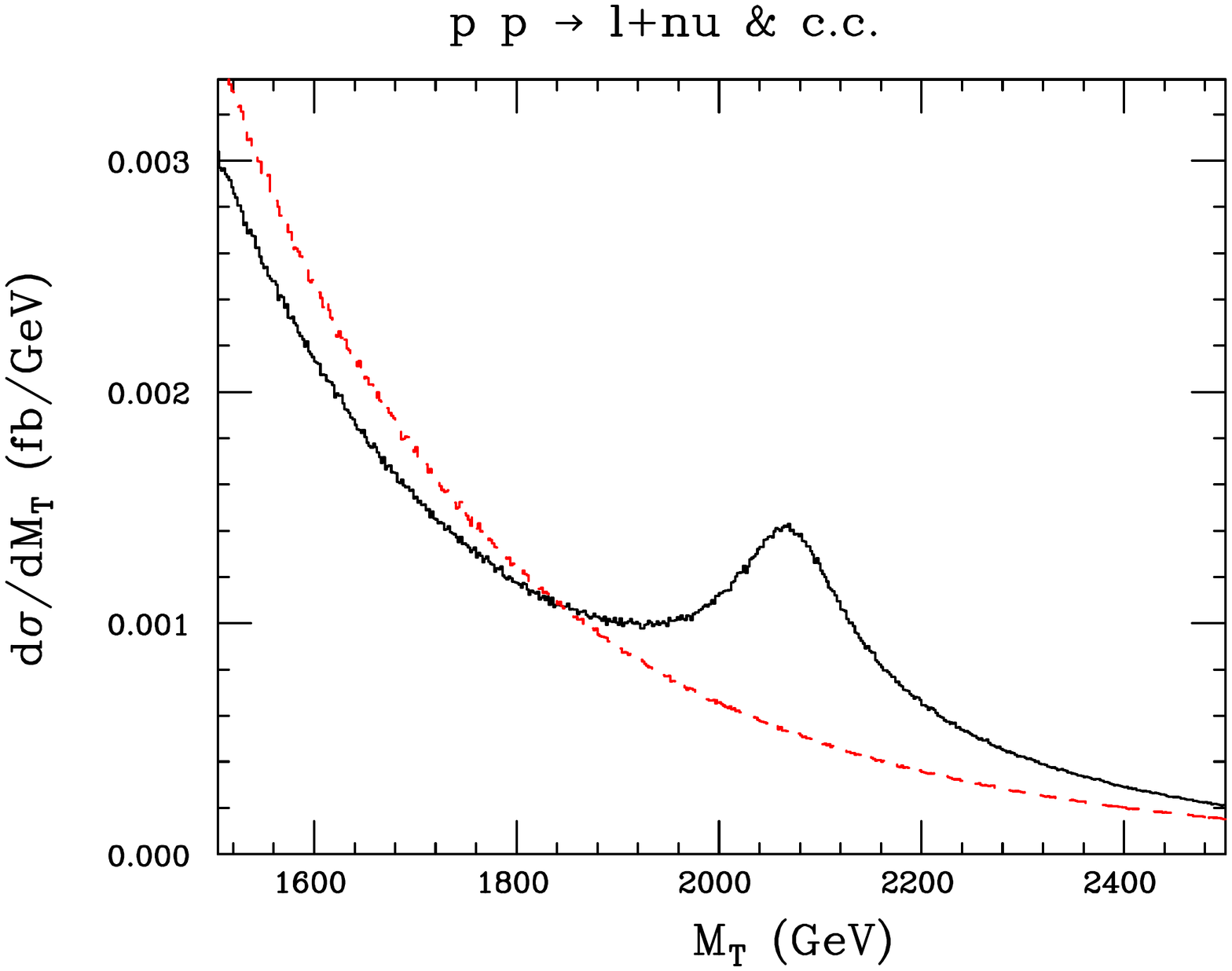}{(b)}
\includegraphics[width=0.35\linewidth,angle=90]{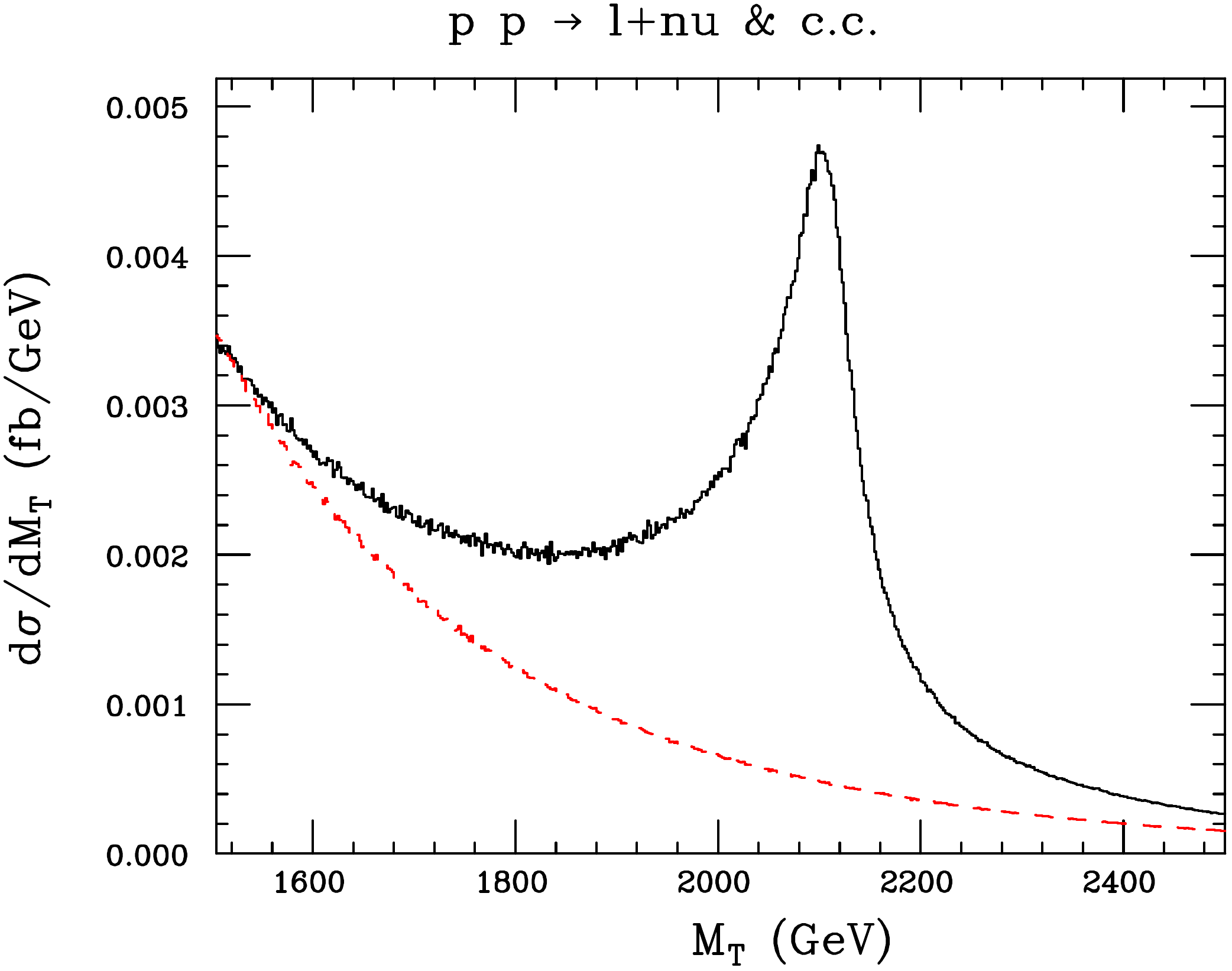}{(c)}
\includegraphics[width=0.35\linewidth,angle=90]{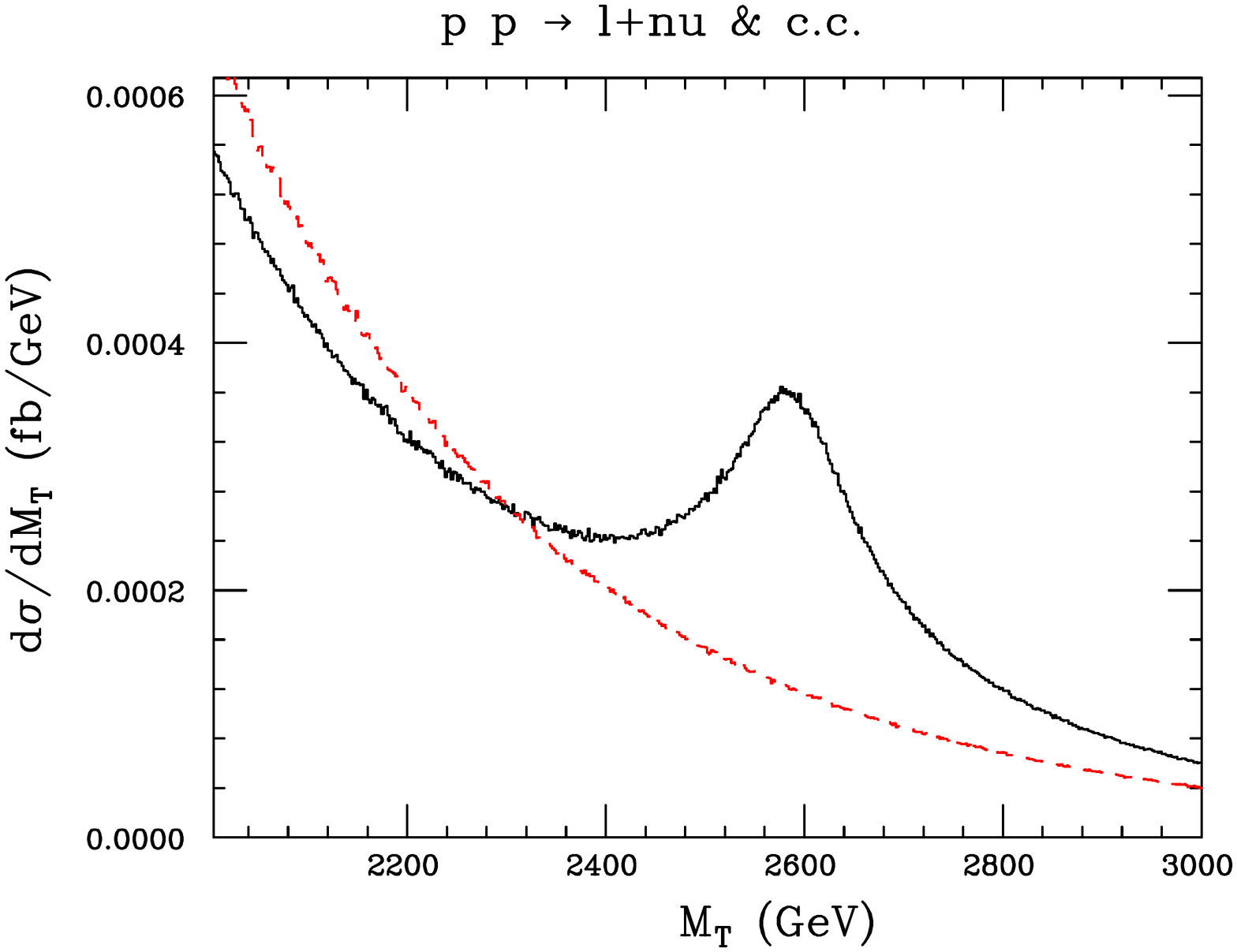}{(d)}
\includegraphics[width=0.35\linewidth,angle=90]{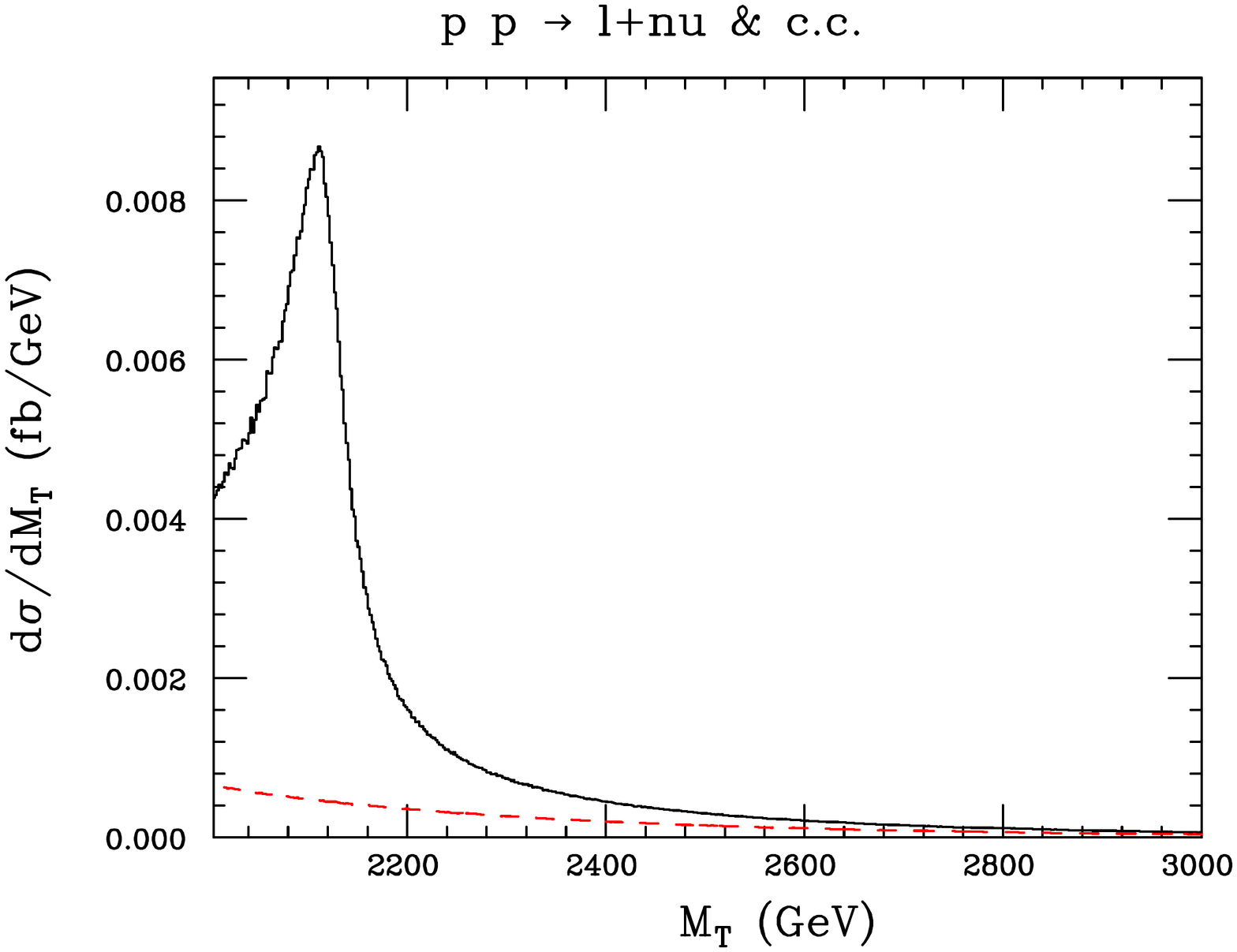}{(e)}
\includegraphics[width=0.35\linewidth,angle=90]{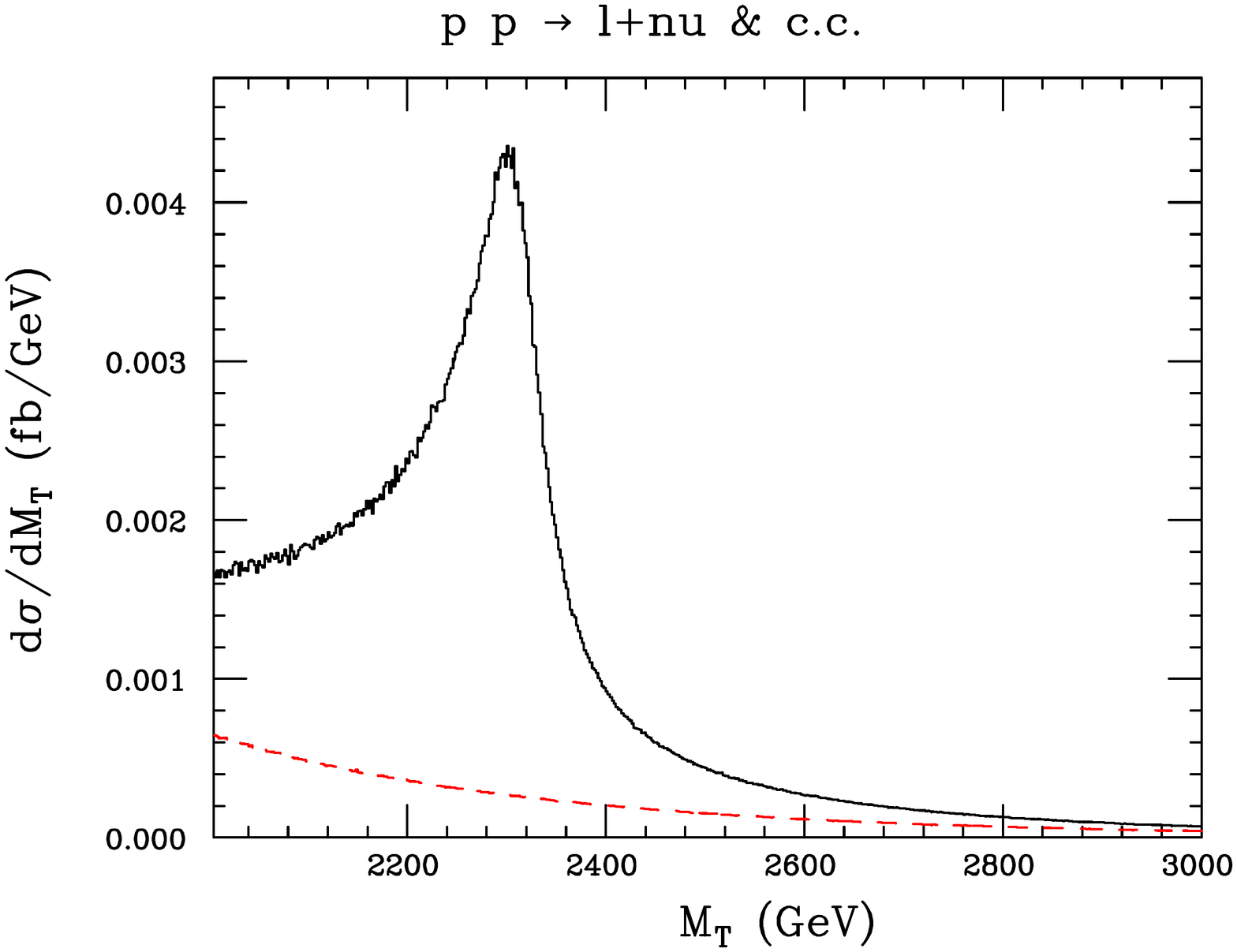}{(f)}
\caption{Differential distributions in transverse mass $M_T$
for the cross section
at the 14 TeV LHC for CC DY in the 4DCHM (solid) for
(a) $f=0.75$ TeV and $g^*=2.0$ 
(b) $f=0.8$ TeV and $g^*=2.5$ 
(c) $f=1$ TeV and $g^*=2$ 
(d) $f=1$ TeV and $g^*=2.5$ 
(e) $f=1.1$ TeV and $g^*=1.8$ 
(f) $f=1.2$ TeV and $g^*=1.8$ 
and in the SM (dashed).
The integrated cross sections are
(a) 13.22[6.96] fb  
(b) 1.19[1.06] fb  
(c) 2.04[1.06] fb  
(d) 0.28[0.23] fb  
(e) 1.33[0.23] fb  
(f) 1.11[0.23] fb
for the 4DCHM[SM]
after the cuts 
$p^T_l>20$ GeV, $|\eta_l|<2.5$ and 
(a) $M_{T}>1.0$ TeV
(b) $M_{T}>1.5$ TeV
(c) $M_{T}>1.5$ TeV
(d) $M_{T}>2.0$ TeV
(e) $M_{T}>2.0$ TeV
(f) $M_{T}>2.0$ TeV.
Bin width is here 2 GeV.}
\protect{\label{fig:CC-Mass}}
\end{figure}

The corresponding case for the CC cross section is displayed in Fig.~\ref{fig:CC-Mass} (again, see the caption for parameter choices, cuts and normalisations). 

Here, only two
resonances are involved, i.e., $W_2$ and $W_3$, though the heaviest one is again unreachable with the foreseen standard LHC configuration. In this channel it is not possible 
to reconstruct the mass of the decaying $W'$ boson from a resonance, rather one uses the transverse mass, as explained, which is much less correlated to the intervening $M_{W'}$ value, so that the visible peaks
are much broader in comparison to the NC case and also somewhat displaced towards smaller masses with respect to the true mass values of the 4DMCH gauge boson. Again though, in all cases, the
enhancement is well visible over the SM noise. Further, event rates are somewhat larger than in the NC case.  

In both the NC and CC channel, it is possible to define the AFB of the cross section, as the direction of the reference incoming quark or antiquark can be inferred from the direction of the boost
onto the final state in the laboratory frame. The  AFB can be sampled in invariant (NC) or transverse (CC) mass bins
\cite{Baur:2001ze}, by defining
\begin{equation}\label{AFB}
\frac{d 
{\rm{AFB}}}{dM}=\frac{d\sigma(\cos\theta>0)/dM-d\sigma(\cos\theta<0)/dM}{d\sigma(\cos\theta>0)/dM+d\sigma(\cos\theta<0)/dM},
\end{equation}
where $\theta$  is the polar angle of the reference outgoing lepton or antilepton relative to the direction of the reference incoming quark or antiquark\footnote{In presence of real QCD radiation, it would become more appropriate to 
define the polar angle in the  Collins-Soper frame \cite{Collins:1977iv}.} and 
$M\equiv M_{l^+l^-}$ for process (\ref{NC})
or
$M\equiv M_{T     }$ for process (\ref{CC}). Notice that in the CC case one can not reconstruct unambiguously the longitudinal component of
the neutrino momentum to define the asymmetry, because of the twofold solution from the mass equation.
So we take both  solutions and plot each of them with half weight, which somewhat dilutes the asymmetry, in order to
individuate the direction of the boost. This is done by assuming that the invariant mass of the final state coincide with the $W$ mass in the
case of the SM hypothesis and with the $W_{2,3}$ mass in the case of the 4DCHM hypothesis (where an indicative
value for the latter can be obtained from the NC, in the spirit  of point 4 discussed in Sect.~\ref{sec:intro} and elaborated
upon later on). 

\begin{figure}[htb]
\centering
\includegraphics[width=0.35\linewidth,angle=90]{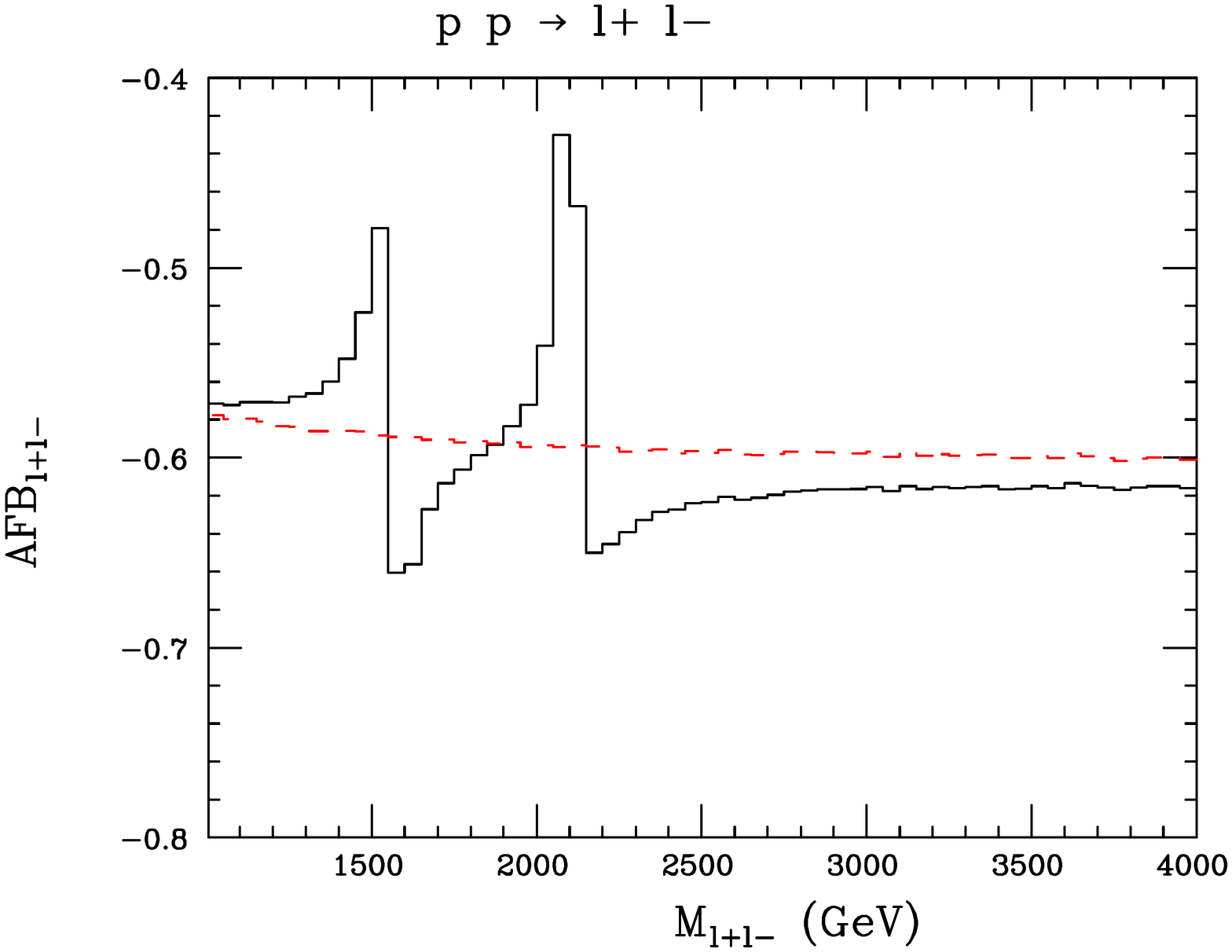}{(a)}
\includegraphics[width=0.35\linewidth,angle=90]{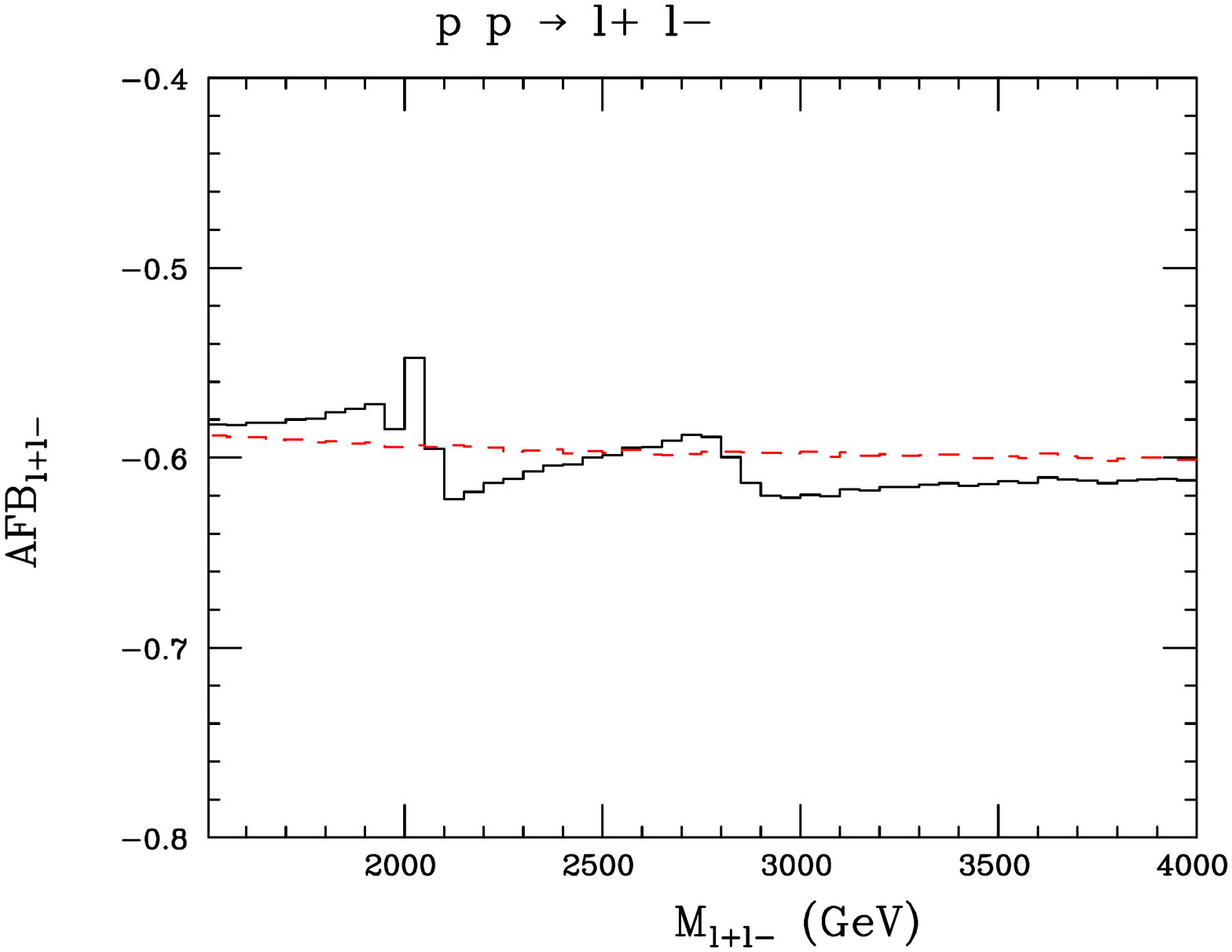}{(b)}
\includegraphics[width=0.35\linewidth,angle=90]{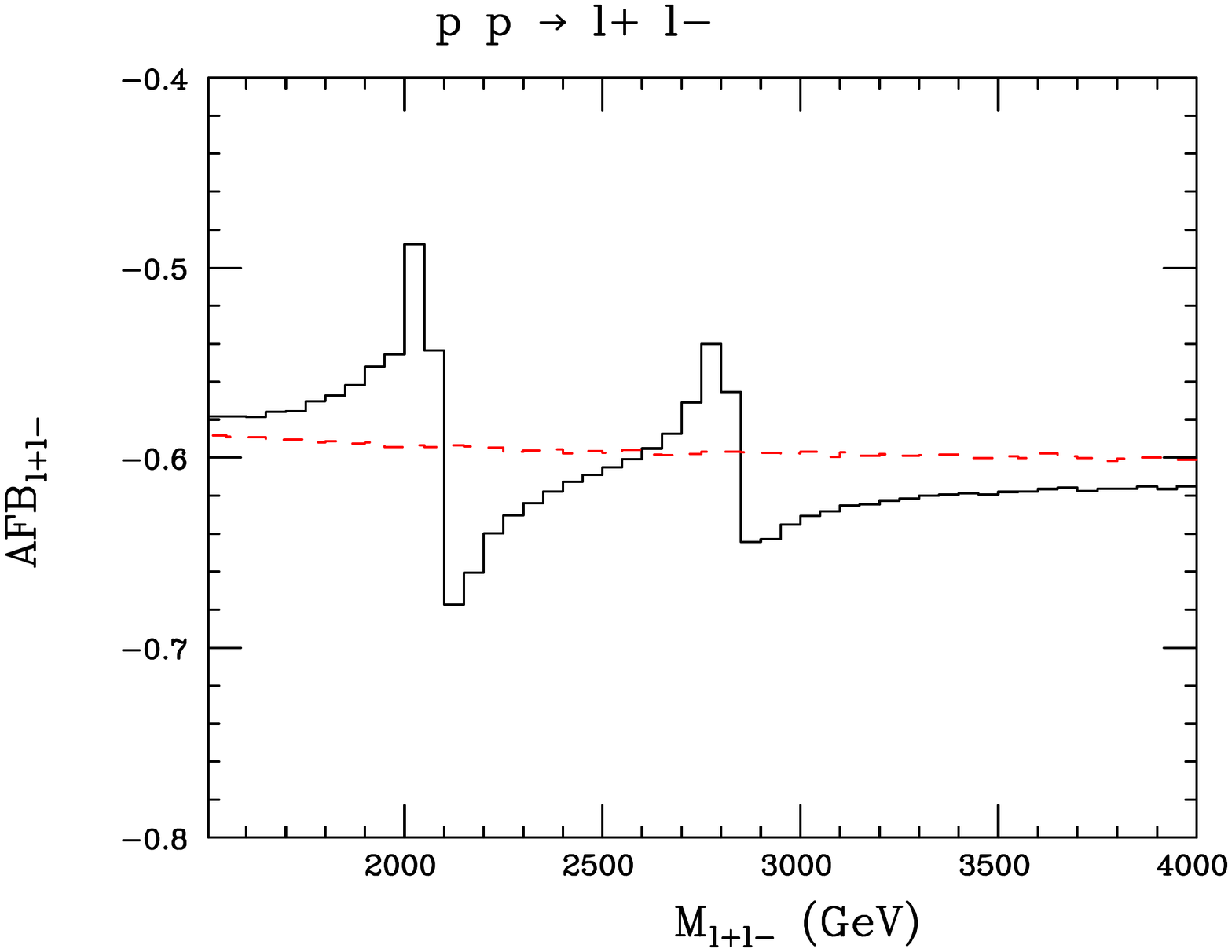}{(c)}
\includegraphics[width=0.35\linewidth,angle=90]{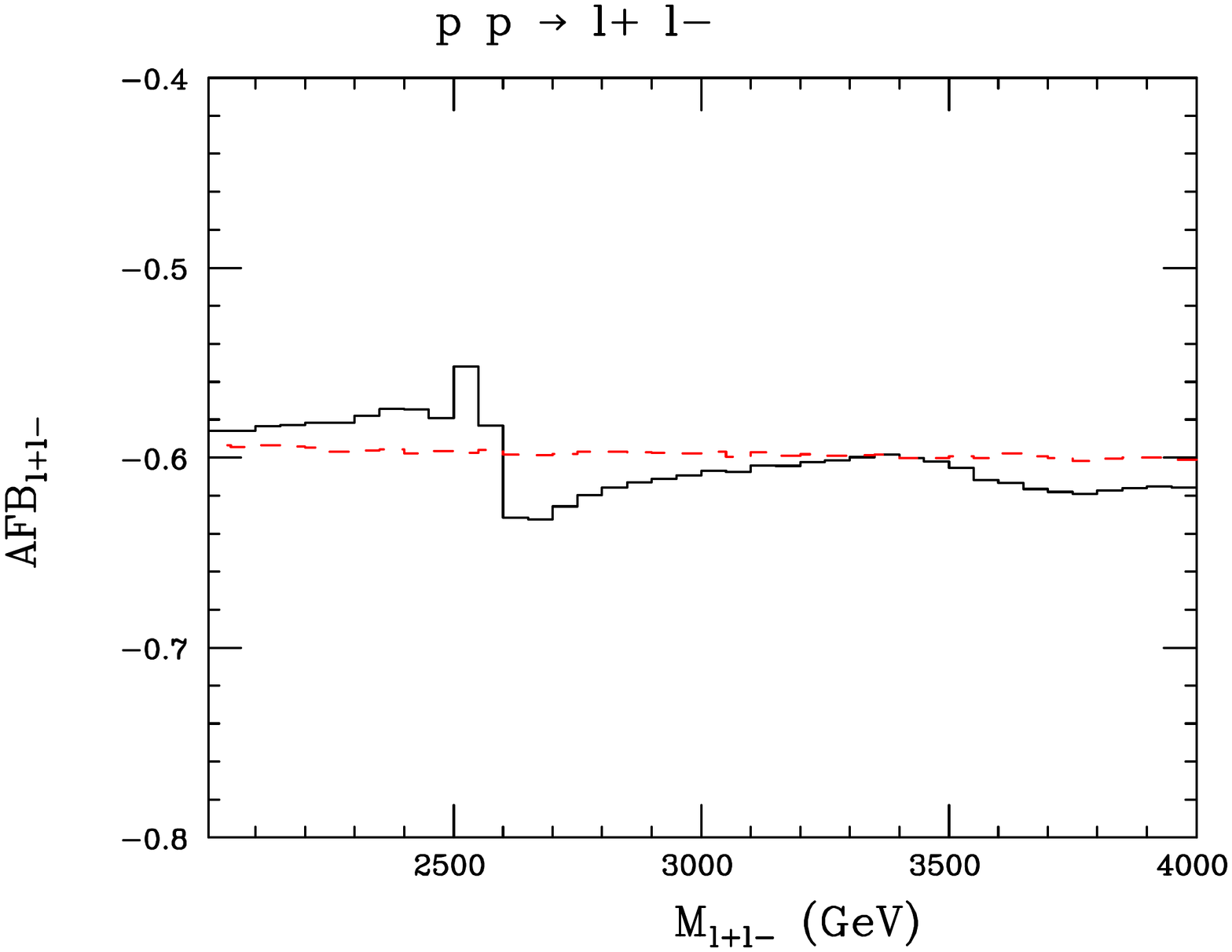}{(d)}
\includegraphics[width=0.35\linewidth,angle=90]{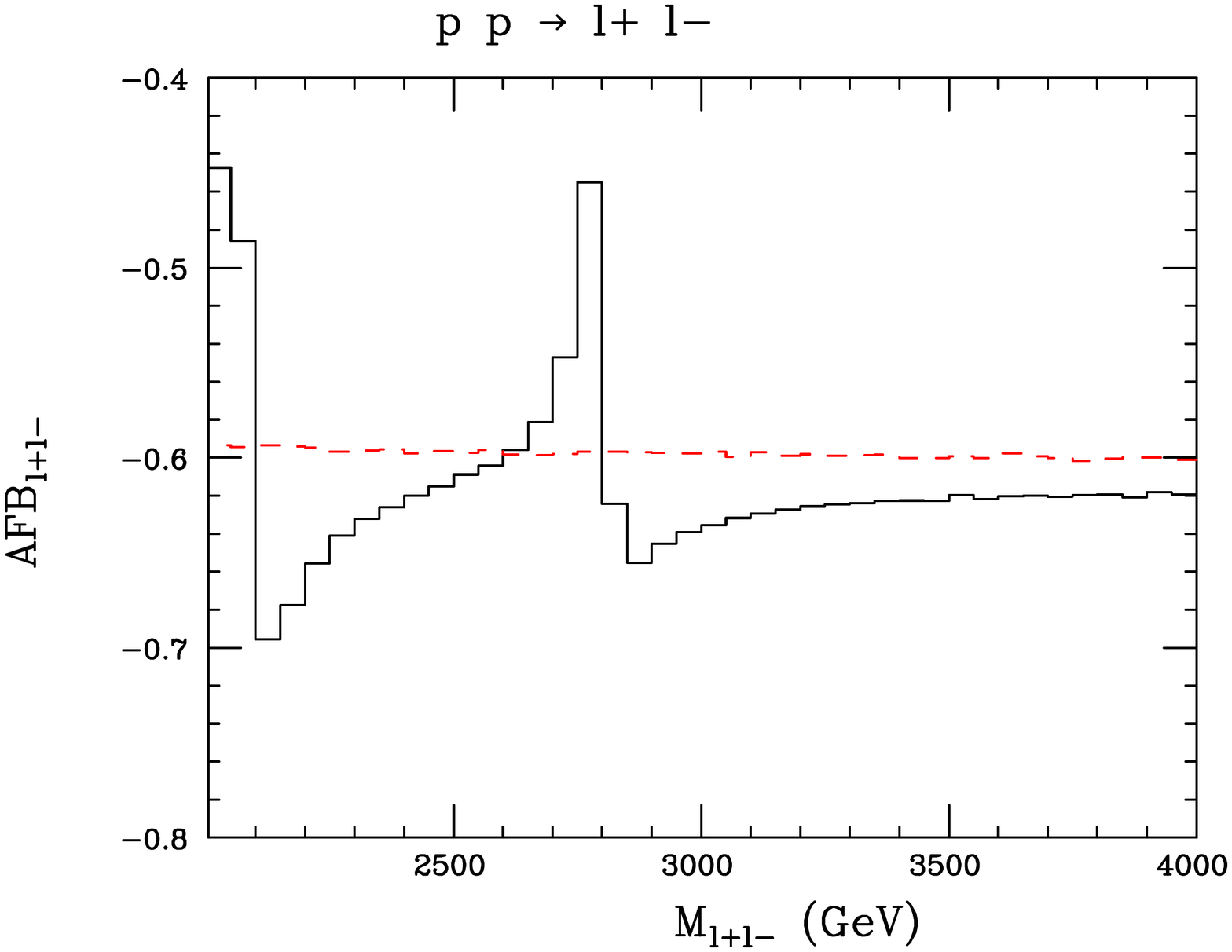}{(e)}
\includegraphics[width=0.35\linewidth,angle=90]{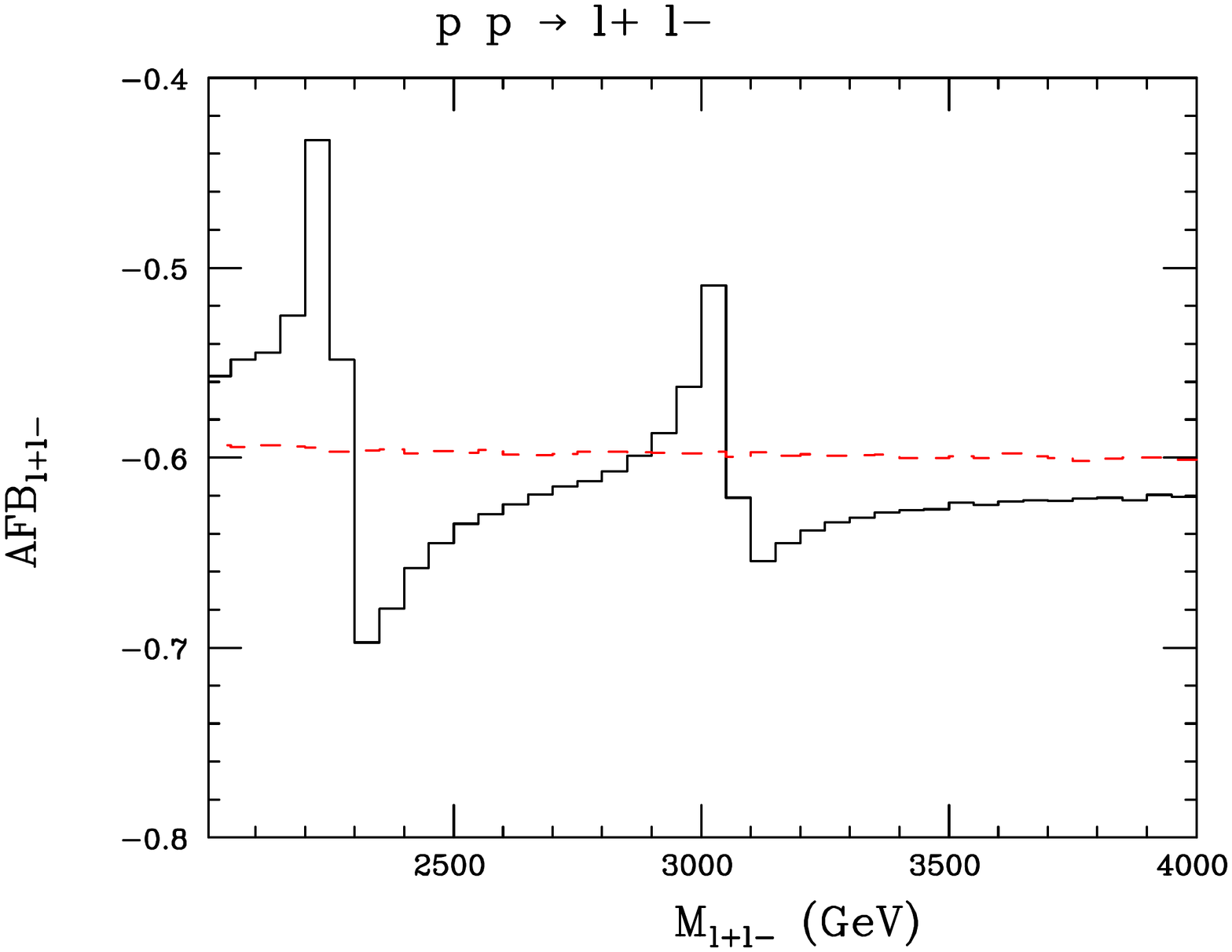}{(f)}
\caption{Differential distributions in invariant mass $M_{l^+l^-}$
for the forward-backward asymmetry
at the 14 TeV LHC for NC DY in the 4DCHM (solid) for
(a) $f=0.75$ TeV and $g^*=2$ 
(b) $f=0.8$ TeV and $g^*=2.5$ 
(c) $f=1$ TeV and $g^*=2$ 
(d) $f=1$ TeV and $g^*=2.5$ 
(e) $f=1.1$ TeV and $g^*=1.8$ 
(f) $f=1.2$ TeV and $g^*=1.8$
and in the SM (dashed).
Cuts, cross sections and mass/width parameters as in the previous plots.
Bin width is here 50 GeV.}
\protect{\label{fig:NC-AFB}}
\end{figure}

Remarkably, for the NC channel, see Fig.~\ref{fig:NC-AFB}, such an observable displays a peculiar dependence in the vicinity of all intermediate $Z'$  masses,
including the heavy one (i.e., $Z_5$), for all $f$ and $g_*$ combinations. Furthermore, these effects should be resolvable no matter the final state, as the reader should notice that
this observable is sampled in the plots over histogram bins which are 50 GeV wide. However, as already mentioned, it remains debatable whether even this observable can be used
to single out the heaviest of the $Z'$'s.

\begin{figure}[htb]
\centering
\includegraphics[width=0.35\linewidth,angle=90]{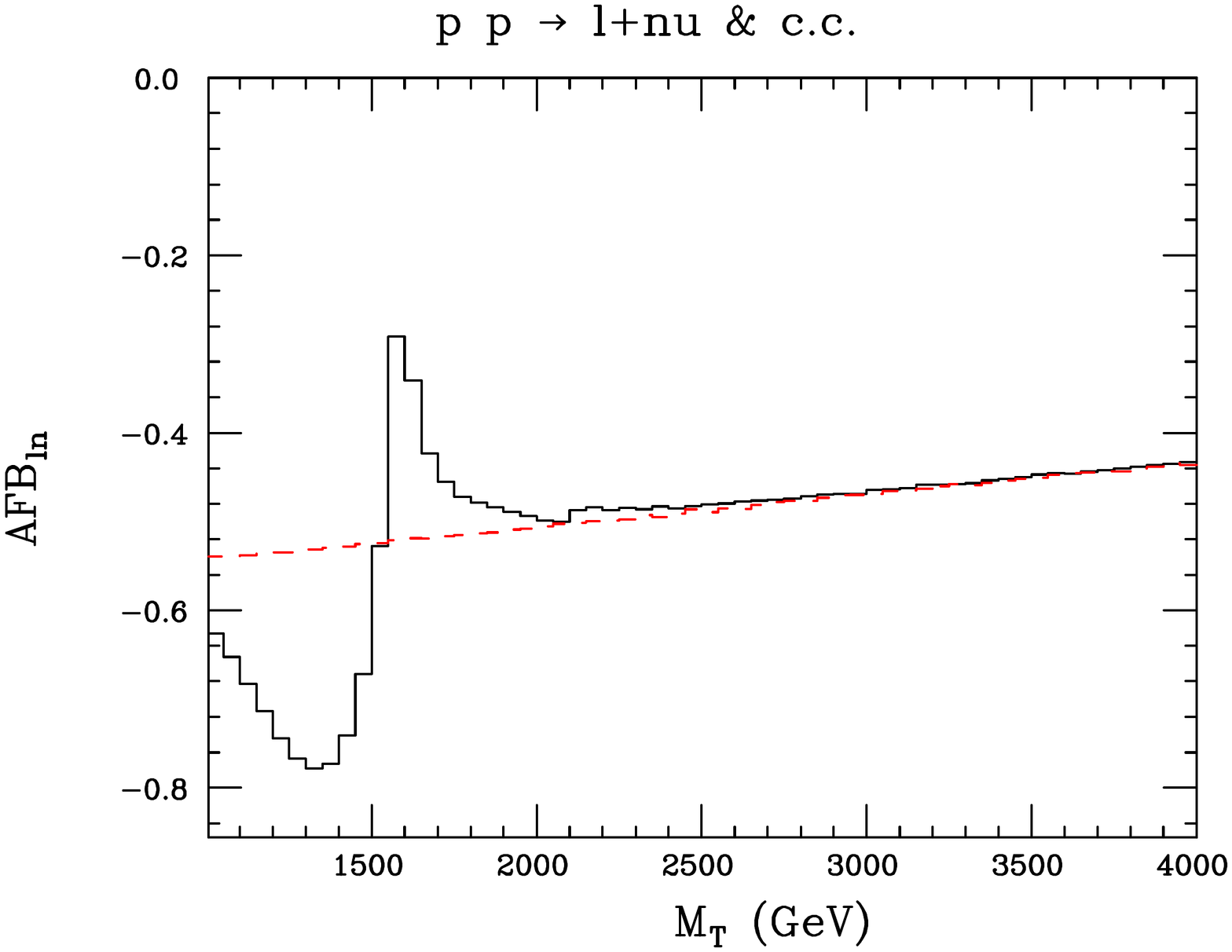}{(a)}
\includegraphics[width=0.35\linewidth,angle=90]{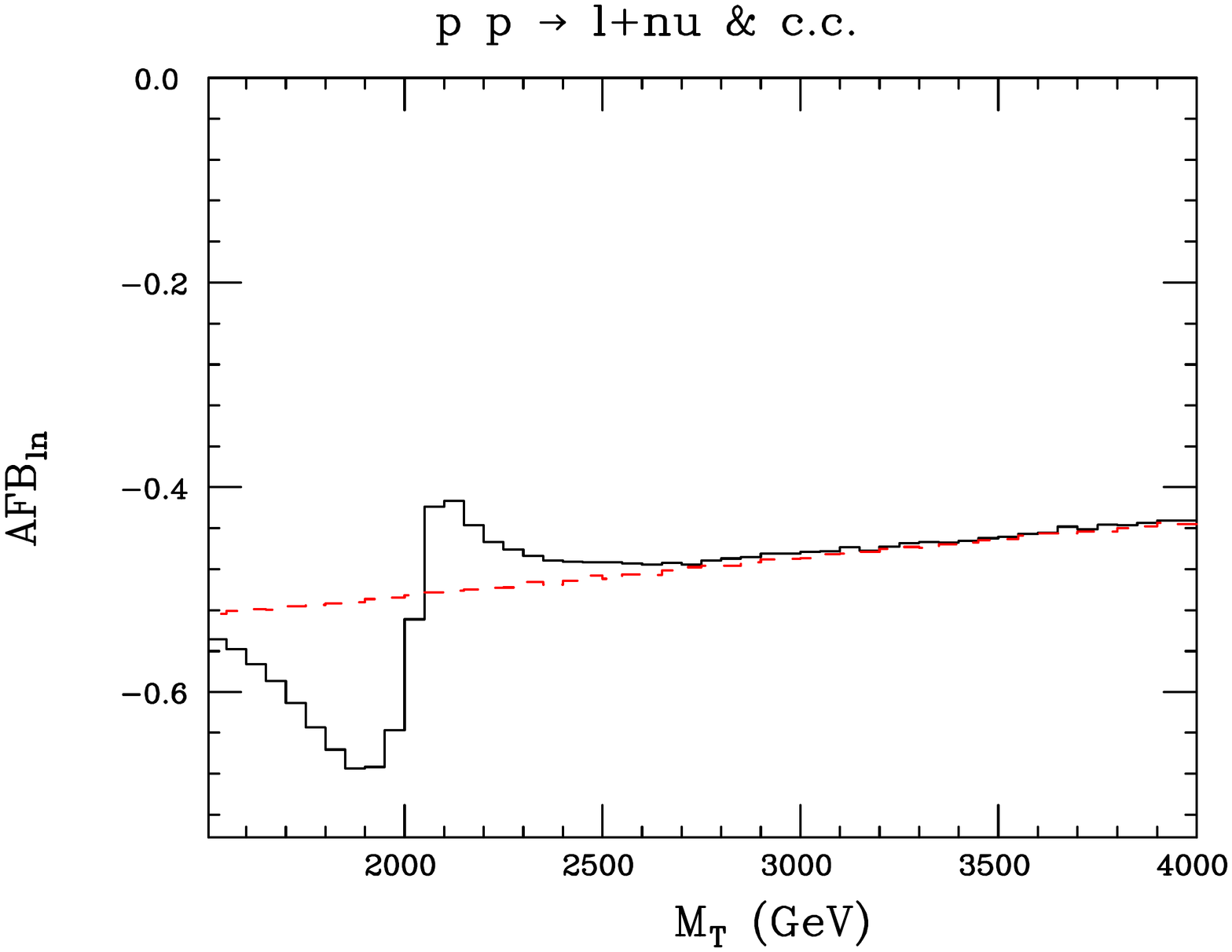}{(b)}
\includegraphics[width=0.35\linewidth,angle=90]{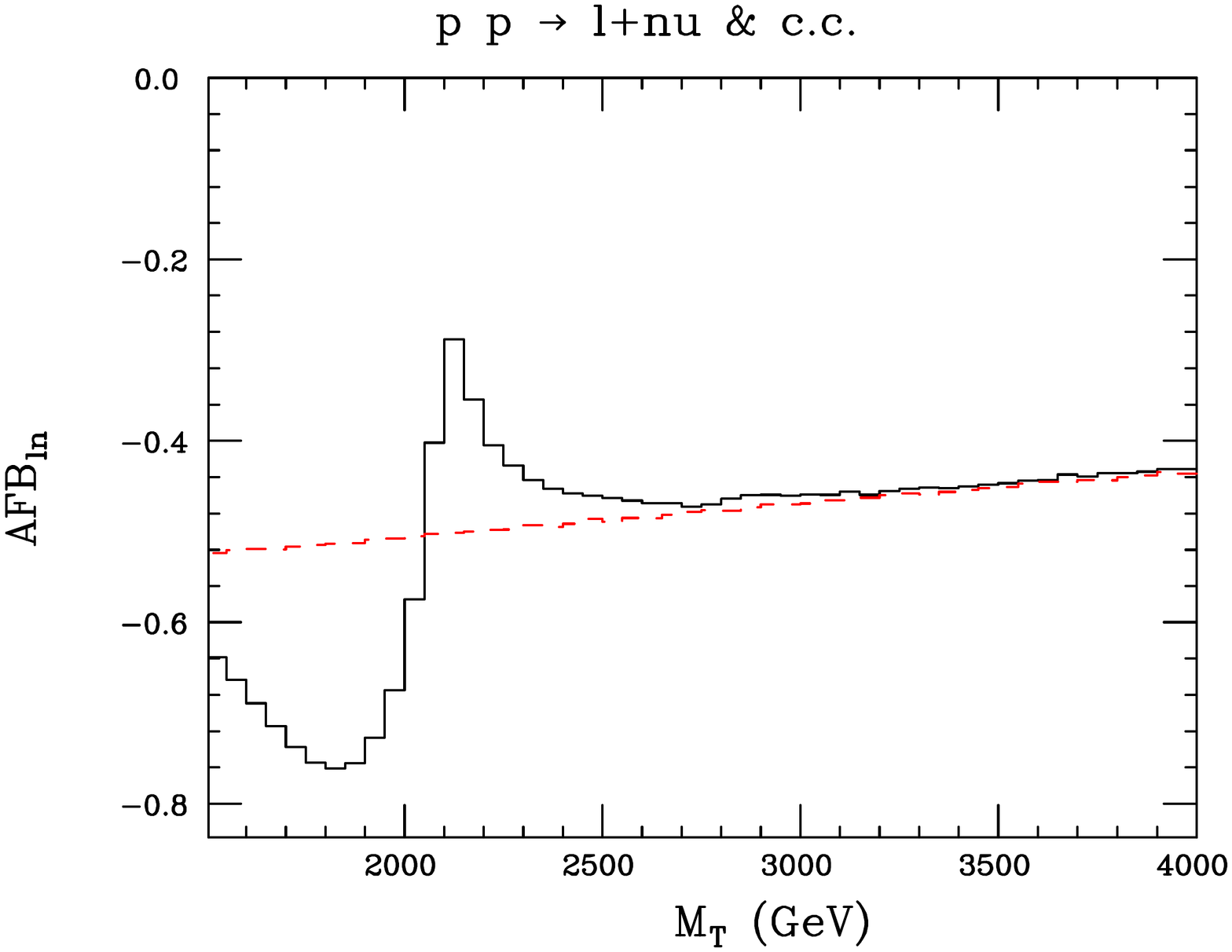}{(c)}
\includegraphics[width=0.35\linewidth,angle=90]{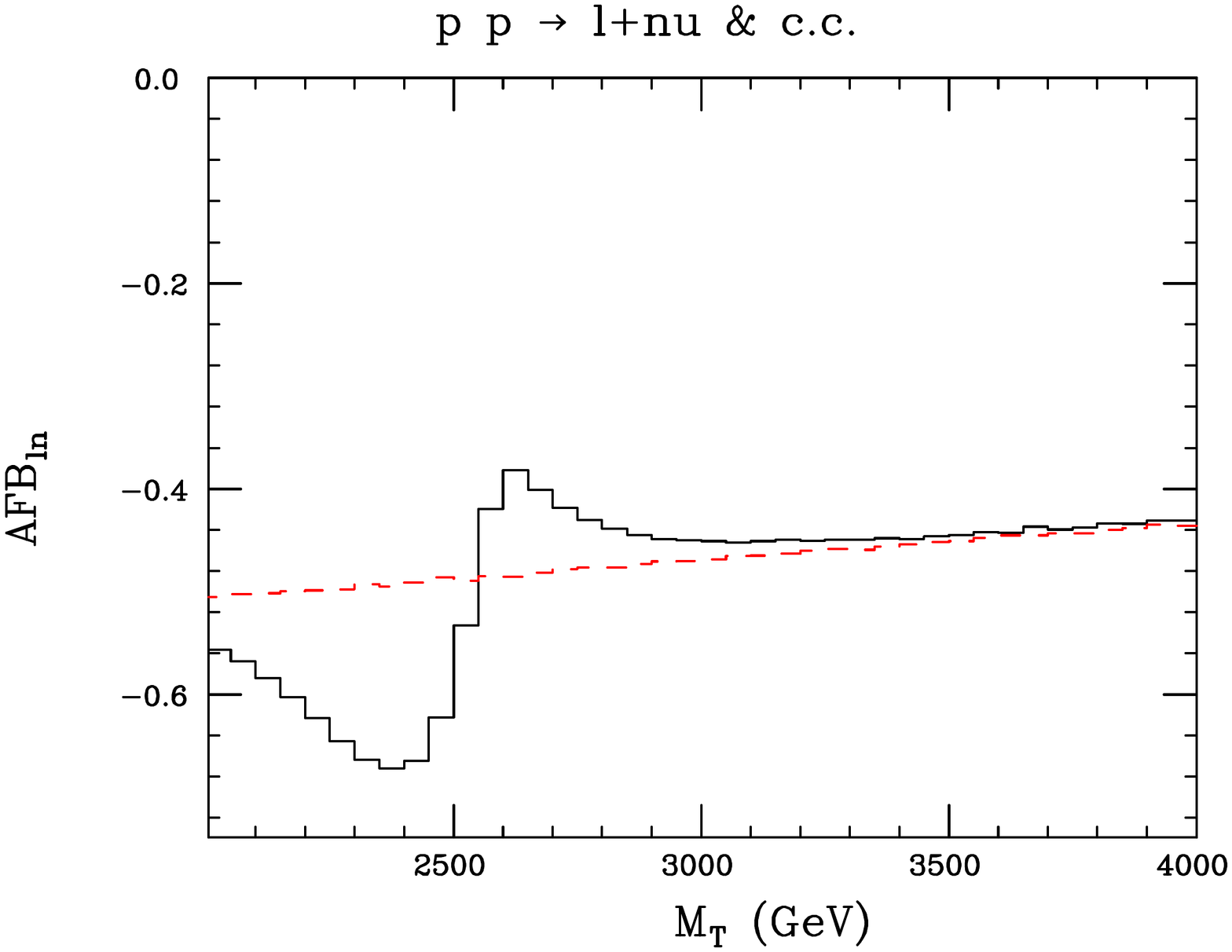}{(d)}
\includegraphics[width=0.35\linewidth,angle=90]{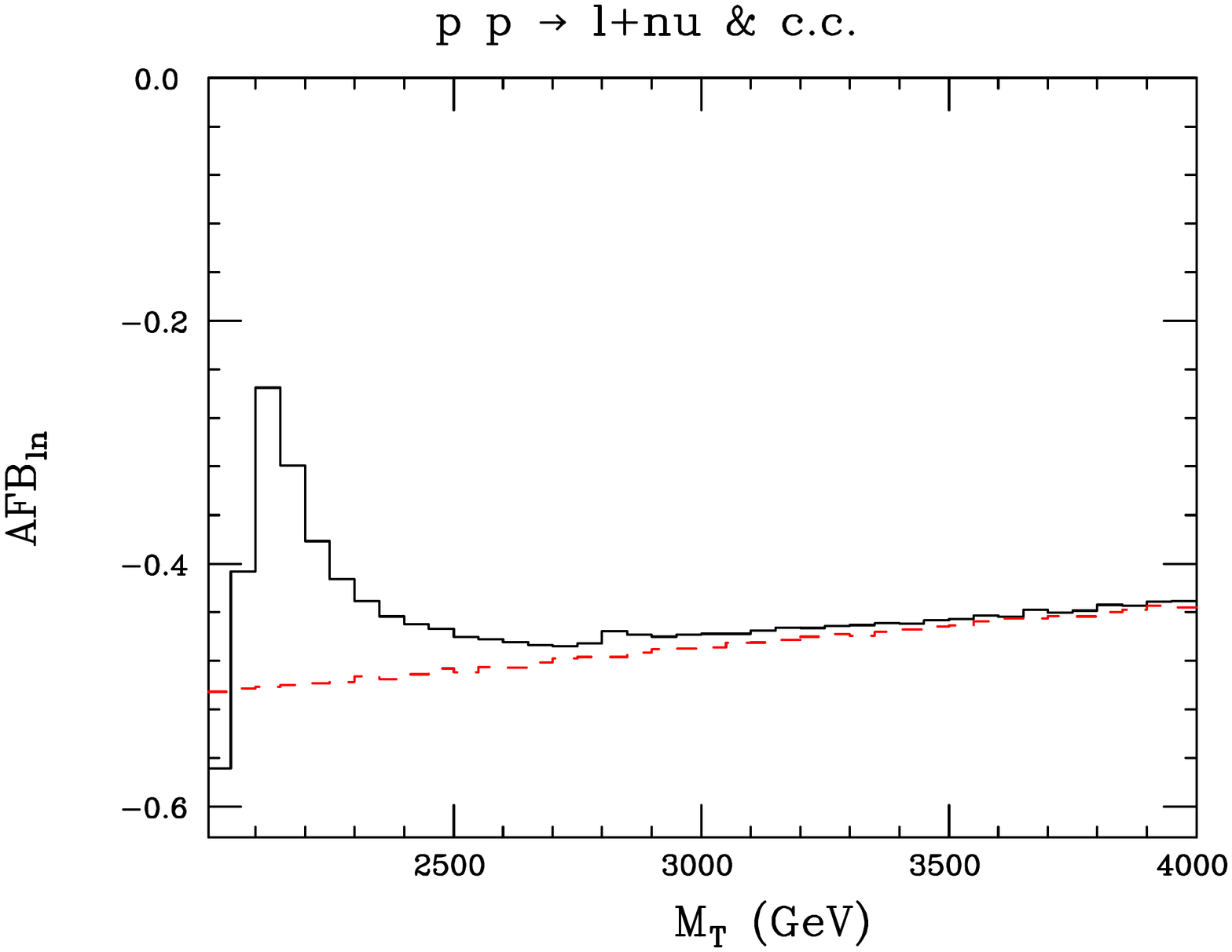}{(e)}
\includegraphics[width=0.35\linewidth,angle=90]{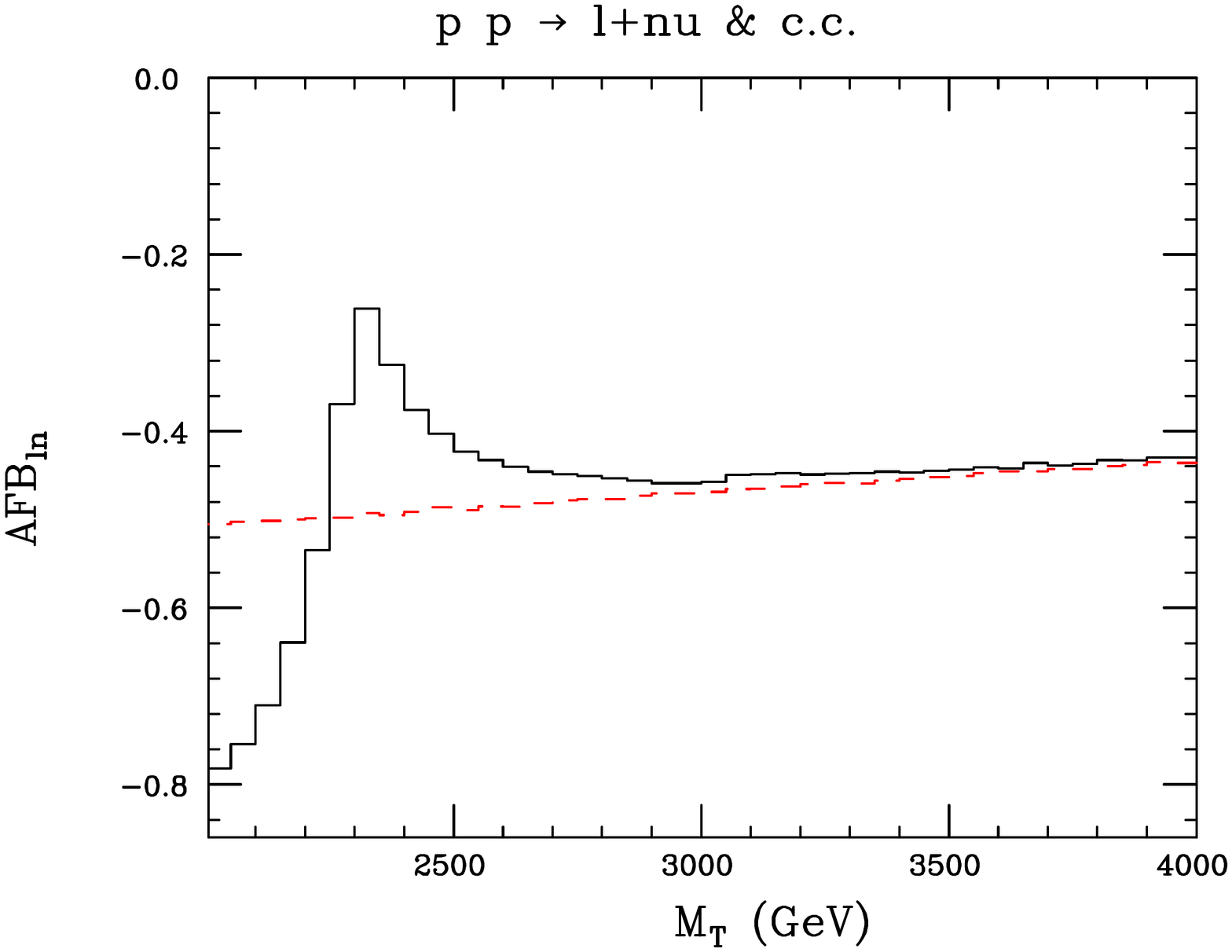}{(f)}
\caption{Differential distributions in transverse mass $M_T$ 
for the forward-backward asymmetry
at the 14 TeV LHC for CC DY in the 4DCHM (solid) for
(a) $f=0.75$ TeV and $g^*=2$ 
(b) $f=0.8$ TeV and $g^*=2.5$ 
(c) $f=1$ TeV and $g^*=2$ 
(d) $f=1$ TeV and $g^*=2.5$ 
(e) $f=1.1$ TeV and $g^*=1.8$ 
(f) $f=1.2$ TeV and $g^*=1.8$
and in the SM (dashed).
Cuts, cross sections and mass/width parameters as in the previous plots.
Bin width is here 50 GeV.}
\protect{\label{fig:CC-AFB}}
\end{figure}

The plots for AFB in the case of the CC process are found in Fig.~\ref{fig:CC-AFB}. Herein, one notices that the resolving power of the resonances in the AFB is diminished, as the 
presence of the heaviest $W'$ boson is hardly visible (just a little `kink' and only in some of the cases).  In this connection, further recall that the transverse mass resolution
cannot afford one with exploring the yield of a smaller bin width.

\begin{figure}[htb]
\centering
\includegraphics[width=0.6\linewidth,angle=90]{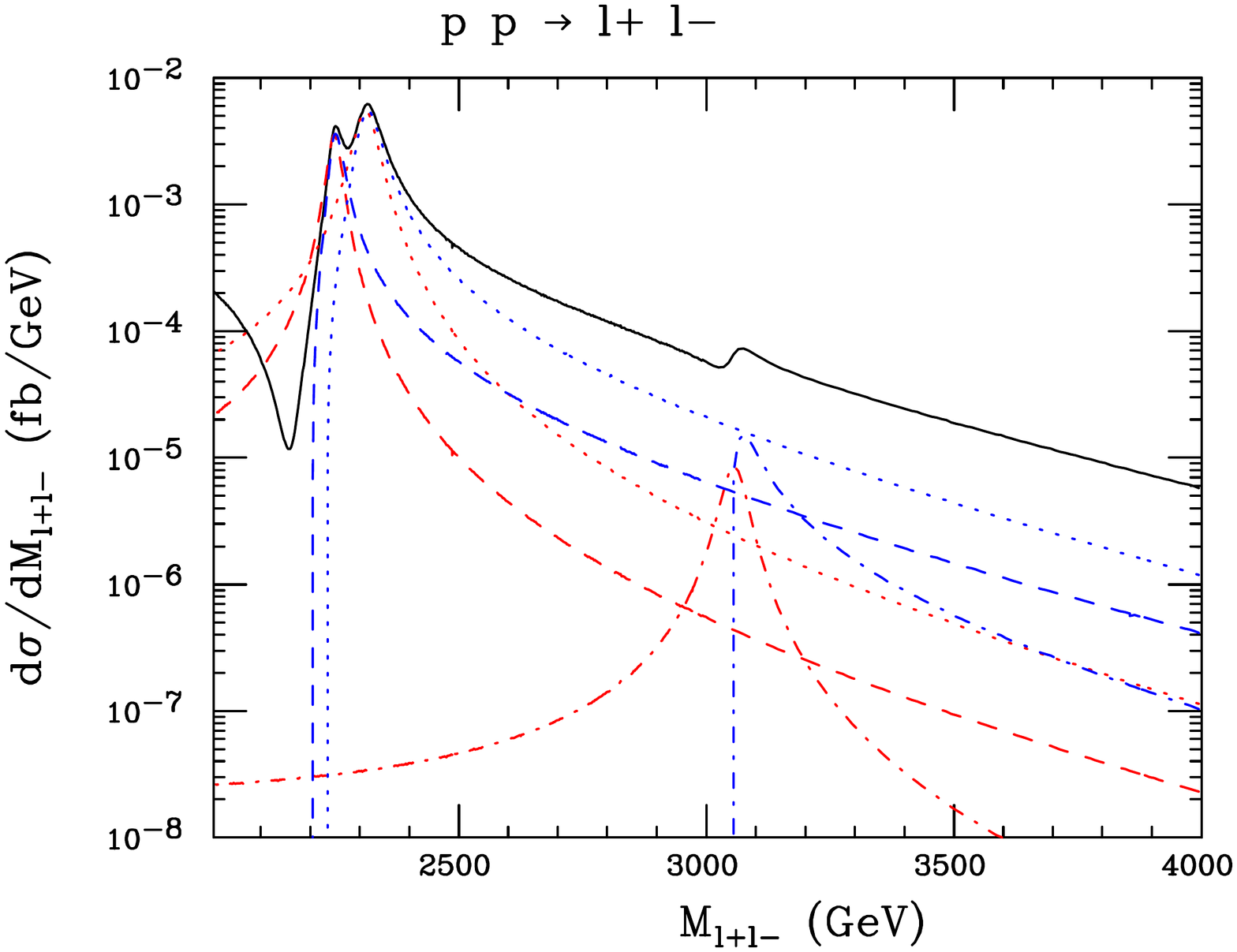}
\includegraphics[width=0.6\linewidth,angle=90]{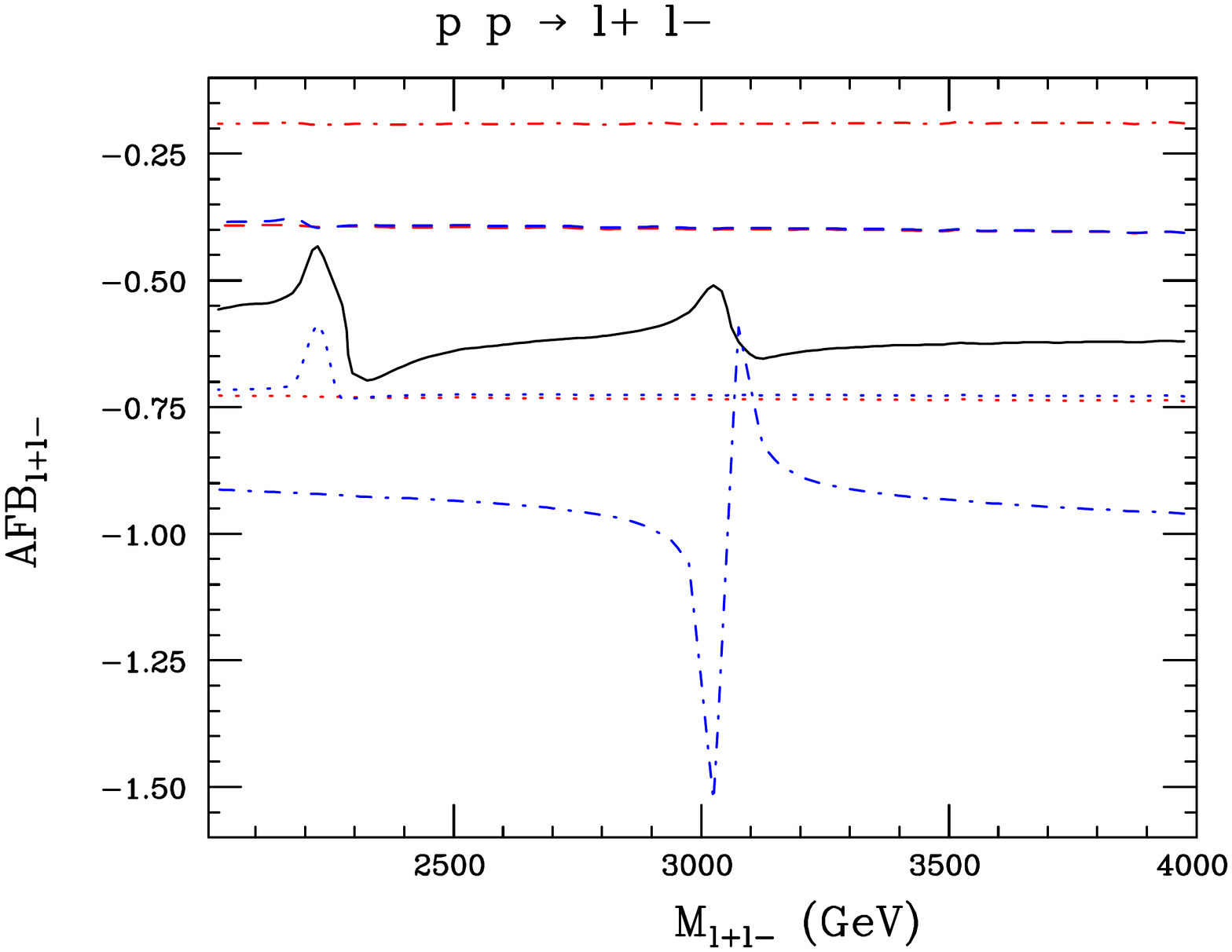}
\caption{Differential distributions in invariant mass $M_{l^+l^-}$
for (top) the cross section and (bottom) the forward-backward asymmetry
at the 14 TeV LHC for NC DY in the 4DCHM 
for $f=1.2$ TeV and $g^*=1.8$
in the case of the complete result (solid), for each resonance squared separately (dashed, dotted and dot-dashed in red)
as well for their intereference with the SM plus the SM squared (dashed, dotted and dot-dashed in blue).
Cuts, cross sections and mass/width parameters as in the previous plots.}
\protect{\label{fig:NC-Split}}
\end{figure}
\begin{figure}[htb]
\centering
\includegraphics[width=0.6\linewidth,angle=90]{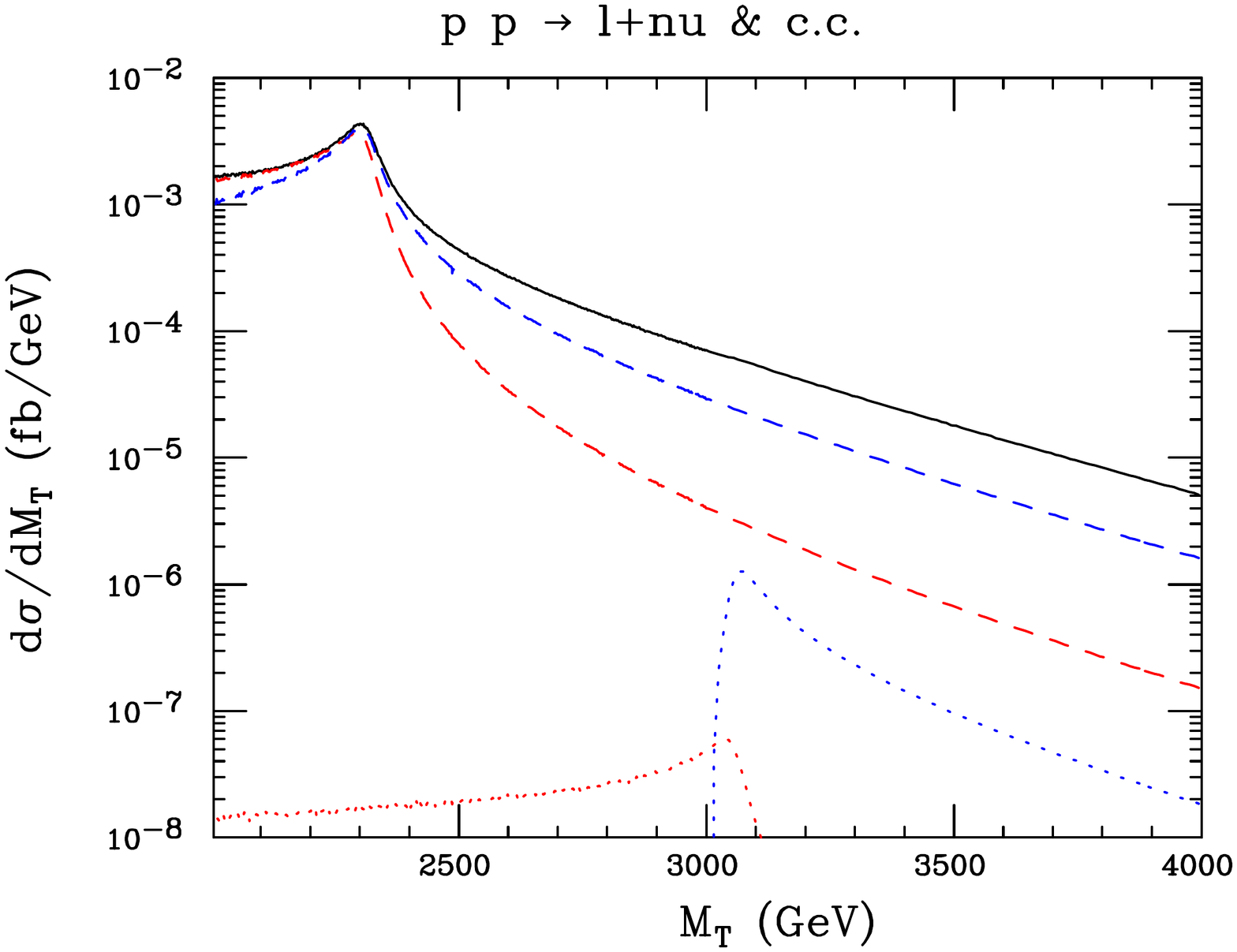}
\includegraphics[width=0.6\linewidth,angle=90]{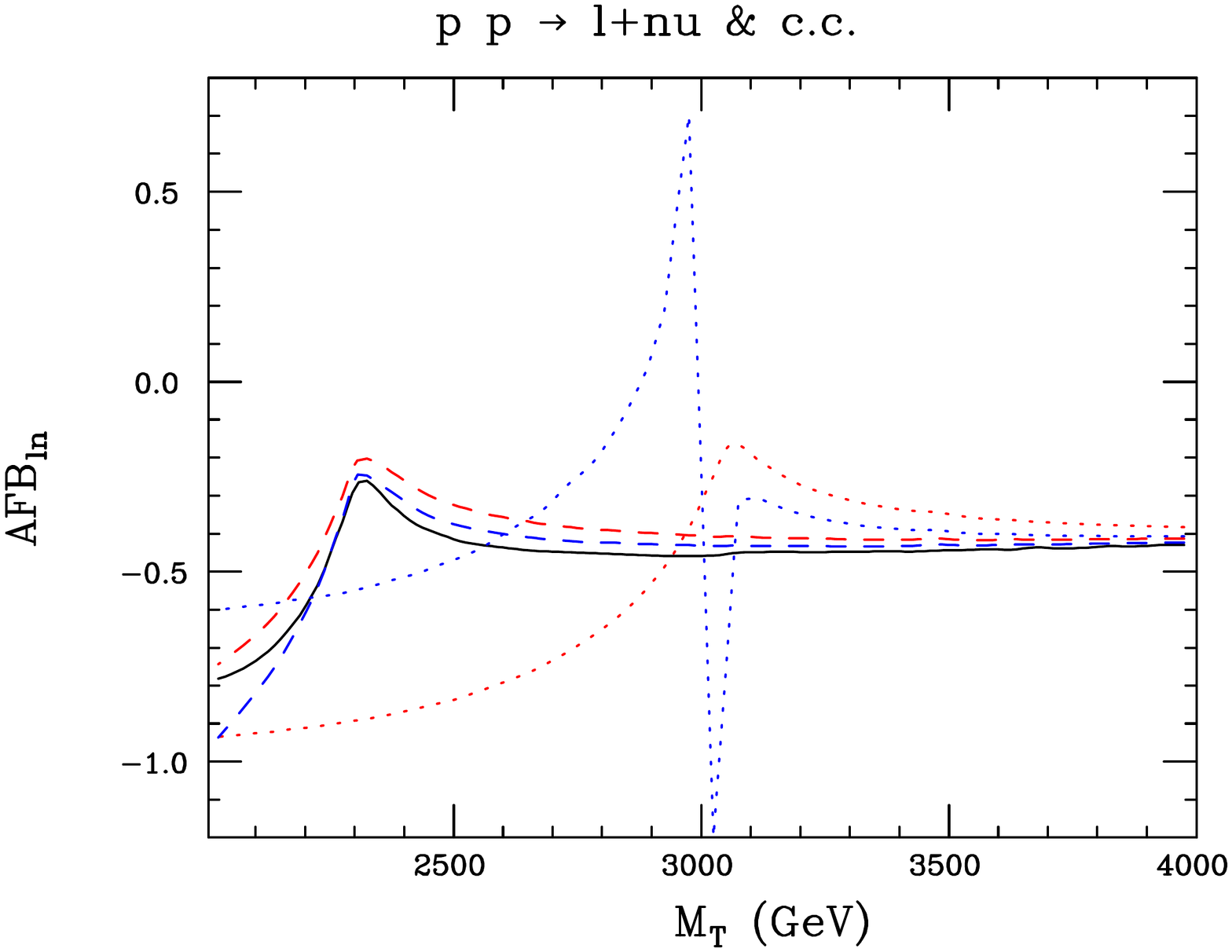}
\caption{Differential distributions in transverse mass $M_T$ 
for (top) the cross section and (bottom) the forward-backward asymmetry
at the 14 TeV LHC for CC DY in the 4DCHM 
for $f=1.2$ TeV and $g^*=1.8$
in the case of the complete result (solid), for each resonance squared separately (dashed and dotted in red)
as well for their intereference with the SM plus the SM squared (dashed and dotted in blue).
Cuts, cross sections and mass/width parameters as in the previous plots.}
\protect{\label{fig:CC-Split}}
\end{figure}

In order to explore the composition of the 4DCHM signals, both in the NC and CC current, we present Figs.~\ref{fig:NC-Split} and \ref{fig:CC-Split}, respectively. In both cases, we show the mass dependence of the cross section and the AFB.
While for both channels it is clear that the shape of the cross section (top frames) is largely due to the contribution of the resonant 4DCHM diagrams, it is
interesting to notice that the shape of the AFB in the NC channel is driven by interference effects between the 4DCHM resonant diagrams ($Z'$) and the SM 
contributions, while in the case of the CC process this is also due to 4DCHM resonances ($W'$).

To be able to quantitatively address the distinguishability between the 4DCHM and SM, the statistical error 
of the predictions ought to be calculated. While we can confirm that the above is not an issue in the case of the cross section,
particular attention has to be given to the AFB distributions.
Given that AFB is defined in terms of the number of events measured in some `forward' ($N_{F}$) and `backward' ($N_{B}$) directions, the statistical error is evaluated by 
propagating the Poisson error on each measured 
quantity (i.e., $\delta N_{F(B)}=\sqrt{N_{F(B)}}$). Per given integrated luminosity $\mathcal{L}$, the measured number of events will be $N_{F(B)}=\varepsilon\mathcal{L}\sigma_{F(B)}$, $\sigma_{F(B)}$ being the integrated or differential `forward(backward)' cross section, yielding an uncertainty on AFB of 
\begin{equation}\label{eqn:error}
	\delta ({\rm AFB})\equiv \delta\left(\frac{N_{F}-N_{B}}{N_{F}+N_{B}}\right)=\sqrt{\frac{4}{\mathcal{L}\varepsilon}	
\frac{\sigma_{F}\sigma_{B}}{(\sigma_{F}+\sigma_{B})^{3}}}.
\end{equation}
 Here, $\varepsilon$  corresponds to the assumed reconstruction efficiency of the $l^+ l^-$ and $l^+\nu$ + c.c. systems.
 
\begin{figure}[htb]
\centering
\includegraphics[width=0.52\linewidth,angle=90]{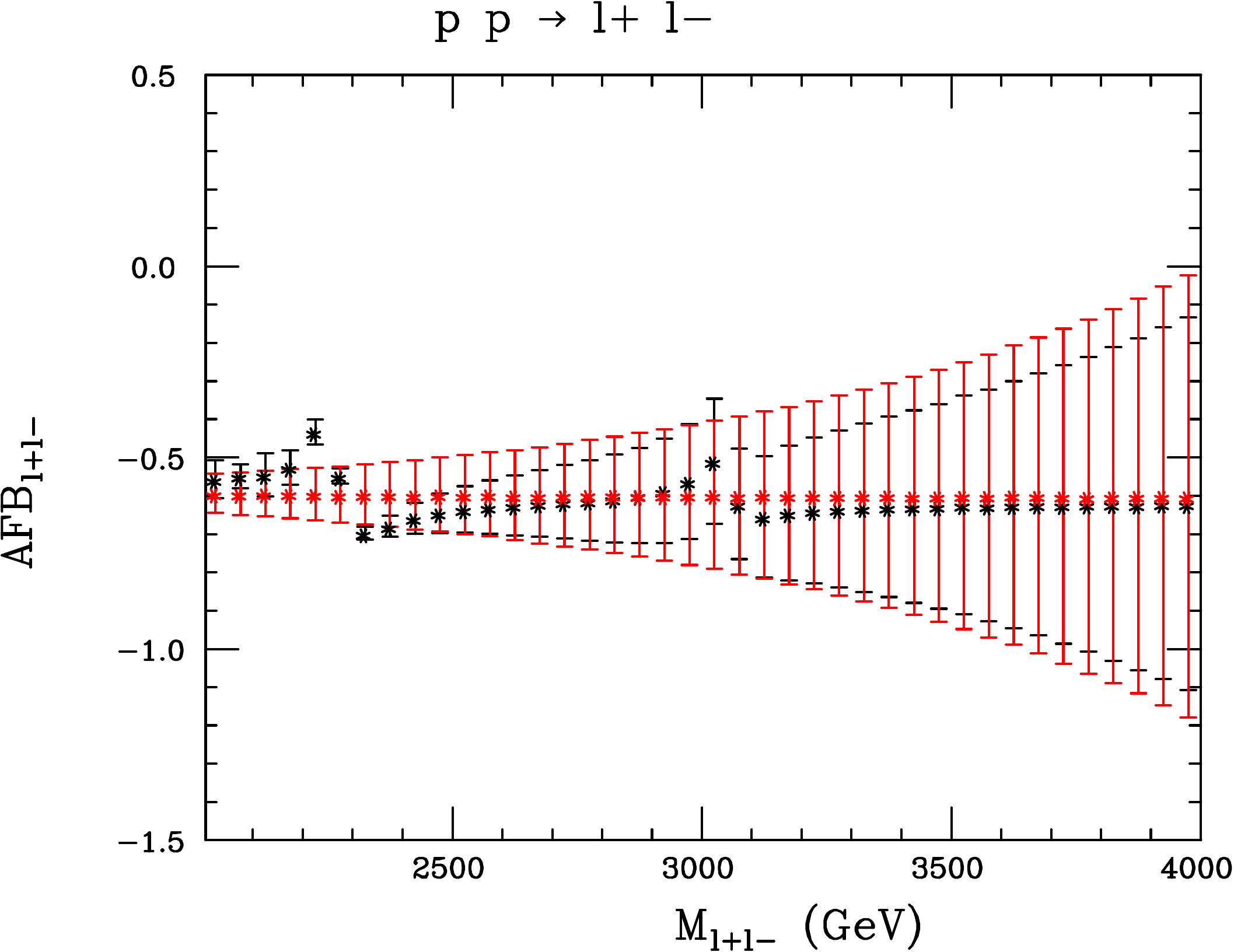}
\caption{Differential distributions in invariant mass $M_{l^+l^-}$
for the forward-backward asymmetry
at the 14 TeV LHC for NC DY in the 4DCHM 
for $f=1.2$ TeV and $g^*=1.8$, our benchmark (f), including error estimates, for ${\cal L}=1500$ fb$^{-1}$ and $\varepsilon=90\%$. 
Cuts, cross sections and mass/width parameters as in the previous plots.
Bin width is here 50 GeV. Further, the 4DCHM(SM) is in black(red). }
\protect{\label{fig:NC-ErrAFB}}
\end{figure}
\begin{figure}[htb]
\centering
\includegraphics[width=0.52\linewidth,angle=90]{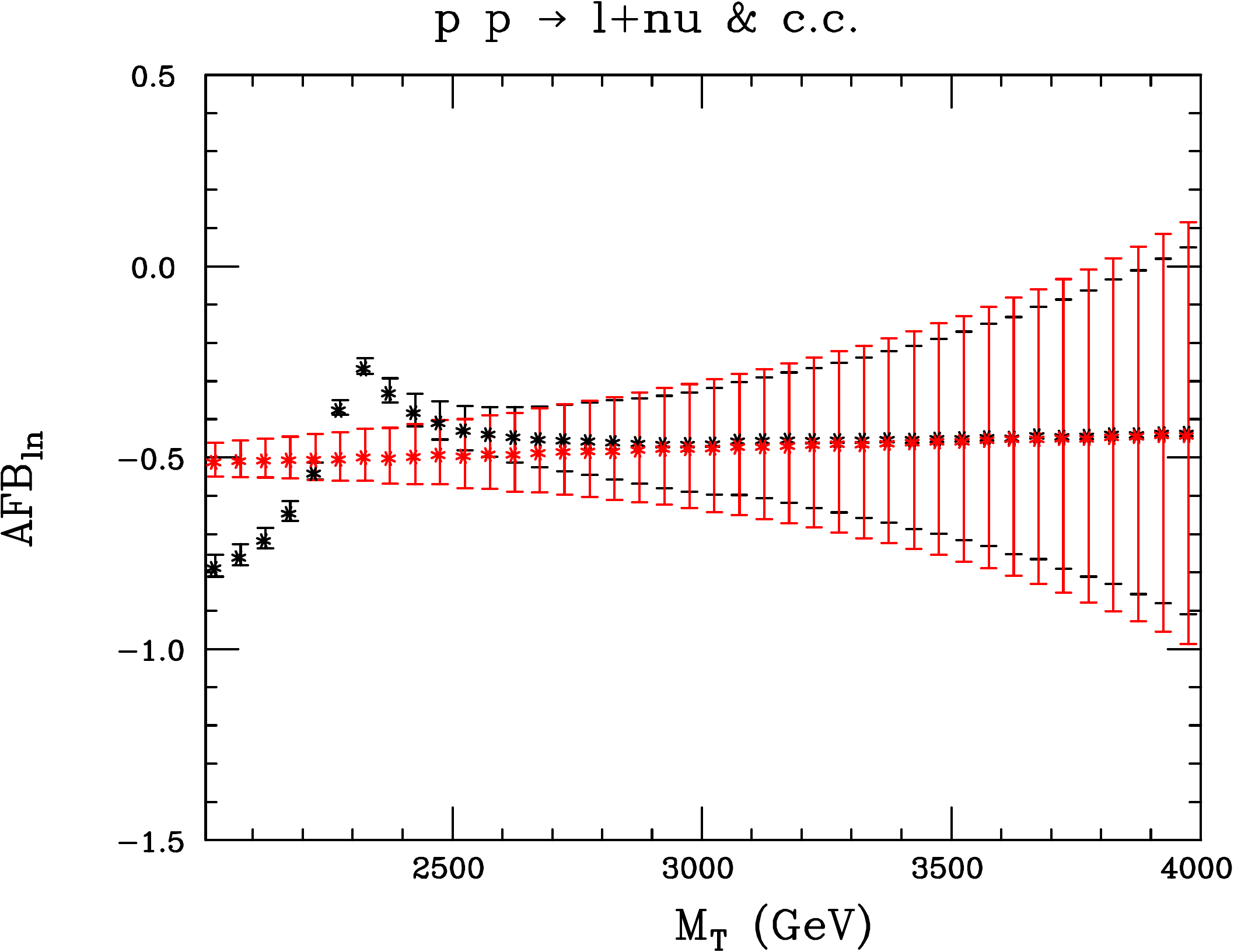}
\caption{Differential distributions in transverse mass $M_T$ 
for the forward-backward asymmetry
at the 14 TeV LHC for CC DY in the 4DCHM 
for $f=1.2$ TeV and $g^*=1.8$, our benchmark (f), including error estimates, for ${\cal L}=1500$ fb$^{-1}$ and $\varepsilon=90\%$.
Cuts, cross sections and mass/width parameters as in the previous plots.
Bin width is here 50 GeV. Further, the 4DCHM(SM) is in black(red). }
\protect{\label{fig:CC-ErrAFB}}
\end{figure}

Figs.~\ref{fig:NC-ErrAFB}--\ref{fig:CC-ErrAFB} show our findings in this connection. Herein,
the starred points are the central values for the given asymmetry, with a statistical error quantified by binning the cross sections in 
$M_{l^+l^-}$ (NC channel) and $M_T$ (CC channel), for an histogram width of $50$ GeV. 
Assuming ${\cal L}=1500$ fb$^{-1}$ and $\varepsilon=90\%$, it is clear that 4DCHM resonant effects in AFB can be discernible with respect
to the SM noise in both the NC and CC case, albeit (probably) limitedly to the lowest lying resonances in either case, i.e., $Z_{2,3}$ and
$W_{2}$, respectively, so long that very high luminosities can be attained (for example, at the Super-LHC). In the NC case, however, for di-electron final states, also the extraction of the heaviest neutral resonance ($Z_5$) remains an open possibility.  

In view of such a high demand of luminosity to extract fully differential effects in AFB, it makes sense looking at integrated
values of the latter, over suitable mass intervals. By inspecting Figs.~\ref{fig:NC-AFB} and \ref{fig:CC-AFB}, it is clear that
differences between the 4DCHM and SM are maximised when one integrates to the left and to the right of each resonance.
In the case of the NC channel, there are two mass points (in $M_{l^+l^-}$) where this can be attempted, again assuming a
conservative 50 GeV resolution, around the overlapping $Z_{2,3}$ resonances and then around $Z_5$. In the case
of the CC channel, there is only one mass point (in $M_T$) to exploit for such a resolution, that is, around the $W_2$ resonance. (In essence, the extremes of each mass interval used in the integration procedure are defined by the crossing 
points between the 4DCHM and SM histograms
in Figs.~\ref{fig:NC-AFB} and \ref{fig:CC-AFB}.)

\begin{table}[htb]
\captionsetup[subfloat]{labelformat=empty,position=top}
\centering
 \begin{tabular}{|l|l|l|}
\hline
\hline
 & (a) $f=0.75$ TeV and $g_*=2$  \\
\hline
 & AFB($l^+ l^-$) & AFB($l^+\nu$ + c.c.) \\
\hline
4DCHM [I]& ($-0.554\pm0.013$)[$-0.627\pm0.029$]~| &  ($-0.698\pm0.011$)[$-0.460\pm0.074$] \\
SM   & ($-0.583\pm0.013$)[$-0.590\pm0.029$]~| &  ($-0.533\pm0.011$)[$-0.478\pm0.074$] \\
4DCHM [II]& ($-0.531\pm0.045$)[$-0.620\pm0.11$] & \\
SM    & ($-0.593\pm0.045$)[$-0.598\pm0.11$] & \\
\hline
\hline
 & (b) $f=0.8$ TeV and $g_*=2.5$  & \\
\hline
4DCHM [I]&  ($-0.578\pm0.025$)[$-0.607\pm0.060$] & ($-0.607\pm0.022$)[$-0.455\pm0.093$] \\
SM    &   ($-0.591\pm0.025$)[$-0.596\pm0.060$] &  ($-0.515\pm0.022$)[$-0.469\pm0.093$]\\
\hline
4DCHM [II]&  ($-0.59\pm0.16$)[$-0.61\pm0.15$] & \\
SM    &  ($-0.60\pm0.16$)[$-0.60\pm0.15$] & \\
\hline
\hline
 & (c) $f=1$ TeV and $g_*=2$  \\
\hline
4DCHM [I]&  ($-0.515\pm0.085$)[$-0.625\pm0.059$]  &  ($-0.57\pm0.16$)[$-0.62\pm0.17$] \\
SM    &  ($-0.593\pm0.085$)[$-0.596\pm0.059$]  &  ($-0.60\pm0.16$)[$-0.60\pm0.17$] \\
\hline
4DCHM    [II]& ($-0.699\pm0.022$)[$-0.443\pm0.093$]& \\
SM       &  ($-0.514\pm0.022$)[$-0.469\pm0.093$]& \\
\hline
\hline
 & (d) $f=1$ TeV and $g_*=2.5$  \\
\hline
4DCHM [I]&($-0.578\pm0.052$)[$-0.61\pm0.11$] &($-0.612\pm0.045$)[$-0.44\pm0.12$] \\
SM   &   ($-0.595\pm0.052$)[$-0.60\pm0.11$] & ($-0.496\pm0.045$)[$-0.46\pm0.12$]\\
\hline
4DCHM [II]&($-0.60\pm0.48$)[$-0.61\pm0.30$] &\\
SM    &	($-0.60\pm0.48$)[$-0.60\pm0.30$] &\\
\hline
\hline
 & (e) $f=1.1$ TeV and $g_*=1.8$  \\
\hline
4DCHM 	[I]&       ($-0.467\pm0.085$)[$-0.634\pm0.059$]&  ($-0.417\pm0.044$)[$-0.45\pm0.12$]        \\
SM 	&       ($-0.594\pm0.085$)[$-0.596\pm0.059$]&  ($-0.496\pm0.044$)[$-0.46\pm0.12$]        \\
\hline
4DCHM 	[II]&       ($-0.55\pm0.16$)[$-0.63\pm0.17$]  &      \\
SM 	&       ($-0.60\pm0.16$)[$-0.60\pm0.17$]  &      \\
\hline
\hline
 & (f) $f=1.2$ TeV and $g_*=1.8$  \\
\hline
4DCHM 	 [I]& ($-0.526\pm0.057$)[$-0.635\pm0.077$]&	    ($-0.507\pm0.045$)[$-0.44\pm0.12$]      \\
SM 	 & ($-0.594\pm0.057$)[$-0.597\pm0.077$]&	    ($-0.496\pm0.045$)[$-0.46\pm0.12$]      \\
\hline
4DCHM 	 [II]& ($-0.55\pm0.24$)[$-0.63\pm0.20$]  &	     \\
SM 	 & ($-0.60\pm0.24$)[$-0.60\pm0.20$]  &	     \\
\hline
\hline
\end{tabular}
\caption{Value of the integrated AFB and relative error for the benchmarks of Tab.~\ref{table:benchmark1}, obtained as described in 
the text, for the first [I] and (limited to the NC case) second resonance [II] in the mass interval to the (left)[right] of it, for both the NC 
($M_{l^+l^-}$) and CC ($M_T$) channel, assuming ${\cal L}=600$ fb$^{-1}$ and $\varepsilon=90\%$.}
\label{table:intAFB}
\end{table}

Tab.~\ref{table:intAFB} shows our findings, now for the complete list
of benchmarks (a)--(f), assuming here a more modest luminosity, of $600$ fb$^{-1}$, again
with a 90\% tagging efficiency. Depending on the benchmark at hand, it is clear that some resolving power between the 4DCHM and the SM exists for the integrated AFB, the more so the lighter the resonant masses involved, in both the
NC and CC case. In fact, taking the difference between the integrate AFB values obtained to the left and to the right of each resonance seems the most effective way to disentangle the two models, as such a quantity is very nearly zero for the
SM in all cases whereas it is not so (even in a statistical sense) for the 4DCHM.

\begin{figure}[htb]
\centering
\includegraphics[width=0.5\linewidth,angle=90]{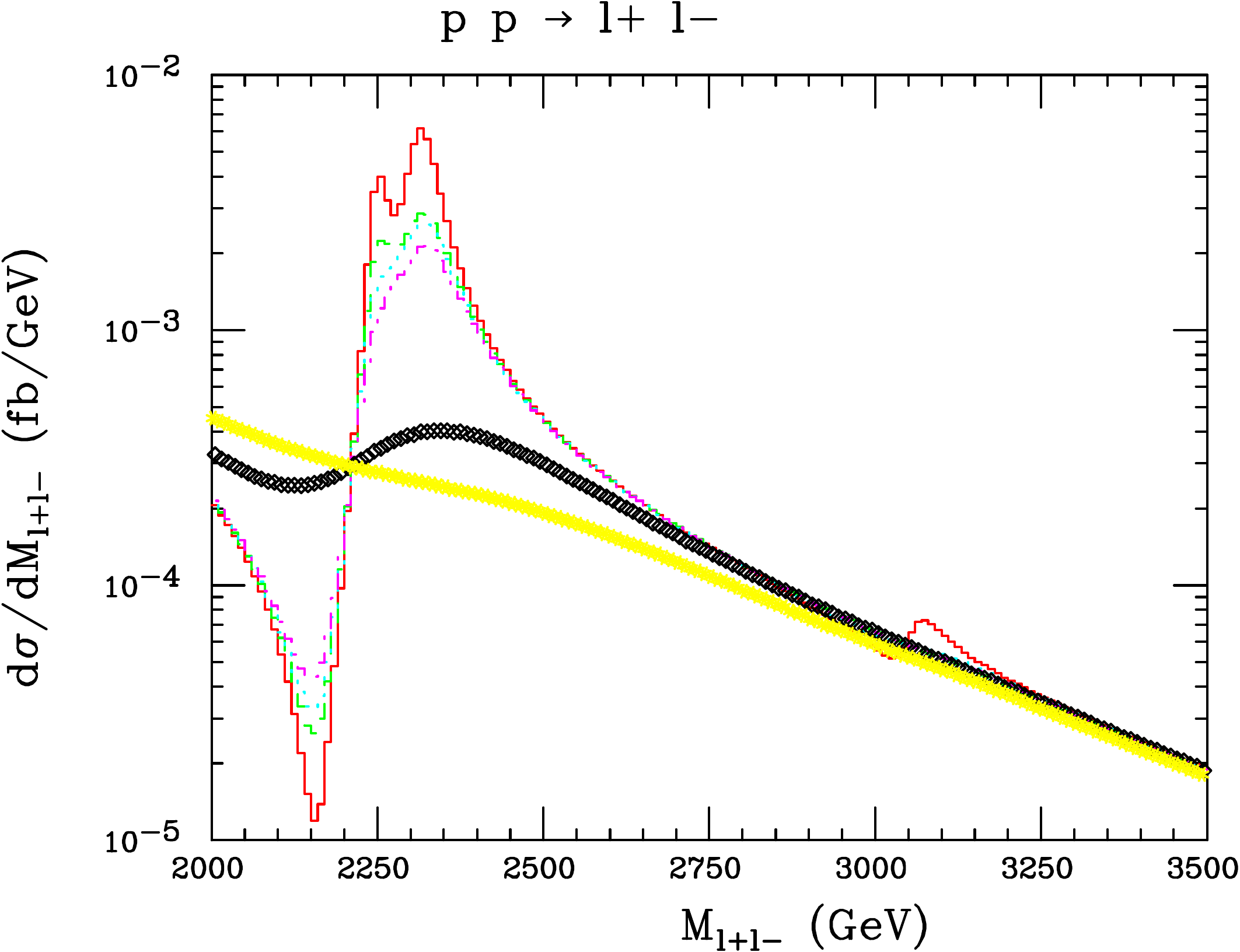}
\caption{Differential distributions in invariant mass $M_{l^+l^-}$
for the cross section
at the 14 TeV LHC for NC DY in the 4DCHM 
for $f=1.2$ TeV and $g^*=1.8$
for same mass 
and different width values
of the resonant gauge bosons:
the colour scheme is as in Fig.~\ref{plot:Hpotential}.
The integrated cross sections are
0.78, 0.56, 0.52, 0.47, 0.26 and 0.23 fb, respectively.
Cuts as in the previous plots.
Bin width is here 10 GeV.}
\protect{\label{fig:NC-Widths}}
\end{figure}
\begin{figure}[htb]
\centering
\includegraphics[width=0.5\linewidth,angle=90]{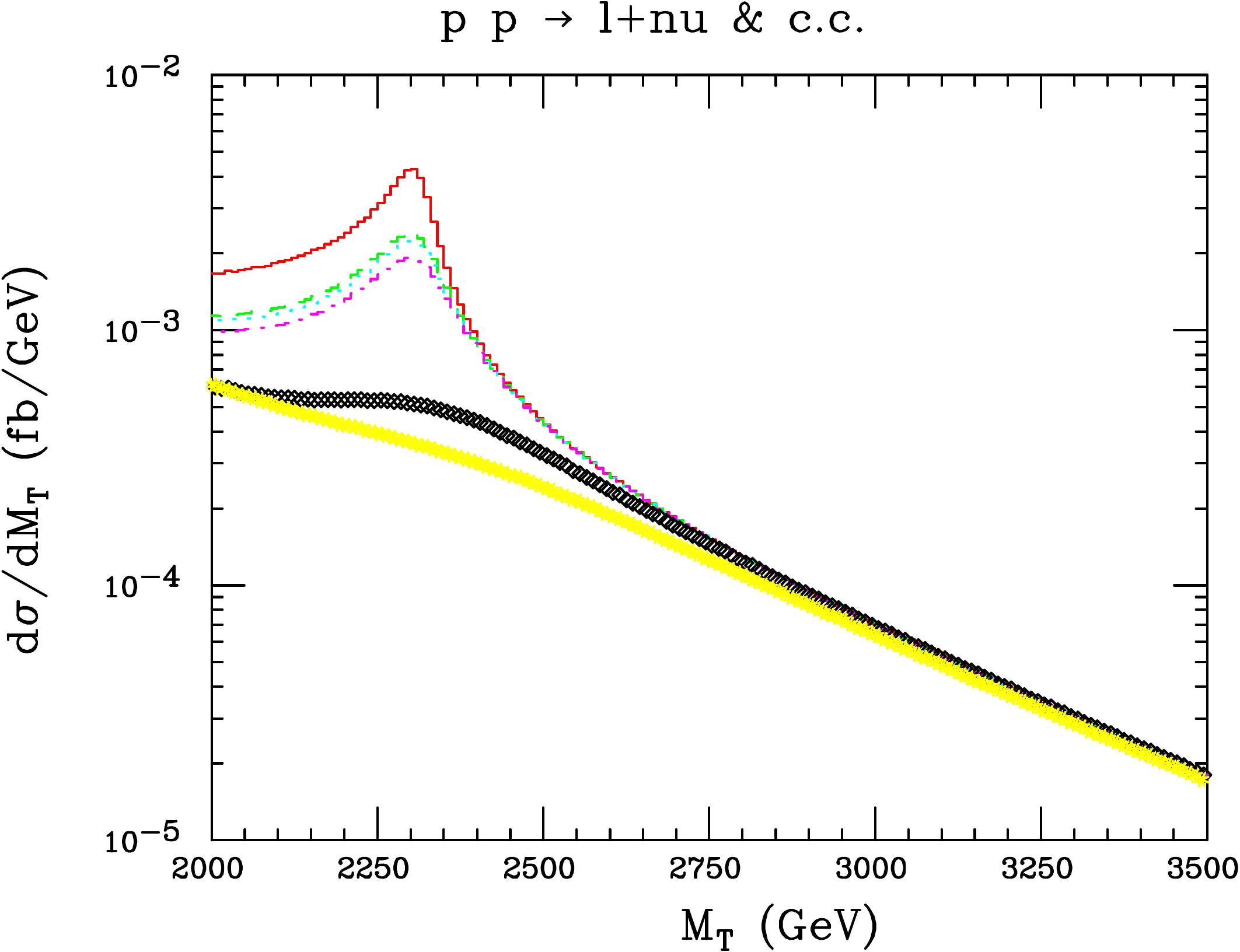}
\caption{Differential distributions in transverse mass $M_T$
for the cross section
at the 14 TeV LHC for CC DY in the 4DCHM 
for $f=1.2$ TeV and $g^*=1.8$
for same mass 
and different width values
of the resonant gauge bosons:
the colour scheme is as in Fig.~\ref{plot:Hpotential}.
The integrated cross sections are
1.11, 0.79, 0.76, 0.70, 0.36 and 0.30 fb, respectively.
Cuts as in the previous plots.
Bin width is here 10 GeV.}
\protect{\label{fig:CC-Widths}}
\end{figure}

If the mass difference between the $Z'$ and $W'$ gauge boson and the masses of the heavy fermion increase, a notable increase occurs in the
width of the gauge bosons, as discussed in detail in sections 4.1.1a--4.1.1c. We study the impact of this phenomenon onto the mass 
distributions of both the NC and CC process in Figs. \ref{fig:NC-Widths} and \ref{fig:CC-Widths}, where we keep the same gauge boson
masses as till now (i.e., for the benchmark $f=1.2$ TeV and $g_*=1.8$), yet we gradually increase the widths of both the $Z'$ and
$W'$ states, by selecting two further points in the low width regime and one each in the medium width regime and large width regime (as detailed
in the captions).

As the values of $\Gamma_{Z'}$ grow larger, the resolution of the separate $Z'$ resonances, $Z_2$ and $Z_3$,
is gradually no longer possible, as they start merging, with the ultimate tendency, for very large widths, to flatten down significantly,
to the extent that the extraction of a narrow resonance (i.e., $\Gamma\ll M$) is no longer possible, rather the signal will appear as an excess over a similarly
shaped SM background. In the case of the heaviest resonance, the same phenomenology occur, though the non resonant limit is reached
earlier on, owning to the fact that the original resonance only marginally emerged from the background. (Recall though that this state is 
hardly accessible at the LHC with standard design.) In the case  of the only visible $W'$ resonance ($W_2$) the pattern is straightforward, the
larger the value of $\Gamma_{W'}$ the less prominent the mass peak, which essentially no longer stems out of the background as soon as one enters
the medium width regime.

\begin{figure}[htb]
\centering
\includegraphics[width=0.53\linewidth,angle=90]{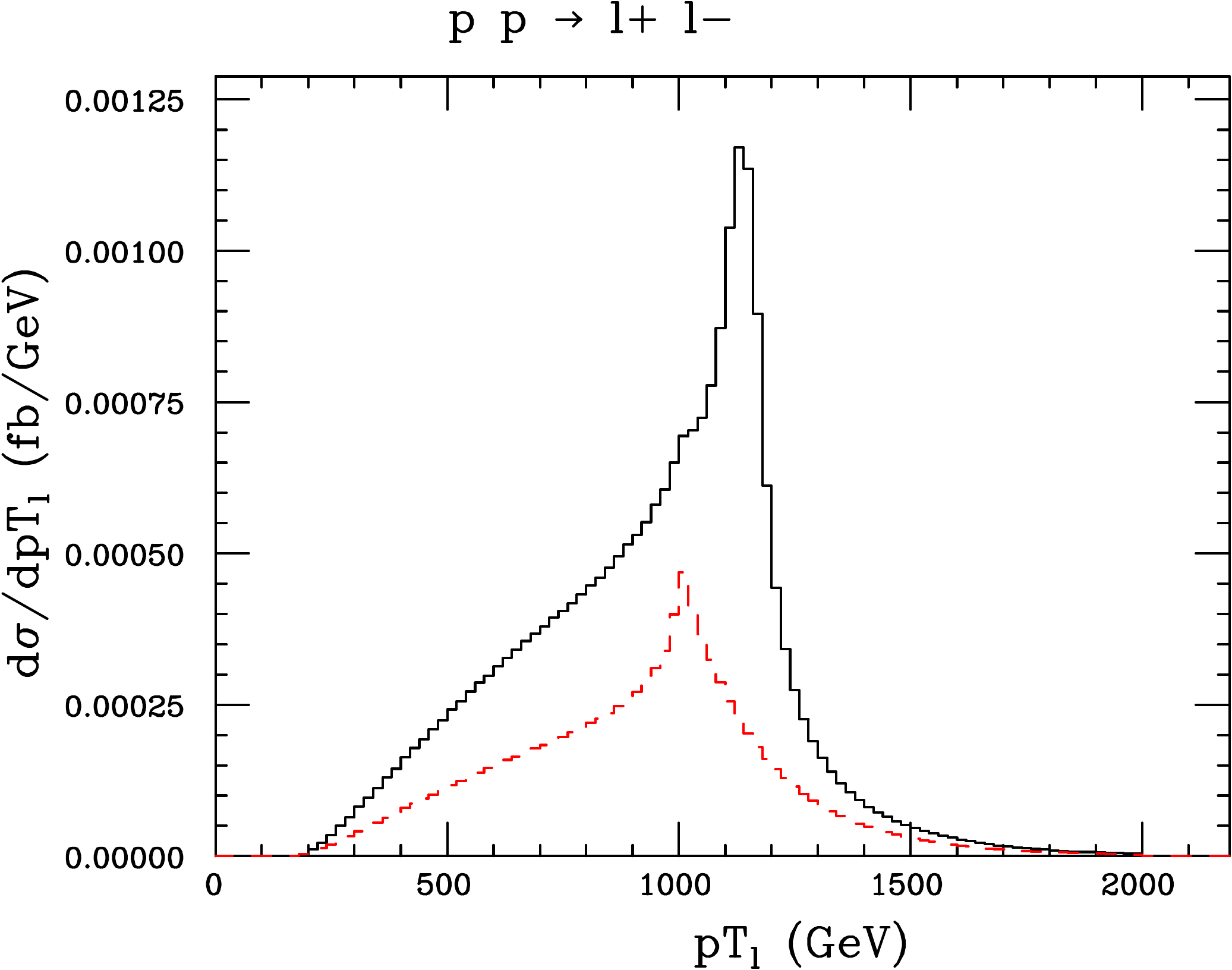}
\caption{Differential distributions in transverse momentum $p^T_{l^\pm}$
of the lepton for the cross section
at the 14 TeV LHC for NC DY in the 4DCHM 
for $f=1.2$ TeV and $g^*=1.8$ with
$M_{Z_{2(3)[5]}}=2249(2312)[3056]$ GeV 
and 
$\Gamma_{Z_{2(3)[5]}}=75(104)[313]$ GeV (solid)
plus the SM (dashed).
Cuts as in the previous plots.
Bin width is here 20 GeV.}
\protect{\label{fig:NC-pT}}
\end{figure}
\begin{figure}[htb]
\centering
\includegraphics[width=0.53\linewidth,angle=90]{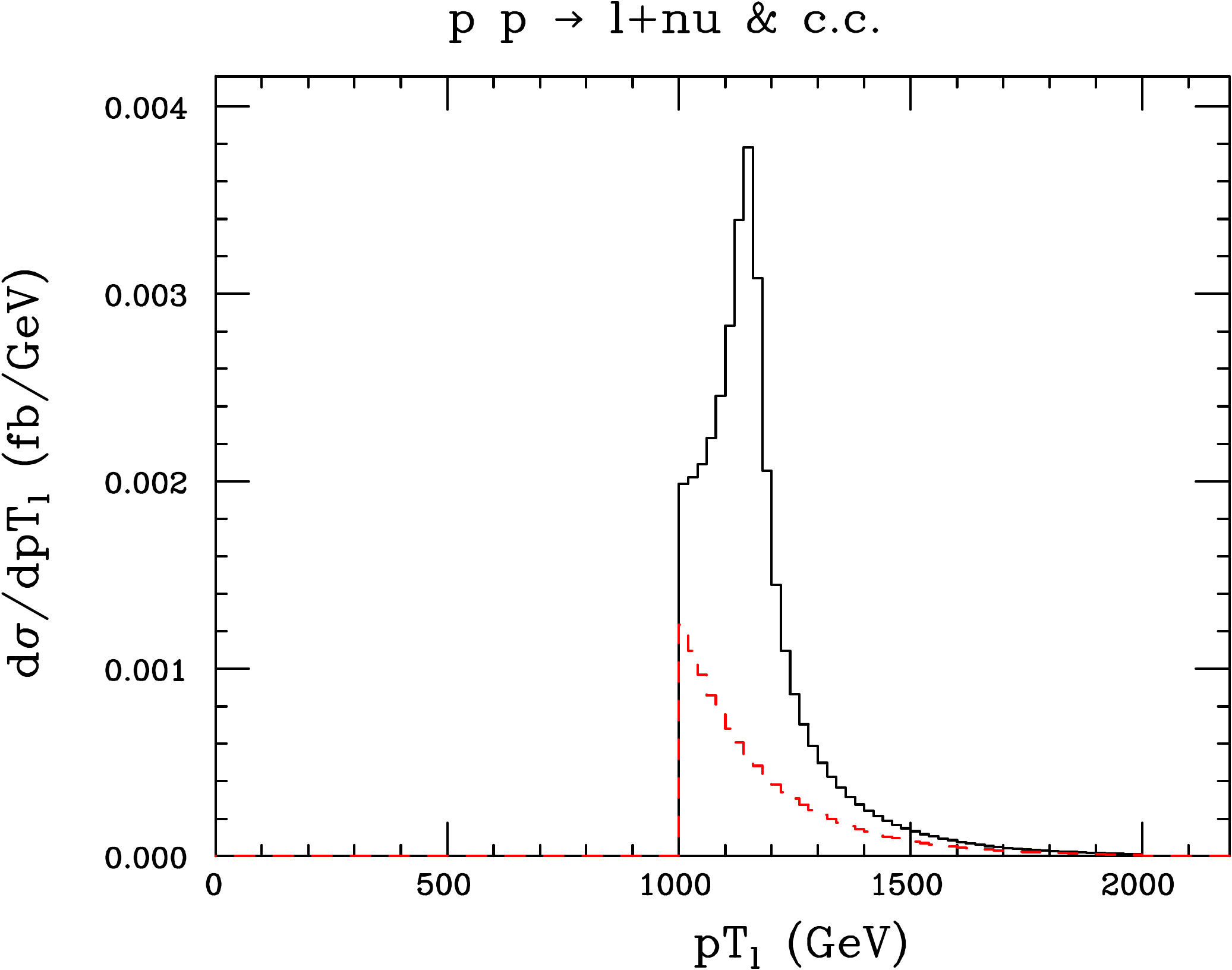}
\caption{Differential distributions in transverse momentum $p^T_{l^\pm}$
of the lepton for the cross section
at the 14 TeV LHC for CC DY in the 4DCHM 
for $f=1.2$ TeV and $g^*=1.8$ with
$M_{W_{2(3)}}=2312.18(3056.1)$ GeV 
and 
$\Gamma_{W_{2(3)}}=104(293)$ GeV (solid)
plus the SM (dashed).
Cuts as in the previous plots.
Bin width is here 20 GeV.}
\protect{\label{fig:CC-pT}}
\end{figure}

One last exercise that we would like to perform before closing is to show how the exploitation of the fact that in the 4DCHM the 
mass of the two lightest $Z'$ ($Z_2$ and $Z_3$) is strongly correlated to the one of the lightest $W'$ state ($W_2$), as previously explained.
In fact, a similar correlation exists in the case of the heaviest states in the two sectors, neutral ($Z_5$) and charged ($W_3$), though
this has no phenomenological relevance in the present context. As intimated, such a correlation can be exploited to improve searches in either 
channel. Assume for example that a $W'$ resonance has been seen in the transverse mass spectrum and nothing appears instead in the
invariant mass one, after the standard cuts in lepton and missing transverse momentum. This could well be justified by the fact that
cross sections in the CC case are higher than in the NC one. Knowledge of $M_{W_2}$ would then imply also knowledge of $M_{Z_2}$ and
$M_{Z_3}$, so that, to exalt the resonances associated to the latter, one may impose onto the di-lepton sample associated to the NC
the cut, e.g., $p^T_l>M_{W_2}/2$, extracted from analysis of the lepton-missing energy sample associated with the CC. (Recall that
one can easily fit the $M_{W_2}$ value to the $M_T$ distribution using simulated data.). The advantage in doing so in order to enrich the candidate
$Z'$ sample of signal events is evident from Fig. \ref{fig:NC-pT}, as most of the SM background would be eliminated whereas the
majority of the signal would be selected. The specular exercise would also be possible, as in the case of very narrow widths
it could well be that it is a $Z'$ signal the first to be seen. The benefit is evident here too, by investigating Fig. \ref{fig:CC-pT}.
Finally, notice that the $M_T$ and $p^T_l$ quantities are directly related in the CC case, unlike the NC one for $M_{l^+l^-}$ and  $p^T_l$, so that
in the former case the lepton transverse momentum distribution displays a hard lower threshold, while this is not true in the latter case.

\section{Conclusions}
\label{sec:summa}

In summary, we have investigated the phenomenological implications of the 4DCHM for the case of DY processes at the LHC, in the case of both the NC and CC channels.
By assuming only the 4DCHM configurations consistent with theoretical requirements of self-consistency and current experimental limits, we have verified that only the 14 TeV stage of the LHC
has the potential to probe such a model (as the 7 and 8 TeV data samples yield event rates below detectable level), assuming standard and Super-LHC luminosities, through both the aforementioned channels. The study of both the cross section and AFB distribution offers 
complementary handles
to extract the underlying model spectrum in the gauge sector, although limited to the case of the lowest lying mass resonances, two in the case of the NC and one in the case of the CC. 

Further, we have isolated a few salient features of the 4DCHM,
with respect to popular models proposed in literature boasting the presence of $Z'$ and/or $W'$ bosons. Firstly, 
we have described how the opening of said decay channels impinges onto the line shape of the visible $Z'$ and $W'$ resonances in a manner that one can correlate the size of the width to the
typical mass scale of the heavy fermions of the 4DCHM.
Secondly, 
we have shown that almost degenerate $Z'$ resonances appearing in the 4DCHM spectrum can be resolved, so long that decays of the additional gauge bosons into pairs of heavy fermions are inaccessible.
 Thirdly, we have explained how to exploit the fact that $Z'$ and $W'$ masses are strongly correlated through the gauge 
structure of the 4DCHM
in order to improve the extraction of the signals over the SM background, by cross-exchanging kinematic information between the NC and CC channel, and vice versa. In essence,
we emphasise that it is the extraction of both the $Z_2$ and $Z_3$ resonances together with the $W_2$ one (either via cross section or asymmetry
measurements), all rather close in mass, that would represent the hallmark signature of this model.
Subsequently, an analysis of the new gauge boson line shapes  would reveal, or otherwise, the presence of light 
additional fermions, shedding further light on the spectrum of this model.

This is the first phenomenological study of the 4DCHM for the case of the LHC, tackling the heavy 
gauge boson sector, limited to the case of DY production. However, the case of di-boson production is now
under way too. Further studies are also ongoing regarding the heavy fermion and Higgs sectors. In fact, unlike more minimal versions of composite Higgs models, the      4DCHM suggests a realistic realisation of the common underslying dynamics of
such scenarios which is fully testable at the LHC.

Instrumental to the accomplishement of the complete investigation of the 4DCHM at the CERN machine and 
as by-product of our study, we have produced and validated several computing modules, enabling the automated implementation and generation of the 4DCHM in the context of public 
tools such as LanHEP and CalcHEP.
We have also generated a Mathematica code allowing one to perform said theoretical and experimental tests efficiently. Finally, the 4DCHM files are being uploaded onto  the HEPMDB (https://hepmdb.soton.ac.uk/) for public use.

\section*{Acknowledgments}
DB and SM are financed in part through the NExT Institute. The work of GMP has been supported by the German Research Foundation DFG through Grant No.\ STO876/2-1 and by BMBF Grant No.\ 05H09ODE, moreover he acknowledges financial support and hospitality by the University of Southampton.
We would all like to thank M. Redi and A. Tesi for sharing with us the algorithm that computes the Higgs' VEV and the Higgs' mass and for continuous advice and encouragement. We finally thank also A. Semenov as well as A. Pukhov for the help given in solving some problems that we have found during the implementation of the model in LanHEP and/or CalcHEP.

\newpage
\appendix
\section{Appendix A: Analytical expressions}
\subsection{Masses of the gauge bosons}
At the leading order in $\xi=v^2/f^2$,  where $v$ is the VEV of the Higgs, the analytical expressions of the neutral gauge boson masses are given by
\begin{eqnarray}\label{eq:MZ}
M^2_{\gamma}= &&0,\nonumber\\
M^2_{Z}\simeq && \frac{f ^2}{4} g_*^2(s^2_\theta+\frac{s^2_\psi}{2})  \xi,\nonumber\\
M^2_{Z_1}=&& f ^2g_*^2,\nonumber\\
M^2_{ Z_2}\simeq&& \frac{f ^2g_*^2}{ c_\psi^2}(1-\frac{s^2_\psi c^4_\psi }{4 c_{2\psi}}\xi),\nonumber\\
M^2_{Z_3}\simeq&& \frac{f ^2g_*^2}{ c_\theta^2}(1-\frac{s^2_\theta c^4_\theta }{4 c_{2\theta}}\xi), \nonumber\\
M^2_{Z_4}=&&2 f ^2g_*^2, \nonumber\\
M^2_{Z_5}\simeq&& 2 f ^2g_*^2(1+\frac 1 {16} (\frac 1 {c_{2\theta}}+\frac 1{2 c_{2 \psi}})\xi)
\end{eqnarray}
with $\tan\theta=(s_\theta/c_\theta)=(g_0/g_*)$ and  $\tan\psi=(s_\psi/c_\psi)=(\sqrt{2} g_{0Y}/g_*)$. 

From these expressions we note that $Z_1$  and $Z_4$ have their masses completely determined by the composite sector and they do not receive any contribution from EWSB.

For the charged sector we have:
\begin{eqnarray}
\label{eq:MW}
M^2_{W}\simeq && \frac{f ^2}{4} g_*^2s^2_\theta  \xi,\nonumber\\
M^2_{W_1}=&& f ^2g_*^2,\nonumber\\
M^2_{ W_2}\simeq&& \frac{f ^2g_*^2}{ c_\theta^2}(1-\frac{s^2_\theta c^4_\theta}{2 c_{2\theta}}\xi),\nonumber\\
M^2_{W_3}\simeq&&2 f ^2g_*^2 (1- \frac{s^2_\theta}{4c_{2\theta}}\xi).
\end{eqnarray}
In the same way as $Z_1$ and $Z_4$ do, also $W_1$ does not receive any mass correction from  EWSB.

\subsection{Masses of the fermions}
The expressions of the top and bottom mass at the leading order in $\xi$ are given by
\begin{equation}
\label{fermionmass-a}
\begin{split}
m_b^2 &\simeq\xi  \frac{m_*^2} 2   \frac{ \tilde \Delta_{b_L}^2 \tilde \Delta_{b_R}^2 \tilde Y_B^2}{(1+F_L) }, \\
m_t^2 &\simeq \xi \frac{m_*^2} 2   \frac{ \tilde \Delta_{t_L} ^2\tilde \Delta_{t_R} ^2\tilde Y_T^2}{(1+F_L)(1+F_R) },\\
\end{split}
\end{equation}
where we have defined $\tilde \Delta_{t_L,t_R,b_L,b_R}=\Delta_{t_L,t_R,b_L,b_R}/m_*$, $\tilde Y_{T,B}= Y_{T,B}/m_*$,  $\tilde M_{Y_{T,B}}=M_{Y_{T,B}}/m_*$, 
\begin{equation}
\label{FLR}
F_L=\tilde\Delta_{t_L}^2(1+\tilde M^2_{Y_T}),\quad\quad F_R=\tilde\Delta_{t_R}^2(1+(\tilde M_{Y_T}+\tilde Y_T)^2)
\end{equation}
and, for simplicity, we have taken $ \Delta_{b_L}=\Delta_{b_R}=0$ except in the bottom mass expression.

At order $\xi=0$ the masses of the lightest fermionic resonances are given by
\begin{equation}
\label{fermionmass-b}
\begin{split}
m^2_{T_1} &\simeq\frac{m_*^2}{2} \left(2 +\tilde{M}_{Y_T}^2-\tilde{M}_{Y_T}\sqrt{4 +\tilde{M}^2_{Y_T}}\right)=m^2_{\tilde T_1} ,\quad |M_{Y_T}|>|M_{Y_B}|\\
m^2_{B_1} &\simeq\frac{m_*^2}{2} \left(2 +\tilde{M}_{Y_B}^2-\tilde{M}_{Y_B}\sqrt{4 +\tilde{M}^2_{Y_B}}\right)=m^2_{\tilde B_1},
\end{split}
\end{equation}
where again we have taken $ \Delta_{b_L}=\Delta_{b_R}=0$.

\subsection{Couplings of the gauge bosons to the fermions}
We give here the analytical expressions of the couplings of both the neutral and charged gauge bosons to the three generations of leptons and the first two generations of quarks at the leading order in  $\xi$.

Starting from the elementary sector, where the neutral gauge fields of $SU(2)_L\times U(1)_Y$ are coupled with the light fermionic currents, we get, after taking into account the mixing among the fields, the following expression for the neutral current Lagrangian:
\begin{equation}\label{LNC}
{\cal L}_{NC}=\sum_f\big[ e\bar\psi^f \gamma_\mu Q^f \psi^f A^\mu+ \sum_{i=0}^5   (\bar\psi^f_L  g_{Z_i}^L(f) \gamma_\mu  \psi^f_L+\bar\psi^f_R  g_{Z_i}^R(f) \gamma_\mu  \psi^f_R ) Z_i^\mu \big], 
\end{equation}
where $\psi_{L,R}=[(1\pm\gamma_5)/2]\psi$ and the label $i=0$ corresponds to the neutral SM gauge boson $Z$. The photon field, $A_\mu$, is coupled to the electromagnetic current in the standard way, i.e., with
\begin{equation}\label{e}
e=\frac{g_L g_Y}{\sqrt{g_L^2+g_Y^2}},\quad\quad g_L=g_0 c_\theta,\quad \quad g_Y=g_{0Y} c_\psi ,
\end{equation}
while the couplings of the $Z_i$'s have the following expressions:
\begin{equation}
g_{Zi}^L(f)= A_{Z_i}T^3_L(f)+ B_{Z_i} Q^f, \quad\quad
g_{Zi}^R(f)=  B_{Z_i}Q^f,
\end{equation}
where $A_{Z_i}=(g_0 \alpha_i - g_{0Y} \beta_i) $, $B_{Z_i}=g_{0Y} \beta_i$,  
with $\alpha_i$ and $\beta_i$ being the diagonalisation matrix elements, namely: 
\begin{equation}
W_3=\sum_{i=0}^5 \alpha_i Z_i, \quad\quad  Y=\sum_{i=0}^5 \beta_i Z_i.
\end{equation}
Here $W_3$ and  $Y$ are the elementary gauge field associated to $SU(2)_L$ and $U(1)_Y$, respectively.
At the leading order in $\xi$ we get:
\begin{eqnarray}
&&A_{Z}= \sqrt{g_L^2+g_Y^2}\big[1+(\frac{g_L^2}{g_L^2+g_Y^2} a_Z+ \frac{g_Y^2}{g_L^2+g_Y^2} b_Z)\xi\big],~
B_{Z}= -\frac{g_Y^2}{\sqrt{g_L^2+g_Y^2}}(1+b_Z\xi), \\
&&A_{Z_2}=  -g_Y \frac{s_\psi}{c_\psi} \Big[1+(\frac{g_L}{g_Y} a_{Z_2}-b_{Z_2})\xi\Big], \quad\quad  B_{Z_2}= g_Y \frac{s_\psi}{c_\psi} \Big[1-b_{Z_2}\xi\Big],\\
&&A_{Z_3}=  -g_L\frac{s_\theta}{c_\theta}\big[1+(a_{Z_3}+\frac{g_Y}{g_L} b_{Z_3})\xi\big], \quad \quad   B_{Z_3}=   g_Y \frac{s_\theta}{c_\theta} b_{Z_3}\xi,\\
&&A_{Z_5}= (g_L a_{Z_5}-g_Y b_{Z_5})\sqrt{\xi},\quad\quad   B_{Z_5}=  g_Y b_{Z_5}\sqrt{\xi},
\end{eqnarray}
with
\begin{eqnarray}
a_{Z}= && (2 s_\theta^2+s_\psi^2)(4 cˆ_\theta^2-1)/32,\quad \quad
b_{Z}=  (2 s_\theta^2+s_\psi^2)(4 cˆ_\psi^2-1)/32, \\
a_{Z_2}=&& \frac{\sqrt{2} s_\theta s_\psi c_\psi^6}{4(c_\psi^2-c_\theta^2)(2 c_\psi^2-1)} ,     \quad \quad
b_{Z_2}= \frac{c_\psi^4(2-7c_\psi^2+9 c_\psi^4-4 c_\psi^6)}{8s_\psi^2 (1-2 c_\psi^2)^2},\\
a_{Z_3}= &&  \frac{-2 c_\theta^4+5 c_\theta^6-4 c_\theta^8}{4(1-2 c_\theta^2)^2},\quad \quad
b_{Z_3}=  \frac{\sqrt{2} s_\theta s_\psi c_\theta^6}{4 (2 c_\theta^2-1)(c_\theta^2-c_\psi^2) },\\
a_{Z_5}= &&  \frac{s_\theta}{2 \sqrt{2}(1-2 c_\theta^2)},\quad \quad
b_{Z_5}= - \frac{s_\psi}{4(1-2 c_\psi^2)}.
\end{eqnarray}
In the same way we can work out the expressions for the charged sector that are:
\begin{equation}
{\cal L}_{CC}=[ g_W^+  W^+ +g_{W_1}^+ W_1^++ g_{W_2}^+ W_2^++ g_{W_3}^+ W_3^+ ]J^- +h.c.
\label{LCC}
\end{equation}
with $J^\pm=(J^1\pm i J^2)/2$, $ J^i_\mu=\bar\psi T^i_L \gamma_\mu[(1-\gamma_5)/2]\psi$,  and

\begin{eqnarray}
g_W^\pm=&&-\frac{g_* s_\theta}{\sqrt 2}(1+\frac  {s_\theta}{4 c_\theta}  a_{12}  \xi)\label{Wff},\\
g_{W_1}^\pm= &&0\label{W1ff},\\
g_{W_2}^\pm= &&\frac{g_* s_\theta^2}{\sqrt{2}c_\theta}    (1+\frac 1 4 (a_{22}-\frac{ c_\theta}{s_\theta}a_{12})\xi )\label{W2ff},\\
 g_{W_3}^\pm= &&\frac{g_* s_\theta^2}{2\sqrt{2}c_\theta} a_{24}\sqrt{\xi},
\label{W3ff}
\end{eqnarray}
where 
\begin{equation}
a_{12}= -\frac 1 4 c_\theta (1-4 c_\theta^2) s_\theta,  \quad \quad
a_{22}=-\frac{ c_\theta^2}{4(1-2 c_\theta^2)^2}, \quad \quad
a_{24}=-\frac{c_\theta }{\sqrt{2}(1-2c_\theta^2)}.
\end{equation}

Because of the non-universality of the couplings of the gauge sectors to the three generations of quarks the expressions of the coupling of the neutral and charged gauge bosons to the top and the bottom quark are different with respect to the first two generations of quarks.
As an example we show the couplings of the neutral gauge bosons to the top quark and, as the expressions turn out to be quite complicated even at the leading order in $\xi$, we only show the terms originating from the elementary-composite mixing before EWSB ($\xi=0$).
After taking into account the mixing among the gauge and fermionic fields the neutral current Lagrangian is the following:

\begin{equation}
\mathcal{L}_{NC}^{t}=\frac{2}{3} e \bar \psi^t \gamma_{\mu} \psi^t A^{\mu}+\sum_{i=0}^5(g^L_{Z_i}(t) \bar\psi^t_L\gamma_{\mu}\psi^t_L+g^R_{Z_i}(t)\bar\psi^t_R\gamma^{\mu}\psi^t_R)Z_i^{\mu}
\label{eq:lag_nc_top}
\end{equation}

where

\begin{equation}
\begin{split}
g_{Z_0}^L(t)=\frac{e}{s_\omega c_\omega}(\frac 1 2 -\frac 2 3 s^2_\omega)+{\cal O}(\xi),\quad 
&g_{Z_0}^R(t)=\frac{e}{s_\omega c_\omega}( -\frac 2 3 s^2_\omega)+{\cal O}(\xi),\\
g_{Z_1}^L(t)\sim{\cal O}(\xi),\quad \quad\quad\quad\quad&g_{Z_1}^R(t)\sim{\cal O}(\xi)\\
g_{Z_2}^L(t)=\frac{e}{6 c_\omega}\frac{s_\psi}{c_\psi}\frac{1}{(1+F_L)}(1-\frac{c_\psi^2}{s_\psi^2} F_L)+{\cal O}(\xi),\quad &g_{Z_2}^R(t)=\frac{2 e}{3 c_\omega}\frac{s_\psi}{c_\psi}\frac{1}{(1+F_R)}(1-\frac{c_\psi^2}{s_\psi^2} F_R)+{\cal O}(\xi),\\
g_{Z_3}^L(t)=-\frac{e}{2 s_\omega}\frac{s_\theta}{c_\theta}\frac{1}{(1+F_L)}(1-\frac{c_\theta^2}{s_\theta^2} F_L)+{\cal O}(\xi),\quad &g_{Z_3}^R(t)\sim{\cal O}(\xi),\\
g_{Z_4}^L(t)\sim g_{Z_4}^R(t)=0,\quad\quad    &g_{Z_5}^L(t)\sim g_{Z_5}^R(t)\sim{\cal O}(\sqrt{\xi}),
\end{split}
\end{equation}
with
\begin{equation}
\tan\omega=\frac{g_Y}{g_L},\quad\quad e=g_L s_\omega=g_Y c_\omega,\quad\quad  \frac e{s_\omega c_\omega}= \sqrt{g_L^2+g_Y^2}. 
\end{equation}

\bibliographystyle{h-physrev5}
\newpage
\bibliography{biblio}

\begin{thebibliography}{10}

\bibitem{Campbell:2006wx}
J.~M. Campbell, J.~Huston, and W.~Stirling,
\newblock Rept. Prog. Phys. {\bf 70}, 89 (2007), arXiv:hep-ph/0611148.

\bibitem{Cacciapaglia:2006pk}
G.~Cacciapaglia, C.~Csaki, G.~Marandella, and A.~Strumia,
\newblock Phys. Rev. {\bf D74}, 033011 (2006), arXiv:hep-ph/0604111.

\bibitem{Langacker:2008yv}
P.~Langacker,
\newblock Rev. Mod. Phys. {\bf 81}, 1199 (2009), arXiv:hep-ph/0801.1345.

\bibitem{Salvioni:2009mt}
E.~Salvioni, G.~Villadoro, and F.~Zwirner,
\newblock JHEP {\bf 11}, 068 (2009), arXiv:hep-ph/0909.1320.

\bibitem{Accomando:2010fz}
E.~Accomando, A.~Belyaev, L.~Fedeli, S.~F. King, and
  C.~Shepherd-Themistocleous,
\newblock Phys. Rev. {\bf D83}, 075012 (2011), arXiv:1010.6058.

\bibitem{Basso:2011na}
L.~Basso, S.~Moretti, and G.~M. Pruna,
\newblock JHEP {\bf 1108}, 122 (2011), arXiv:1106.4762.

\bibitem{Athron:2011wu}
P.~Athron, S.~King, D.~Miller, S.~Moretti, and R.~Nevzorov,
\newblock Phys. Rev. {\bf D84}, 055006 (2011), arXiv:1102.4363.

\bibitem{Athron:2012sq}
P.~Athron, S.~King, D.~Miller, S.~Moretti, and R.~Nevzorov,
\newblock (2012), arXiv:1206.5028.

\bibitem{Athron:2012pw}
P.~Athron, D.~Stockinger, and A.~Voigt,
\newblock (2012), arXiv:1209.1470.

\bibitem{Basso:2012gz}
L.~Basso, B.~O'Leary, W.~Porod, and F.~Staub,
\newblock (2012), arXiv:1207.0507.

\bibitem{O'Leary:2011yq}
B.~O'Leary, W.~Porod, and F.~Staub,
\newblock JHEP {\bf 1205}, 042 (2012), arXiv:1112.4600.

\bibitem{Hirsch:2012kv}
M.~Hirsch, W.~Porod, L.~Reichert, and F.~Staub,
\newblock (2012), arXiv:1206.3516.

\bibitem{Schmaltz:2005ky}
M.~Schmaltz and D.~Tucker-Smith,
\newblock Ann. Rev. Nucl. Part. Sci. {\bf 55}, 229 (2005),
  arXiv:hep-ph/0502182.

\bibitem{Perelstein:2005ka}
M.~Perelstein,
\newblock Prog. Part. Nucl. Phys. {\bf 58}, 247 (2007), arXiv:hep-ph/0512128.

\bibitem{Cheng:2007bu}
H.-C. Cheng,
\newblock (2007), arXiv:0710.3407.

\bibitem{Contino:2003ve}
R.~Contino, Y.~Nomura, and A.~Pomarol,
\newblock Nucl. Phys. {\bf B671}, 148 (2003), arXiv:hep-ph/0306259.

\bibitem{Agashe:2004rs}
K.~Agashe, R.~Contino, and A.~Pomarol,
\newblock Nucl. Phys. {\bf B719}, 165 (2005), arXiv:hep-ph/0412089.

\bibitem{Contino:2010rs}
R.~Contino,
\newblock (2010), arXiv:1005.4269.

\bibitem{DeCurtis:2011yx}
S.~De~Curtis, M.~Redi, and A.~Tesi,
\newblock JHEP {\bf 1204}, 042 (2012), arXiv:1110.1613.

\bibitem{Kaplan:1983fs}
D.~B. Kaplan and H.~Georgi,
\newblock Phys.Lett. {\bf B136}, 183 (1984).

\bibitem{Georgi:1984ef}
H.~Georgi, D.~B. Kaplan, and P.~Galison,
\newblock Phys.Lett. {\bf B143}, 152 (1984).

\bibitem{Georgi:1984af}
H.~Georgi and D.~B. Kaplan,
\newblock Phys.Lett. {\bf B145}, 216 (1984).

\bibitem{Dugan:1984hq}
M.~J. Dugan, H.~Georgi, and D.~B. Kaplan,
\newblock Nucl.Phys. {\bf B254}, 299 (1985).

\bibitem{:2012gk}
ATLAS Collaboration, G.~Aad {\em et~al.},
\newblock Phys. Lett. {\bf B716}, 1 (2012), arXiv:1207.7214.

\bibitem{:2012gu}
CMS Collaboration, S.~Chatrchyan {\em et~al.},
\newblock Phys. Lett. {\bf B716}, 30 (2012), arXiv:1207.7235.

\bibitem{Agashe:2007ki}
K.~Agashe {\em et~al.},
\newblock Phys. Rev. {\bf D76}, 115015 (2007), arXiv:0709.0007.

\bibitem{Agashe:2008jb}
K.~Agashe, S.~Gopalakrishna, T.~Han, G.-Y. Huang, and A.~Soni,
\newblock Phys. Rev. {\bf D80}, 075007 (2009), arXiv:0810.1497.

\bibitem{Panico:2011pw}
G.~Panico and A.~Wulzer,
\newblock JHEP {\bf 1109}, 135 (2011), arXiv:1106.2719.

\bibitem{Contino:2006nn}
R.~Contino, T.~Kramer, M.~Son, and R.~Sundrum,
\newblock JHEP {\bf 0705}, 074 (2007), arXiv:hep-ph/0612180.

\bibitem{Contino:2006qr}
R.~Contino, L.~Da~Rold, and A.~Pomarol,
\newblock Phys. Rev. {\bf D75}, 055014 (2007), arXiv:hep-ph/0612048.

\bibitem{Redi:2012ha}
M.~Redi and A.~Tesi,
\newblock (2012), arXiv:1205.0232.

\bibitem{Semenov:2010qt}
A.~Semenov,
\newblock (2010), arXiv:1005.1909.

\bibitem{Pukhov:1999gg}
A.~Pukhov {\em et~al.},
\newblock (1999), arXiv:hep-ph/9908288.

\bibitem{Belyaev:2012qa}
A.~Belyaev, N.~D. Christensen, and A.~Pukhov,
\newblock (2012), arXiv:1207.6082.

\bibitem{Belanger:2010st}
G.~Belanger, N.~D. Christensen, A.~Pukhov, and A.~Semenov,
\newblock Comput. Phys. Commun. {\bf 182}, 763 (2011), arXiv:1008.0181.

\bibitem{Brooijmans:2012yi}
M.~Bondarenko {\em et~al.},
\newblock (2012), in arXiv:1203.1488.

\bibitem{mathematica}
{\em {Mathematica Edition: Version 7.0}} (Wolfram Research, Inc., Champaign,
  Illinois, 2008).

\bibitem{Aaltonen:2012qt}
CDF Collaboration, D0 Collaboration, T.~Aaltonen {\em et~al.},
\newblock Phys. Rev. Lett. {\bf 109}, 071804 (2012), arXiv:1207.6436.

\bibitem{Murayama:1992gi}
H.~Murayama, I.~Watanabe, and K.~Hagiwara,
\newblock {\em {HELAS: HELicity amplitude subroutines for Feynman diagram
  evaluations}}, 1992.

\bibitem{Stelzer:1994ta}
T.~Stelzer and W.~Long,
\newblock Comput. Phys. Commun. {\bf 81}, 357 (1994), arXiv:hep-ph/9401258.

\bibitem{Kharraziha:1999iw}
H.~Kharraziha and S.~Moretti,
\newblock Comput. Phys. Commun. {\bf 127}, 242 (2000), arXiv:hep-ph/9909313.

\bibitem{Kleiss:1985gy}
R.~Kleiss, W.~J. Stirling, and S.~Ellis,
\newblock Comput. Phys. Commun. {\bf 40}, 359 (1986).

\bibitem{Lepage:1977sw}
G.~P. Lepage,
\newblock J. Comput. Phys. {\bf 27}, 192 (1978).

\bibitem{Lepage:1980dq}
G.~P. Lepage,
\newblock {\em {VEGAS: AN ADAPTIVE MULTIDIMENSIONAL INTEGRATION PROGRAM}},
  1980.

\bibitem{Adam:2008pc}
N.~E. Adam, V.~Halyo, S.~A. Yost, and W.~Zhu,
\newblock JHEP {\bf 0809}, 133 (2008), arXiv:0808.0758.

\bibitem{Adam:2008ge}
N.~E. Adam, V.~Halyo, and S.~A. Yost,
\newblock JHEP {\bf 0805}, 062 (2008), arXiv:0802.3251.

\bibitem{Balossini:2007zzb}
G.~Balossini {\em et~al.},
\newblock Acta Phys.Polon. {\bf B38}, 3407 (2007).

\bibitem{Balossini:2009sa}
G.~Balossini {\em et~al.},
\newblock JHEP {\bf 1001}, 013 (2010), arXiv:0907.0276.

\bibitem{Lai:1999wy}
CTEQ Collaboration, H.~Lai {\em et~al.},
\newblock Eur. Phys. J. {\bf C12}, 375 (2000), arXiv:hep-ph/9903282.

\bibitem{Ball:2011uy}
NNPDF Collaboration, R.~D. Ball {\em et~al.},
\newblock Nucl.Phys. {\bf B855}, 153 (2012), arXiv:1107.2652.

\bibitem{Aad:2011fe}
ATLAS Collaboration, G.~Aad {\em et~al.},
\newblock Phys. Lett. {\bf B701}, 50 (2011), arXiv:1103.1391.

\bibitem{Chatrchyan:2012qk}
CMS Collaboration, S.~Chatrchyan {\em et~al.},
\newblock (2012), arXiv:1204.4764.

\bibitem{Hayden:2012gc}
ATLAS Collaboration, D.~Hayden {\em et~al.},
\newblock EPJ Web Conf. {\bf 28}, 12003 (2012), arXiv:1201.4721.

\bibitem{Chatrchyan:2012it}
CMS Collaboration, S.~Chatrchyan {\em et~al.},
\newblock (2012), arXiv:1206.1849.

\bibitem{925327}
CMS Collaboration,
\newblock CMS-PAS-EXO-08-009  (2009).

\bibitem{925338}
CMS Collaboration,
\newblock CMS-PAS-EXO-09-012  (2009).

\bibitem{Aad:2012bb}
ATLAS Collaboration, G.~Aad {\em et~al.},
\newblock JHEP {\bf 1204}, 069 (2012), arXiv:1202.5520.

\bibitem{ATLAS:2012aw}
ATLAS Collaboration, G.~Aad {\em et~al.},
\newblock Phys. Rev. Lett. {\bf 109}, 032001 (2012), arXiv:1202.6540.

\bibitem{Marzocca:2012zn}
D.~Marzocca, M.~Serone, and J.~Shu,
\newblock JHEP {\bf 1208}, 013 (2012), arXiv:1205.0770.

\bibitem{Matsedonskyi:2012ym}
O.~Matsedonskyi, G.~Panico, and A.~Wulzer,
\newblock (2012), arXiv:1204.6333.

\bibitem{Gianotti:2002xx}
F.~Gianotti {\em et~al.},
\newblock Eur. Phys. J. {\bf C39}, 293 (2005), arXiv:hep-ph/0204087.

\bibitem{Baur:2001ze}
U.~Baur, O.~Brein, W.~Hollik, C.~Schappacher, and D.~Wackeroth,
\newblock Phys.Rev. {\bf D65}, 033007 (2002), arXiv:hep-ph/0108274.

\bibitem{Collins:1977iv}
J.~C. Collins and D.~E. Soper,
\newblock Phys.Rev. {\bf D16}, 2219 (1977).

\end{thebibliography}

\end{document}